\documentclass[twocolumn,twocolappendix]{aastex701}    
 
\hypersetup{linkcolor=red,citecolor=blue,urlcolor=magenta}

\usepackage{rotating}
\usepackage{booktabs}
\usepackage{graphicx}
\graphicspath{{../}}
\usepackage{amsmath}

\usepackage{newtxtext,newtxmath} 
\usepackage{comment}
\usepackage[T1]{fontenc}

\NewDocumentCommand\michele{O{Rewritten}m}{\textsf{\color{green}[Michele: #1]\(\rightarrow\)[``\emph{#2}'']}}

\begin{document}

\title{\Large The Structural Anatomy of Galaxies: Environmental Dependence of Morphological and Quenching Evolution at $0.1 < z < 0.5$}

\author[0000-0002-8437-6659]{Can Xu}
\affiliation{School of Astronomy and Space Science, Nanjing University, Nanjing 210093, China}
\affiliation{Key Laboratory of Modern Astronomy and Astrophysics, Nanjing University, Nanjing 210093, China}
\affiliation{Kavli Institute for the Physics and Mathematics of the Universe (Kavli IPMU, WPI), UTIAS, Tokyo Institutes for Advanced Study, University of Tokyo, Chiba, 277-8583, Japan}
\affiliation{Center for Data-Driven Discovery, Kavli IPMU (WPI), UTIAS, The University of Tokyo, Kashiwa, Chiba 277-8583, Japan}
\email[show]{canxu@smail.nju.edu.cn}

\author[0000-0002-0000-6977]{John D. Silverman}
\affiliation{Kavli Institute for the Physics and Mathematics of the Universe (Kavli IPMU, WPI), UTIAS, Tokyo Institutes for Advanced Study, University of Tokyo, Chiba, 277-8583, Japan}
\affiliation{Center for Data-Driven Discovery, Kavli IPMU (WPI), UTIAS, The University of Tokyo, Kashiwa, Chiba 277-8583, Japan}
\affiliation{Department of Astronomy, Graduate School of Science, The University of Tokyo, 7-3-1 Hongo, Bunkyo, Tokyo 113-0033, Japan}
\affiliation{Center for Astrophysical Sciences, Department of Physics \& Astronomy, Johns Hopkins University, Baltimore, MD 21218, USA}
\email{john.silverman@ipmu.jp}

\author{Qiusheng Gu}
\affiliation{School of Astronomy and Space Science, Nanjing University, Nanjing 210093, China}
\affiliation{Key Laboratory of Modern Astronomy and Astrophysics, Nanjing University, Nanjing 210093, China}
\email[show]{qsgu@nju.edu.cn}

\author[]{Daniela Galárraga-Espinosa}
\affiliation{Kavli Institute for the Physics and Mathematics of the Universe (Kavli IPMU, WPI), UTIAS, Tokyo Institutes for Advanced Study, University of Tokyo, Chiba, 277-8583, Japan}
\affiliation{Center for Data-Driven Discovery, Kavli IPMU (WPI), UTIAS, The University of Tokyo, Kashiwa, Chiba 277-8583, Japan}
\email{daniela.galarraga@ipmu.jp}

\author[]{Andreas L Faisst}
\affiliation{Caltech/IPAC, 1200 E. California Boulevard, Pasadena, CA 91125, USA}
\email{afaisst@ipac.caltech.edu}

\author{Emanuele Daddi}
\affiliation{Université Paris-Saclay, Université Paris Cité, CEA, CNRS, AIM, 91191, Gif-sur-Yvette, France}
\email{emanuele.daddi@cea.fr}

\author{Hassen M. Yesuf}
\email{yesufh@shao.ac.cn}
\affiliation{Shanghai Astronomical Observatory, Chinese Academy of Sciences, 80 Nandan Road, Shanghai 200030, People's Republic of China}
\email{}

\author{Angelo George}
\email{ageorge@asiaa.sinica.edu.tw}
\affiliation{Academia Sinica Institute of Astronomy and Astrophysics, AS/NTU Astronomy-Mathematics Building, No. 1, Sec. 4, Roosevelt Road, Taipei 10617, Taiwan}
\email{}

\author{Esra Bulbul}
\email{ebulbul@mpe.mpg.de}
\affiliation{Max-Planck-Institut f\"ur extraterrestrische Physik, Gie\ss enbachstra\ss e 1, 85748 Garching, Germany}
\email{}

\author[0000-0002-0182-1973]{Zeyu Gao}
\affiliation{Kavli Institute for Astronomy and Astrophysics, Peking University, Beijing 100871, People's Republic of China}
\affiliation{Kavli Institute for the Physics and Mathematics of the Universe (Kavli IPMU, WPI), UTIAS, Tokyo Institutes for Advanced Study, University of Tokyo, Chiba, 277-8583, Japan}
\email{zygao@stu.pku.edu.cn}

\author[0000-0003-4032-2445]{Lalitwadee Kawinwanichakij}
\email{lkawinwanichakij@swin.edu.au}
\affiliation{Centre for Astrophysics and Supercomputing, Swinburne University of Technology, Hawthorn, VIC 3122, Australia}

\author[0009-0003-4742-7060]{Takumi S. Tanaka}
\affiliation{Kavli Institute for the Physics and Mathematics of the Universe (Kavli IPMU, WPI), UTIAS, Tokyo Institutes for Advanced Study, University of Tokyo, Chiba, 277-8583, Japan}
\affiliation{Department of Astronomy, Graduate School of Science, The University of Tokyo, 7-3-1 Hongo, Bunkyo-ku, Tokyo, 113-0033, Japan}
\affiliation{Center for Data-Driven Discovery, Kavli IPMU (WPI), UTIAS, The University of Tokyo, Kashiwa, Chiba 277-8583, Japan}
\email{}

\author[0000-0003-0111-8249]{Yunjing Wu}
\affiliation{Kavli Institute for the Physics and Mathematics of the Universe (Kavli IPMU, WPI), UTIAS, Tokyo Institutes for Advanced Study, University of Tokyo, Chiba, 277-8583, Japan}
\affiliation{Center for Data-Driven Discovery, Kavli IPMU (WPI), UTIAS,
The University of Tokyo, Kashiwa, Chiba 277-8583, Japan}
\email{yunjing.wu@ipmu.jp}

\author[0009-0001-9170-3363]{Jialai Wang}
\affiliation{Department of Astronomy, University of Science and Technology of China, Hefei 230026, China}
\affiliation{School of Astronomy and Space Science, University of Science and Technology of China, Hefei 230026, China}
\affiliation{Kavli Institute for the Physics and Mathematics of the Universe (Kavli IPMU, WPI), UTIAS, Tokyo Institutes for Advanced Study, University of Tokyo, Chiba, 277-8583, Japan}
\affiliation{Center for Data-Driven Discovery, Kavli IPMU (WPI), UTIAS, The University of Tokyo, Kashiwa, Chiba 277-8583, Japan}
\email{}

\author[0009-0005-4704-8401]{Michiru Maruyama}
\affiliation{Kavli Institute for the Physics and Mathematics of the Universe (Kavli IPMU, WPI), UTIAS, Tokyo Institutes for Advanced Study, University of Tokyo, Chiba, 277-8583, Japan}
\affiliation{Department of Astronomy, Graduate School of Science, The University of Tokyo, 7-3-1 Hongo, Bunkyo-ku, Tokyo, 113-0033, Japan}
\email{michiru.maruyama@ipmu.jp}

\makeatletter
\gdef\@@email{qsgu@nju.edu.cn, canxu@smail.nju.edu.cn}
\makeatother

\makeatletter
\gdef\@@email{%
  qsgu@nju.edu.cn,canxu@smail.nju.edu.cn%
}
\makeatother

\begin{abstract}
We investigate how cluster environments affect galaxy star formation, structure, and stellar populations using spectroscopically-confirmed galaxies in X-ray-selected eROSITA clusters, that have deep HSC-SSP imaging and DESI DR1 spectroscopy. At $0.1<z<0.5$, we compare dust-corrected H$\alpha$ star formation rates (SFRs), effective radii ($R_{\rm e}$), bulge-to-total light ratios ($B/T$), and $D_{\rm n}4000$ between cluster satellites and a stellar-mass- and redshift-matched field control sample. We identify UVJ-quiescent galaxies and divide the UVJ star-forming population at $\mathrm{sSFR}_{\mathrm{H}\alpha}=10^{-10.5}\,\mathrm{yr}^{-1}$ into actively star-forming and H$\alpha$-low systems, the latter referred to as green-valley (GV) galaxies. Cluster--field differences depend strongly on stellar mass, projected cluster-centric radius, and galaxy population. Cluster star-forming galaxies show modest SFR suppression but weak structural offsets. The strongest coupled signatures occur among H$\alpha$-low UVJ-star-forming galaxies, particularly low-mass satellites at $R<0.5R_{500}$, which have lower SFRs, smaller light-weighted sizes, larger $B/T$, and larger $D_{\rm n}4000$ than matched field counterparts. Quiescent galaxies show weaker structural offsets but retain a positive $D_{\rm n}4000$ residual. We quantify the joint signal using a Mahalanobis distance based on matched residuals in SFR, $R_{\rm e}$, $B/T$, and $D_{\rm n}4000$ with a field--field reference covariance. The multivariate offset is largest for low-mass GV galaxies and decreases toward larger projected radius. The broader radial extent of SFR suppression relative to structural offsets suggests that star-formation suppression has a broader environmental reach than structural transformation. These results show that environmental quenching and structural transformation are most strongly coupled in the H$\alpha$-low transition population.
\end{abstract}


\section{Introduction}
Understanding how environment regulates galaxy evolution is a central problem in extragalactic astronomy. Galaxies are embedded within the large-scale cosmic web, a network of clusters, filaments, and voids \citep{Bond1996}, and their star-formation activity, stellar populations, and morphology depend systematically on the surrounding environment \citep{Dressler1980,BlantonMoustakas2009}. Quantifying these environmental dependencies is therefore essential for disentangling the interplay between internal secular evolution and externally driven processes.

The galaxy population exhibits a well-known bimodality, separating into a ``red sequence'' dominated by quiescent, bulge-dominated galaxies (QGs) and a ``blue cloud'' populated primarily by star-forming, disk-dominated galaxies (SFGs) \citep{Strateva2001,Baldry2004}. This dichotomy is closely linked to quenching processes that suppress star formation and move galaxies away from the star-forming sequence. Environmental mechanisms are widely regarded as important contributors to this transformation, particularly in high-density regions \citep{Peng2010,Wetzel2012,vanDerBurg2018}.

Galaxy clusters, which occupy the densest nodes of the cosmic web, provide powerful laboratories for studying environmental influences on galaxy evolution. A variety of physical processes operate within clusters, including ram-pressure stripping by the hot intracluster medium (ICM; \citealt{GunnGott1972}; \citealt{BoselliGavazzi2006}), high-velocity gravitational encounters commonly referred to as harassment \citep{Moore1996}, and tidal interactions with other galaxies and the cluster potential \citep{Merritt1983}. These mechanisms can remove or redistribute gas, suppress star formation, and alter galaxy structure and morphology \citep[e.g.,][]{Boselli2022}. Their efficiency depends on galaxy properties and orbital history, including stellar mass, infall time, and trajectory through the host halo, naturally producing a diversity of evolutionary responses within cluster populations \citep{Jaffe2015,OmanHudson2016}.

Because environmental processes can affect star formation, stellar populations, and galaxy structure on different timescales, their signatures need not emerge simultaneously or with equal strength across different observables. A joint analysis of multiple galaxy properties and scaling relations therefore provides a powerful framework for characterizing environmental effects and for distinguishing changes in star-formation activity from structural transformation. The stellar mass--size ($M_*{-}R_{\rm e}$) relation is a particularly important diagnostic of galaxy structural evolution. Star-forming galaxies and quiescent galaxies occupy distinct mass--size sequences, both of which evolve with redshift, with especially strong evolution among massive systems. At $z\gtrsim1$--2, massive galaxies are substantially more compact than their present-day counterparts \citep{Shen2003,Daddi2005,Trujillo2006,vanDerWel2014,Lange2015}.

However, whether environment systematically alters the stellar mass--size relation remains unsettled. Some studies find little or no residual size dependence after controlling for stellar mass, morphology, or star-formation state \citep{HuertasCompany2013,Kelkar2015,Kuchner2017}, whereas others report significant environmental size differences \citep{Valentinuzzi2010,Poggianti2013,Raichoor2012,Delaye2014,George2025}. The sign and strength of these differences are not uniform across the literature. At higher redshift, some studies of massive galaxies report larger sizes in dense environments, potentially associated with accelerated merger-driven evolution \citep{Papovich2012,Lani2013}, while other studies find smaller galaxies or weaker environmental differences. These apparently conflicting results span different redshifts, stellar-mass ranges, population selections, membership criteria, environmental definitions, and structural measurements, making direct comparison difficult.

Environmental structural effects may also depend strongly on stellar mass. Several studies suggest that the magnitude of cluster--field structural differences varies across the stellar-mass distribution \citep{CebrianTrujillo2014,Sweet2017}, indicating that a single population-averaged mass--size relation may obscure differential environmental responses. More generally, galaxy size alone provides only one projection of structural evolution and cannot determine whether an observed environmental difference is accompanied by changes in star formation, central light concentration, or stellar populations. A more complete picture therefore requires complementary observables that trace these different aspects of galaxy evolution.

Furthermore, the bulge-to-total light ratio ($B/T$) provides a measure of the relative prominence of bulge and disk components and is closely connected to galaxy morphology and star-formation state \citep{Dressler1980,Vulcani2011,Bluck2014}. Emission-line SFRs provide a comparatively direct tracer of ongoing star formation, while the 4000~\AA\ break strength ($D_n4000$) is sensitive to recent star-formation history and stellar-population properties \citep{Kauffmann2003,Muzzin2012}. Jointly examining $R_{\rm e}$, $B/T$, SFR, and $D_n4000$ therefore enables environmental differences in star formation, stellar populations, and structure to be characterized within a common framework and across different stellar-mass regimes.

The combination of recent wide-area surveys makes such a joint analysis possible. The Hyper Suprime-Cam Subaru Strategic Program (HSC-SSP) provides deep optical imaging suitable for structural measurements, while the Dark Energy Spectroscopic Instrument (DESI) survey supplies precise spectroscopic redshifts together with emission-line and stellar-population diagnostics \citep{DESI2023}. The eROSITA X-ray survey identifies galaxy clusters through extended emission from their hot intracluster gas, providing a complementary and physically motivated selection of dense environments \citep{Bulbul2024,Kluge2024,Liu2022eFEDS}. Their overlapping coverage therefore enables structural and spectroscopic environmental trends to be studied for spectroscopically confirmed galaxies associated with X-ray-selected clusters.

\begin{figure*}[htp]
    \centering
    \includegraphics[width=\textwidth]{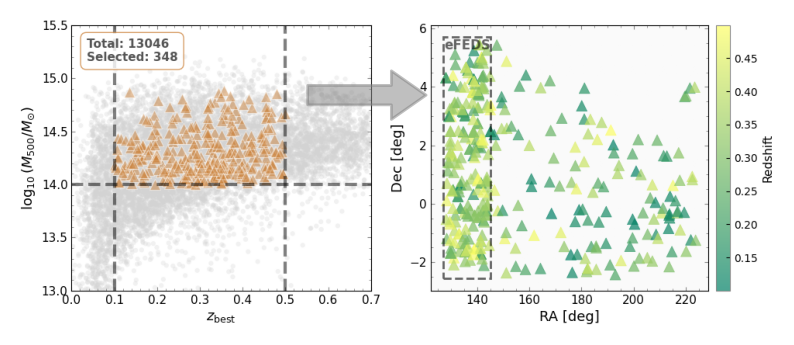} 
    \caption{Physical properties and sky distribution of the eROSITA cluster sample. Left: Cluster mass $M_{500}$ as a function of redshift $z$. The gray circles represent the parent eROSITA catalog, while the Peru-colored triangles highlight the final selected sample used in this analysis. Selection cuts are applied to include only clusters within $0.1 \leq z \leq 0.5$ and $M_{500} \geq 10^{14}\,M_{\odot}$ that are successfully matched with DESI spectroscopic members. Right: Sky distribution (R.A., Dec.) of the selected cluster sample, where triangles are color-coded by redshift. The survey footprint corresponds to the overlap between HSC-SSP Wide and DESI coverage. A prominent overdensity of clusters is observed in the eFEDS region, reflecting the significantly deeper X-ray exposure depth of this field compared to the surrounding eRASS survey areas.}
    \label{fig:cluster_sky_distribution}
\end{figure*}

In this work, we combine HSC-SSP PDR3 imaging, DESI DR1 spectroscopy, and eROSITA X-ray cluster catalogs to investigate environmental galaxy evolution at $0.1\leq z\leq0.5$. Using spectroscopically confirmed cluster satellites and a field control sample matched in stellar mass and redshift, we compare SFR, $R_{\rm e}$, $B/T$, and $D_n4000$ as functions of stellar mass, projected cluster-centric radius, and galaxy population. In addition to UVJ star-forming and quiescent galaxies, we isolate an H$\alpha$-low UVJ-star-forming transition population to examine where environmental effects are most strongly expressed. We further quantify the joint environmental displacement using a multivariate framework based on matched cluster--field residuals, allowing us to assess whether star-formation and structural responses show the same dependence on stellar mass and projected cluster-centric radius.

This paper is organized as follows. Section~\ref{sec:data} describes the datasets and sample construction, and Section~\ref{sec:methods} presents the measurements of galaxy properties and the validation of the structural analysis. Section~\ref{sec:results} presents the environmental trends, which are interpreted in Section~\ref{sec:discussion}. Our main conclusions are summarized in Section~\ref{sec:conclusion}. Throughout this work, we adopt a flat $\Lambda$CDM cosmology with $\Omega_{\rm m}=0.3$, $\Omega_{\Lambda}=0.7$, and $h=0.67$. Stellar masses and SFRs are derived assuming a Chabrier (2003) initial mass function.

\begin{table*}[htbp]
\centering
\caption{Galaxy number counts in cluster environments}
\label{tab:galaxy_counts_all}
\begin{tabular}{lcccccccc}
\hline
Environment & Region & Total & High-mass SF & High-mass GV & High-mass QG & Low-mass SF & Low-mass GV & Low-mass QG \\
\hline
Cluster & $<0.5\,R_{500}$     & 877  & 31  & 145 & 455  & 66  & 51  & 129 \\
Cluster & $0.5$--$1\,R_{500}$ & 1328 & 114 & 159 & 596  & 167 & 88  & 204 \\
Cluster & $1$--$2\,R_{500}$   & 2425 & 269 & 316 & 1040 & 414 & 138 & 248 \\
\hline

\end{tabular}
\begin{flushleft}
\footnotesize
\textbf{Notes.}
Galaxies are divided into high-mass and low-mass populations using a stellar mass threshold of
$\log_{10}(M_{\ast}/M_{\odot}) = 10.5$.
High-mass galaxies satisfy $\log_{10}(M_{\ast}/M_{\odot}) > 10.5$, while low-mass galaxies have
$\log_{10}(M_{\ast}/M_{\odot}) \leq 10.5$.
For cluster galaxies, regions are defined by the projected cluster-centric radius in units of $R_{500}$.
\end{flushleft}
\end{table*}

\section{Data}\label{sec:data}

\subsection{Photometry: HSC-SSP Public Data Release 3} \label{subsec:hsc}

Our primary source of optical photometry and morphological measurements is the third Public Data Release (PDR3) of the HSC-SSP survey \citep{Aihara2022}. The HSC-SSP utilizes the wide-field prime-focus camera \citep[1.77 deg$^2$;][]{Miyazaki2018} on the 8.2 m Subaru Telescope to conduct a deep, multi-band imaging survey in five broad bands \citep[$g, r, i, z, Y$;][]{Kawanomoto2018}. We adopt data from the Wide field of PDR3. The PDR3 imaging achieves excellent median seeing conditions \citep[$\sim$0.6''--0.8'' FWHM in the $i$-band;][]{Aihara2022} and reaches nominal 5$\sigma$ point-source depths of $i \sim 26.2$ mag. This unique combination of depth and image quality enables precise structural measurements—such as effective radii ($R_e$) and bulge--disk decompositions (yielding B/T ratios)—particularly for bright galaxies ($i < 21$ mag). All imaging data are processed with the dedicated \texttt{hscPipe} reduction pipeline \citep{Bosch2018}, which performs astrometric and photometric calibration tied to the Pan-STARRS1 system \citep{Schlafly2012,Tonry2012}.

Following procedures similar to those in \citet{Kawinwanichakij2021}, we construct our parent sample from HSC-PDR3 by requiring galaxies to have $i_{\rm cmodel} < 23$ mag. This ensures sufficiently high signal-to-noise ratios for reliable morphological analyses and is well matched to the typical magnitude limit of the DESI spectroscopic sample used in this work. We further restrict our selection to primary detections (\texttt{detect\_is\_primary=True}) located in regions covered by all five $grizy$ filters (verified via the corresponding \texttt{countinputs} flags). To avoid sources with problematic measurements, we exclude objects flagged for saturation (\texttt{flags\_pixel\_saturated\_any}), proximity to bright stars (\texttt{flags\_pixel\_bright\_object\_center}), image edges (\texttt{flags\_pixel\_edge}), or poor deblending (\texttt{cmodel\_flux\_flags}). These criteria yield a high-quality photometric sample suitable for robust structural and cross-survey analyses.

\subsection{Spectroscopy: DESI Data Release 1} \label{subsec:desi}

Spectroscopic redshifts, essential for determining cluster membership and deriving physical parameters such as D$_{\rm n}$4000 and star-formation rates (SFRs), are obtained from the first Data Release (DR1) of the DESI survey\citep{DESIDR1_2025}. DESI employs a robotic multi-fiber positioner at the prime focus of the Mayall 4-m telescope at Kitt Peak National Observatory, capable of simultaneously observing 5000 targets over an 8 deg$^2$ field of view \citep{DESI2022,Silber2023}. The fibers feed ten three-arm spectrographs covering 3600--9800~\AA, with a spectral resolution of $R = \lambda/\Delta\lambda \sim 2000$--5200 \citep{DESI2022,Guy2023}. 

DESI DR1 comprises data from the Survey Validation (SV) phases \citep{DESI2024a,DESI2024b} and the first 13 months of the main survey (May 2021--June 2022), providing reliable redshifts for over 18 million unique objects \citep{DESIDR1_2025}. We primarily use redshifts and spectral measurements from the Bright Galaxy Survey (BGS) component \citep{Hahn2023}, specifically the BGS\_BRIGHT sample ($r < 19.5$) and the BGS\_FAINT sample ($19.5 < r < 20.175$), which overlaps significantly with our HSC photometric sample at $z \lesssim 0.5$.

We cross-match the HSC PDR3 catalog with the DESI DR1 ``Iron'' spectroscopic production. Following the DESI DR1 definition of unique objects with confident redshifts, we adopt catalog entries with \texttt{ZCAT\_PRIMARY=True} and \texttt{ZWARN=0} \citep{DESIDR1_2025}. The former identifies the recommended best redshift determination for each target, while the latter indicates that no known issues are flagged in the spectroscopic data or the redshift fit.


\begin{figure}[htp]
    \centering
    \includegraphics[width=0.98\columnwidth]{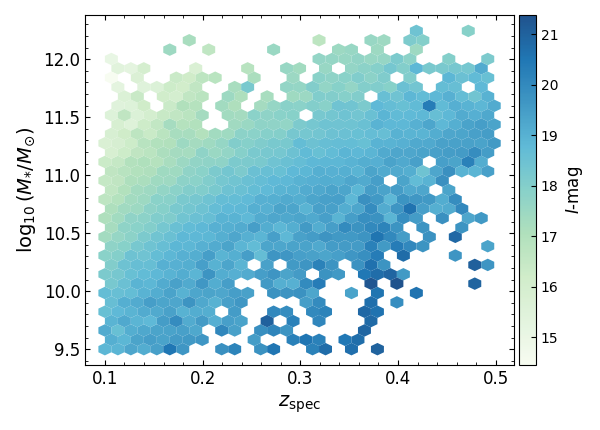}
    \caption{Distribution of cluster galaxies in the stellar mass–redshift ($M_* - z$) plane. The data points are color-coded by their $i$-band magnitude. As shown, our sample has a high degree of completeness down to $\log M_*/M_{\odot} \approx 9.5$ within the redshift range $0.1 \leq z \leq 0.5$.}
    \label{fig:mzmag}
\end{figure}

\begin{figure}[htp]
    \centering
    \includegraphics[width=0.98\columnwidth]{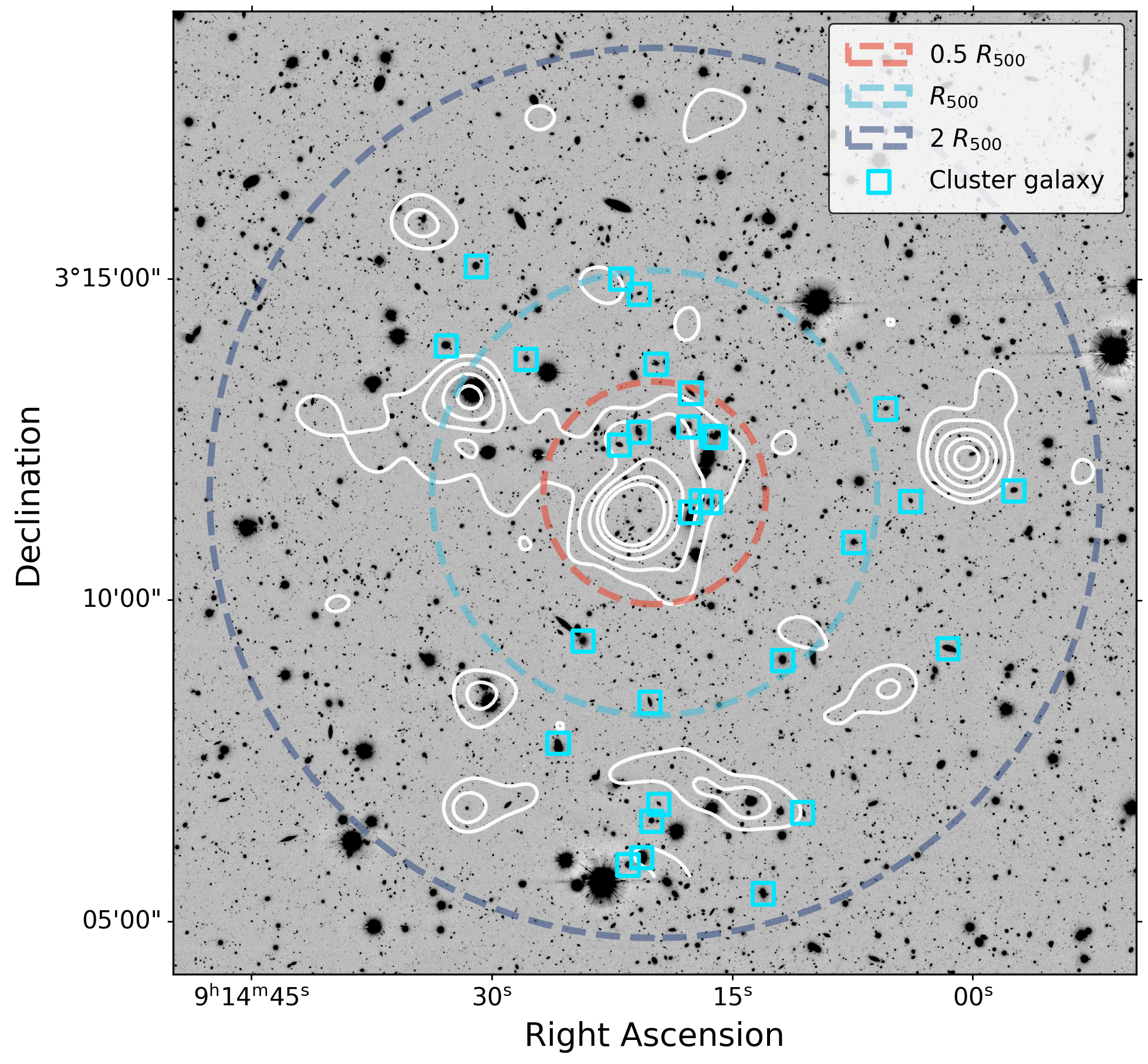}
    \caption{Example eROSITA-selected cluster at $z\simeq0.22$. The grayscale background is the HSC-SSP $i$-band image and the colored contours trace the eROSITA X-ray emission. The circles mark projected radii of $0.5R_{500}$, $R_{500}$, and $2R_{500}$; cyan squares identify DESI spectroscopic members selected with the velocity criterion in Section~\ref{subsec:final_sample_env}. The X-ray peak and galaxy distribution need not be perfectly concentric because clusters can be dynamically disturbed and because the adopted catalog center has finite uncertainty.}
    \label{fig:xray}
\end{figure}

\subsection{Cluster Catalogs: eROSITA eFEDS and eRASS1}\label{subsec:erosita}

We identify cluster environments using X-ray-selected galaxy clusters detected by eROSITA aboard the Spectrum-Roentgen-Gamma (SRG) mission \citep{Sunyaev2021,Predehl2021}. Our cluster sample combines systems from the eROSITA Final Equatorial-Depth Survey (eFEDS; \citealt{Brunner2022,Liu2022eFEDS}) and the first eROSITA All-Sky Survey (eRASS1; \citealt{Bulbul2024,Kluge2024}). The deeper eFEDS observations cover $\sim140$ deg$^2$ overlapping the HSC Wide footprint and yielded 542 X-ray-selected cluster candidates with extent likelihood $\mathcal{L}{\rm ext}>6$ and detection likelihood $\mathcal{L}{\rm det}>5$ in the 0.2--2.3 keV band. The deep X-ray coverage and overlap with HSC make eFEDS ideal for detailed studies \citep[e.g.,][]{Klein2022,Chiu2022,Kawinwanichakij2021,bahar2022,bahar2024}.
To increase the sample size and sky coverage, we supplement eFEDS with the eRASS1 primary catalog, comprising 12,247 optically identified clusters and groups over $\sim13,100$ deg$^2$. The eRASS1 catalog adopts a lower extent likelihood threshold ($\mathcal{L}_{\rm ext}>3$), allowing the inclusion of lower-significance X-ray sources and expanding the discovery space toward low-mass groups and higher-redshift clusters \citep{Bulbul2024}. We additionally require a contamination probability of $P_{\rm cont}<0.1$ to reduce the contribution from spurious or incorrectly associated systems. Optical identification and redshift estimates are obtained using the eROMaPPer algorithm applied to the DESI Legacy Imaging Surveys \citep{Dey2019,Kluge2024}. 

We restrict the combined catalog to $0.1\leq z\leq0.5$ and $10^{14}\leq M_{500}/M_{\odot}\leq10^{15}$. To ensure that each system can be used for our galaxy analysis, we further require overlap with the HSC-SSP PDR3 and DESI DR1 galaxy catalogs and at least two DESI spectroscopic members within $R_{500}$. After these selections, our final sample contains 348 clusters.

Figure~\ref{fig:cluster_sky_distribution} summarizes the cluster selection. The left panel shows the parent eROSITA systems in the $M_{500}$--redshift plane, with the final sample highlighted after applying the redshift and halo-mass cuts together with the HSC--DESI coverage and spectroscopic-member requirements. The right panel shows the sky distribution of the selected clusters, color-coded by redshift. The higher surface density of selected clusters in the eFEDS region mainly reflects its deeper X-ray coverage relative to the surrounding eRASS1 survey area.

We adopt the catalog $M_{500}$ measurements, which are based on eROSITA X-ray observables and calibrated using weak-lensing measurements \citep{Chiu2022,grandis2024,Kleinebreil2025,Okabe2025}. The corresponding $R_{500}$, enclosing a mean overdensity 500 times the critical density of the Universe at $z_{\rm cl}$, is calculated from
\begin{equation}
R_{500}=\left[\frac{3M_{500}}{4\pi\,500\,\rho_{\rm c}(z)}\right]^{1/3},
\end{equation}
where $\rho_{\rm c}(z)$ is the critical density of the Universe at the cluster redshift.

\section{Methodology and Sample Selection} \label{sec:methods}

In this section, firstly we describe the criteria for establishing spectroscopic cluster membership and the construction of the field control sample. And we describe the methodology used to derive galaxy physical properties and define our representative samples of cluster and field galaxies, building upon the multi-wavelength datasets introduced in Section~\ref{sec:data}. We first outline the derivation of rest-frame colors and the subsequent classification of galaxies into quiescent and star-forming populations. This is followed by a description of the estimation of stellar masses and star formation rates (SFRs). We then detail the measurement of structural parameters, including a dedicated suite of image simulations designed to assess the reliability and systematic uncertainties of our bulge-to-total ($B/T$) decompositions.

\subsection{Sample Selection and Environmental Classification} \label{subsec:final_sample_env}

Our final parent galaxy sample comprises all objects satisfying the HSC photometric quality criteria (Section~\ref{subsec:hsc}) and possessing secure DESI spectroscopic redshifts (Section~\ref{subsec:desi}) within the redshift range $0.1 < z < 0.5$. This range ensures substantial overlap with the DESI BGS sample and the redshift distribution of eROSITA-detected clusters (see Figure~\ref{fig:cluster_sky_distribution}), while providing a sufficiently large cosmological volume and lookback time baseline. Additionally, all galaxies in the final sample are required to have valid measurements for the key physical properties used in our analysis, as described in the preceding subsections. 

We classify galaxies into cluster and field according to their large-scale environment using the eROSITA cluster catalogs (Section~\ref{subsec:erosita}). A galaxy is considered a cluster member if it lies within twice the projected virial radius ($2R_{500}$) of an eROSITA-detected cluster and has a line-of-sight velocity relative to the cluster systemic redshift, $z_{\rm cl}$, consistent with gravitational association to the cluster halo. Specifically, following established practice \citep{Ruel2014,Capasso2019}, we define cluster members as galaxies satisfying both

\begin{equation}
R_{\rm proj} < 2R_{500}
\quad \text{and} \quad
|\Delta v| =
\frac{c|z_{\rm gal}-z_{\rm cl}|}
{1+z_{\rm cl}}
< 2000~{\rm km~s^{-1}}.
\end{equation}

We divide members into $R_{\rm proj}<0.5R_{500}$, $0.5\leq R_{\rm proj}/R_{500}<1$, and $1\leq R_{\rm proj}/R_{500}<2$. Brightest cluster galaxies are excluded, so the cluster sample analyzed below is dominated by satellites. The fixed $\pm2000\ {\rm km\,s^{-1}}$ window is deliberately inclusive but can admit interlopers, especially at large projected radius; such contamination would generally dilute rather than create a cluster--field difference.

Figure~\ref{fig:xray} illustrates the X-ray morphology and the projected distribution of selected members for one system. It is an example rather than a membership validation: association is set by the projected-radius and spectroscopic-velocity cuts above. Galaxies out to $2R_{500}$ are retained to probe the core, intermediate region, and outskirts consistently.


Galaxies in the parent sample that do not satisfy the membership criteria for any eROSITA cluster are considered as potential field candidates. To minimize contamination from cluster environments, we exclude all galaxies located within $4R_{500}$ of any eROSITA cluster center within the surveyed area. From the remaining galaxies, we construct a field control sample matched to the cluster sample in stellar mass and redshift, retaining only objects with well-constrained stellar masses. This procedure ensures that the field sample is largely free from cluster contamination and provides a fair comparison sample with similar mass and redshift distributions, without imposing additional selection cuts on other galaxy properties.

\begin{figure}[htp]
    \centering 
    \includegraphics[width=0.9\linewidth]{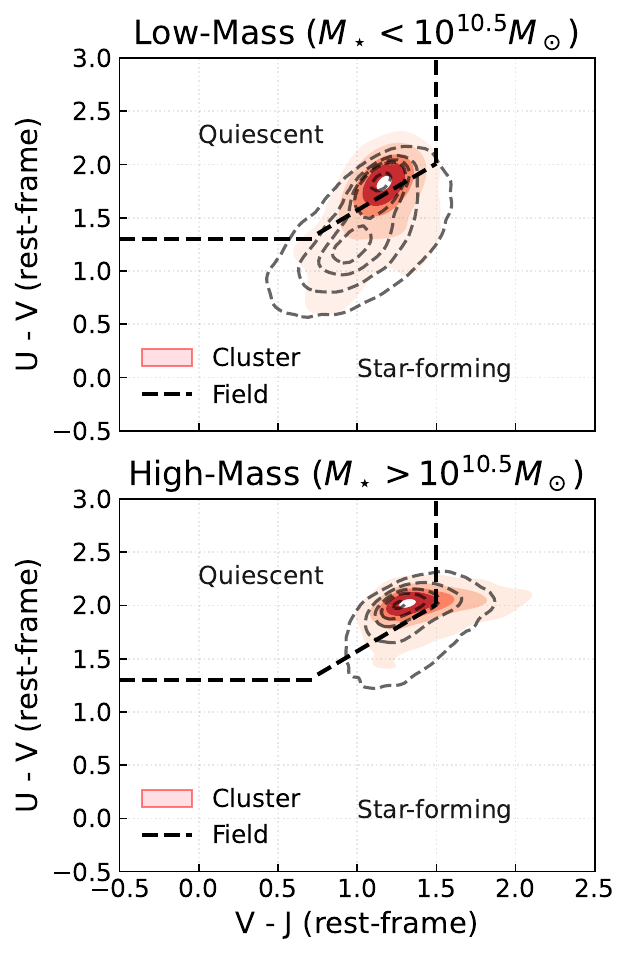} 

    \caption{Rest-frame $U-V$ versus $V-J$ diagrams (Williams et al.2009) for galaxies at $0.1 < z < 0.5$, binned by stellar mass. In the high-mass bin ($\log M_*/M_{\odot} > 10.5$), star-forming cluster galaxies are predominantly located in the upper-right region, indicative of a significant population of dust-obscured systems.}
    \label{fig:uvj_classification}
\end{figure}

The properties of the cluster sample are shown in Figure~\ref{fig:mzmag}, and the environmental differences are illustrated in the UVJ diagrams presented in Figure~\ref{fig:uvj_classification}. The galaxy number counts of the cluster and field samples are listed in Table~\ref{tab:galaxy_counts_all}.

\subsection{Rest-frame Colors and Galaxy Classification} \label{subsec:rest_frame_colors}

Rest-frame properties are essential for classifying galaxies and comparing their intrinsic characteristics. We derive rest-frame magnitudes in standard filters (e.g., $U$, $V$, and $J$) from the observed HSC $grizy$ photometry using the \texttt{EAZY} code \citep{Brammer2008}. The code fits spectral templates to the observed fluxes while fixing each redshift to its DESI spectroscopic value. Following \citet{Williams2009}, \citet{Muzzin2013}, and \citet{Kawinwanichakij2021}, the rest-frame UVJ diagram first separates quiescent galaxies (QGs) from the UVJ star-forming population, including dust-reddened star-forming systems.

Within the UVJ-star-forming population with reliable H$\alpha$ and H$\beta$ detections, we make a second, operational split at $\mathrm{sSFR}_{\mathrm{H}\alpha}=10^{-10.5}\,\mathrm{yr}^{-1}$. Objects above this threshold are labeled actively star-forming (SF), while those below it are labeled H$\alpha$-low UVJ-SF galaxies. For brevity we call the latter the ``green-valley'' (GV) sample throughout this paper. This is our analysis-specific definition, not a claim that every object lies in a conventional optical-color green valley: it identifies UVJ-SF galaxies whose dust-corrected nebular emission is weak relative to their stellar mass. 
Figure~\ref{fig:uvj_classification} shows the parent UVJ classification.

\subsection{Stellar Mass and Star Formation Rate Estimation} \label{subsec:sed_fitting}

Stellar masses ($M_\ast$) and star formation rates (SFRs) are estimated through spectral energy distribution (SED) fitting using \texttt{CIGALE} (Code Investigating GALaxy Emission; \citealt{Boquien2019}; \citealt{Yang2020}; \citealt{Yang2022}). 
The SED-derived SFRs are used only for galaxies classified as quiescent based on the UVJ diagram.

Our fitting incorporates the five HSC optical bands ($grizy$) combined with mid-infrared photometry from the unWISE catalog ($W1$ at 3.4 $\mu$m and $W2$ at 4.6 $\mu$m; \citealt{Schlafly2019}), which helps constrain both older stellar populations and dust emission, improving the robustness of $M_\ast$ and SFR estimates, especially for dusty or quiescent systems. Redshifts are fixed to the DESI spectroscopic values.

We adopt a delayed-$\tau$ star formation history,
\begin{equation}
{\rm SFR}(t) \propto t,\exp(-t/\tau),
\end{equation}
stellar population synthesis models from \citet{Bruzual2003} with a Chabrier initial mass function, and dust attenuation following \citet{Calzetti2000}. Dust emission is modeled using the templates of \citet{Dale2014}, and possible AGN contributions are accounted for using the templates of \citet{Fritz2006}.

For each galaxy, \texttt{CIGALE} outputs the full posterior probability distribution function (PDF) for each parameter. We adopt the median of the PDF as the best estimate for $M_\ast$ and SFR, with uncertainties derived from the 16th and 84th percentiles.
While \texttt{CIGALE} provides SFR estimates for all galaxies, only those for the QG sample are used in the subsequent analysis. For SFGs, we instead adopt SFRs derived from the H$\alpha$ emission line, as described in the following section.


\begin{figure*}[htp]
    \centering
    \includegraphics[width=1\textwidth]{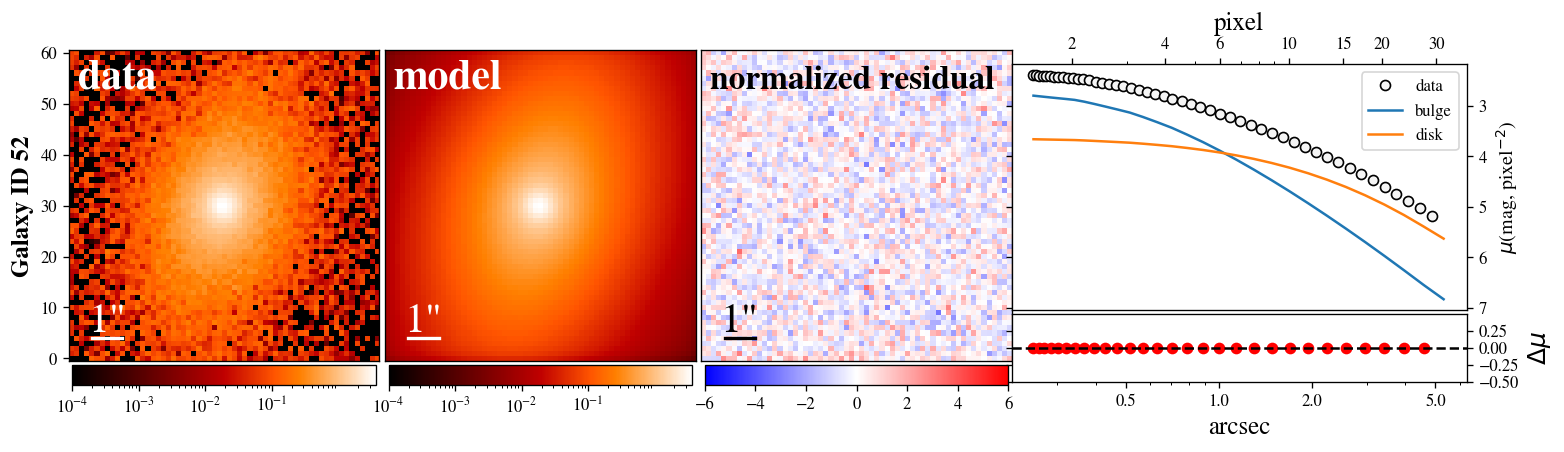}  
    \caption{Example of a double Sérsic decomposition using galight. The panels from left to right show the observed $i$-band image (Data), the best-fit Model, the corresponding Residual map, and the radial surface brightness Profiles. In the rightmost panel, the individual Sérsic components (bulge and disk) and the total model are indicated by the blue, orange, and black lines, respectively.}
    \label{fig:galight}
\end{figure*}

\begin{figure}[htp]
    \centering
    \includegraphics[width=0.98\columnwidth]{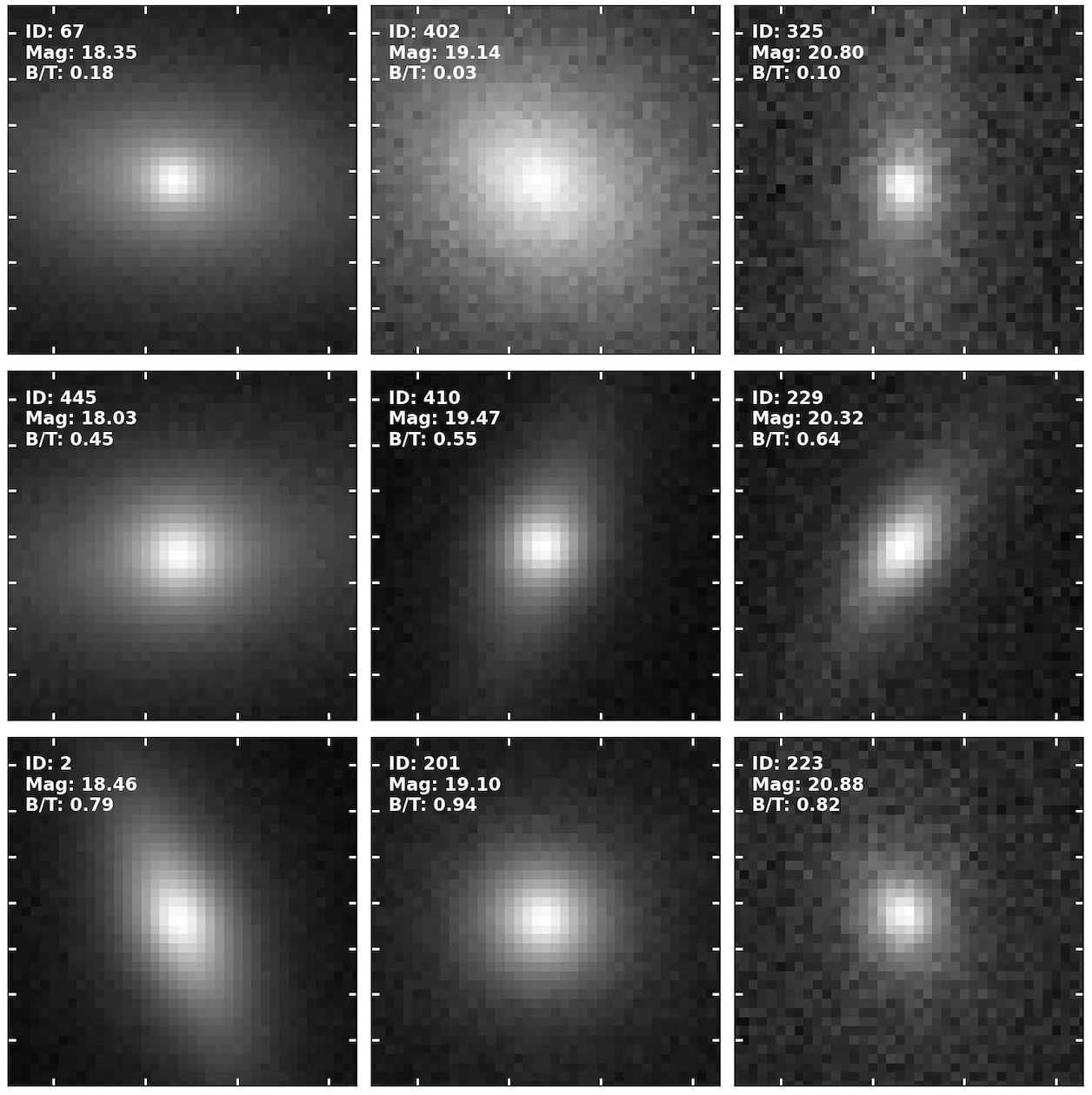}
    \caption{Examples of simulated mock galaxies. The images are generated with double Sérsic profiles and have been convolved with the HSC-SSP $i$-band PSF and add sky noise to match the depth of our observational data.}
    \label{fig:simulate_mock}
\end{figure}

\begin{figure}[htp]
    \centering
    \includegraphics[width=0.98\columnwidth]{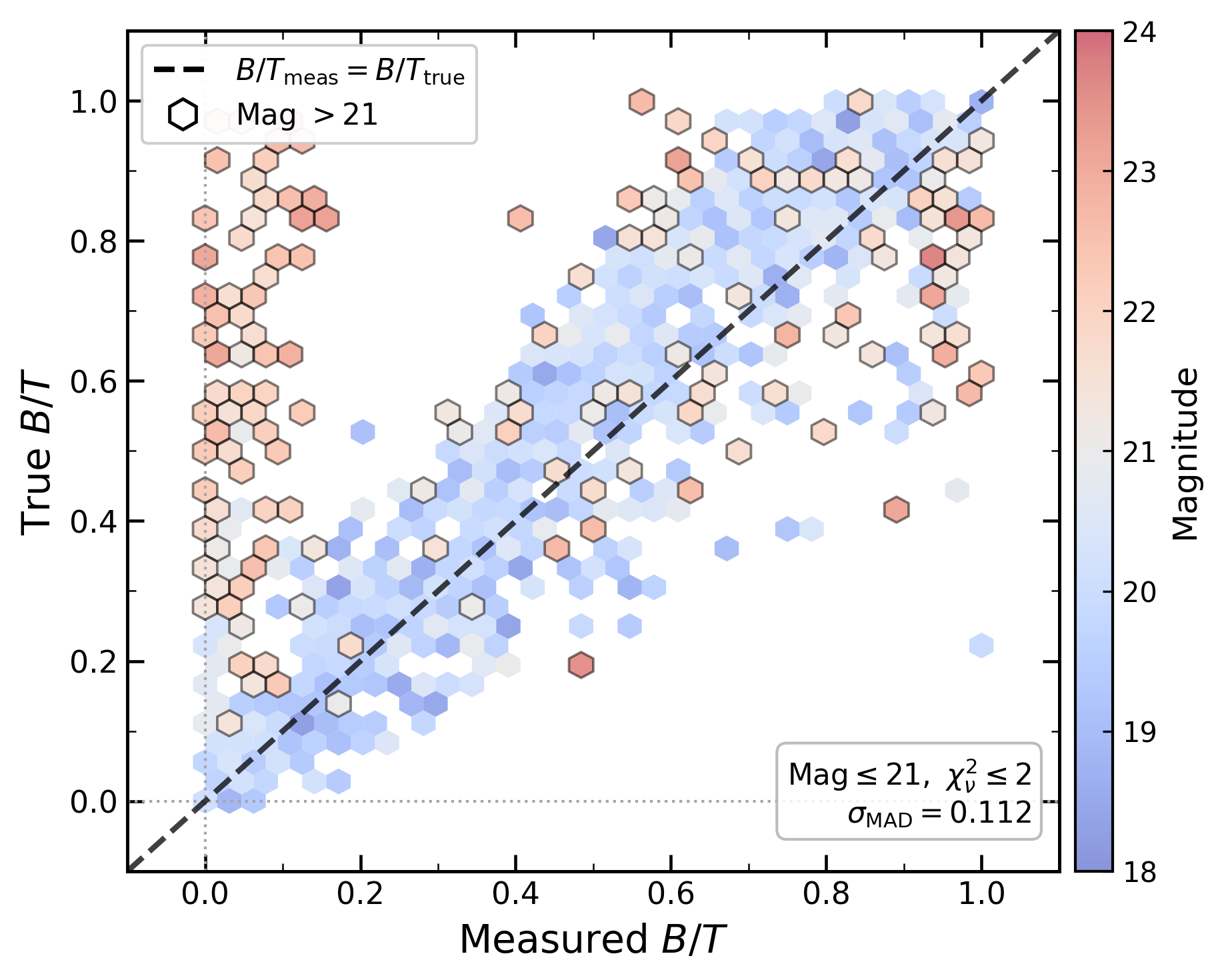}
    \caption{Recovery of $B/T$ in HSC-like mock images. The horizontal axis is the value measured by \texttt{galight} and the vertical axis is the input (true) value; the dashed line is equality. Points are colored by apparent magnitude, and open symbols identify the fainter ($i>21$) subset. For $i\leq21$ and reduced $\chi^2\leq2$, the normalized median absolute deviation of $B/T_{\rm true}-B/T_{\rm measured}$ is $0.112$. The increased scatter toward faint magnitudes motivates the conservative interpretation of small $B/T$ offsets.}
    \label{fig:real-sim-bt}
\end{figure}

\begin{table*}
\centering
\caption{Parameter ranges adopted for single- and double-component galaxy models.}
\label{tab:bt}
\begin{tabular}{lccccc}
\hline
Component Name &
S\'ersic Index &
Half-Light Radius (arcsec) &
Flux (ADUs) &
Axis Ratio &
Position Angle (degrees) \\
\hline
\multicolumn{6}{c}{\textbf{Single-Component Galaxies}} \\
\hline
 &
0.8--1.2 -- 3.5--5.0 &
0.1--5.0 &
900--$1.35\times10^{5}$ &
0.25--1.0 &
$-90.0$ -- $90.0$ \\
 & & &
($m_g \sim 18$--23) & & \\
\hline
\multicolumn{6}{c}{\textbf{Double-Component Galaxies}} \\
\hline
Disk &
0.8--1.2 &
0.1--5.0 &
0.0--1.0 &
0.25--1.0 &
$-90.0$ -- $90.0$ \\
Bulge &
3.5--5.0 &
0.1--3.0 &
1.0--Disk Comp &
0.25--1.0 &
Disk Comp. $\pm[0,15]$ \\
\hline
\end{tabular}
\end{table*}

\subsection{Structural Parameter Measurements} \label{subsec:structural_params}

Galaxy structural parameters are derived from HSC $i$-band imaging using \texttt{GALIGHT} (Galaxy Light Profile Analysis;  \citealt{Ding2020}), a Python package built on \texttt{lenstronomy} \citep{Birrer2015,Birrer2018}. The $i$ band offers the best combination of depth and seeing in HSC Wide \citep{Aihara2022} and probes approximately rest-frame $5100$--$7000$~\AA \ over $0.1<z<0.5$. 
For each galaxy we extract a $12''\times12''$ image cutout and adopt the position-dependent PSF model provided by the HSC-SSP PDR3 data reduction pipeline, which are constructed using PSFEx \citep{Bertin2011} based on nearby stars.

We first perform a single-component Sérsic fit \citep{Sersic1968}, convolved with the PSF, to obtain the major-axis effective radius ($R_{\rm e}$), Sérsic index, axis ratio, and position angle. To account for wavelength-dependent variations in galaxy size, we correct the measured $R_{\rm e}$ to a common rest-frame wavelength of $5000$~\AA\ following the description of \citet{vanDerWel2014}.
We then fit a two-component model with a de Vaucouleurs-like bulge ($n_{\rm bulge}\equiv4$) and exponential disk ($n_{\rm disk}\equiv1$). Fixing the indices stabilizes a large-sample decomposition and makes $B/T$ comparisons uniform, but it does not represent bars, pseudo-bulges, tidal features, or arbitrary Sérsic shapes; accordingly, $B/T$ is treated as a model-dependent light-concentration proxy rather than a unique physical bulge mass fraction. It is computed as

\begin{equation}
B/T = \frac{F_{\rm bulge}}
{F_{\rm bulge}+F_{\rm disk}},
\end{equation}

\noindent serving as a quantitative morphological measure that correlates well with visual classifications \citep{Simard2011,Meert2015}.

\subsubsection{Simulated Galaxies}

To quantify the robustness of our bulge--disk decomposition, we generate realistic mock galaxy cutouts using \texttt{GalSim} \citep{Rowe2015}, designed to mimic the HSC PDR3 $i$-band observations. We create 3000 simulated galaxies with bulge-to-total light ratios uniformly distributed between 0 and 1. For each mock galaxy, the structural parameters of the Sérsic components are drawn from uniform distributions within the ranges listed in Table~\ref{tab:bt}; we allow the Sérsic indices of both the bulge and disk to vary (rather than fixing them), thereby spanning a broad diversity of intrinsic light profiles. The total flux and effective radius ($R_{\rm e}$) are chosen to match the distributions of HSC galaxies at $z\sim0.3$.

To reproduce observational systematics, we convolve each model with a PSF randomly selected from a set of 100 empirical HSC PDR3 Wide-field PSFs. We also add sky noise constructed from 100 sky cutouts drawn from the same data release. After confirming visually that these regions are free of detectable sources, we pool their pixel values to build an empirical noise distribution. We then generate a noise image for each simulation by randomly sampling this distribution into a two-dimensional array with the same dimensions as the mock cutout and add it to the PSF-convolved model.

Each simulated cutout has a size of $178\times178$ pixels, corresponding to approximately $30''\times30''$ for the HSC pixel scale of $0.168'',{\rm pixel}^{-1}$. Figure~\ref{fig:simulate_mock} presents an example of a mock galaxy after PSF convolution and sky-noise addition.

Figure~\ref{fig:real-sim-bt} compares the input and recovered $B/T$. Recovery is tightest for $i\leq21$, which contains most of the spectroscopic sample, and degrades toward fainter magnitudes. These simulations incorporate representative PSFs and blank-sky noise, but they do not explicitly stratify recovery by cluster-centric radius, local surface brightness, crowding, blending with neighbors, or intracluster light. Dust-corrected H$\alpha$ fluxes address attenuation in the emission-line measurement, not these imaging systematics. Excluding BCGs and analyzing primarily satellites reduces the most extreme background and profile-complexity cases, but does not by itself eliminate a radius-dependent structural bias. We therefore interpret $B/T$ and $R_{\rm e}$ trends jointly with the spectroscopic indicators and avoid treating small structural offsets as definitive in isolation.


\begin{figure*}[htbp]
    \centering
    \includegraphics[width=0.96\textwidth]{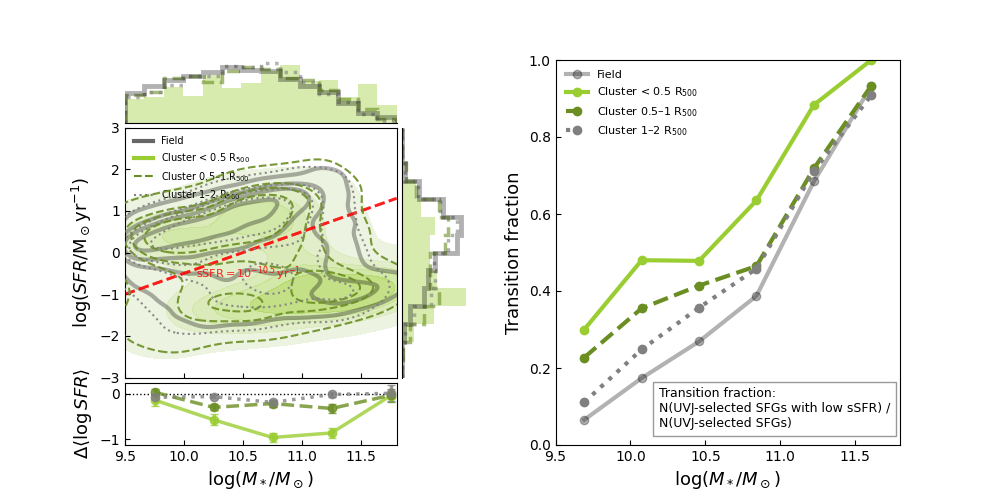}
    \caption{
    {\bf Left:} Dust-corrected H$\alpha$ SFR versus stellar mass for the UVJ-SF, Balmer-line-detected sample in the field and three projected cluster-centric bins. The dashed line marks our operational division at $\mathrm{sSFR}_{\mathrm{H}\alpha}=10^{-10.5}\,\mathrm{yr}^{-1}$: galaxies above it are labeled SF and those below it are labeled H$\alpha$-low UVJ-SF (``GV'' for brevity). The lower panel shows the cluster-minus-field median SFR residual for the SF subset. {\bf Right:} Fraction $f_{\rm GV}=N_{\rm GV}/(N_{\rm GV}+N_{\rm SF})$ in the same parent selection. This is a transition-sample fraction under our operational definition, not a complete color-selected green-valley fraction. Error bars show the uncertainties used in the analysis.
    }
    \label{fig:mass_sfr_offset_sf_results}
\end{figure*}

\subsection{Spectroscopic Parameters} \label{subsec:spec_params}

Complementary physical parameters related to stellar populations are derived directly from the DESI spectra using the publicly available \texttt{fastspecfit} value-added catalog associated with DESI DR1 \citep{DESIDR1_2025,Moustakas2023}. \texttt{fastspecfit} performs rapid fitting of stellar population synthesis models combined with emission-line templates to the DESI spectra.

From these fits, we utilize measurements of the D$*{\rm n}4000$ index, defined as the ratio of the average flux density in the $4000$--$4100$~\AA\ and $3850$--$3950$~\AA\ rest-frame bands \citep{Bruzual1983,Balogh1999}. D$*{\rm n}4000$ is a robust tracer of the mean stellar age of the population dominating the optical light \citep{Kauffmann2003}.

In addition to the SFR derived from SED fitting, we estimate the instantaneous SFR corrected for dust attenuation using the Balmer decrement for galaxies where both H$\alpha$ and H$\beta$ emission lines are reliably detected. We adopt emission-line fluxes from the \texttt{fastspecfit} catalog. Following standard procedures \citep{Osterbrock2006,Dominguez2013}, we select galaxies with H$\alpha$ and H$\beta$ detected at signal-to-noise ratio (SNR) $\ge 3$.
The observed flux ratio, \begin{equation} R_{\rm obs} = \frac{F_{{\rm H}\alpha}}{F_{{\rm H}\beta}}\Big|_{\rm obs}, \end{equation} is compared to the intrinsic Case B recombination ratio, $R_{\rm int}=2.86$, appropriate for typical H,{\sc ii} regions ($T = 10^4$ K and $n_e = 100~{\rm cm^{-3}}$; \citealt{Osterbrock2006}).

The color excess due to dust attenuation is then \begin{equation} E(B-V) = \frac{2.5}{k(\lambda_{{\rm H}\beta}) - k(\lambda_{{\rm H}\alpha})} \log_{10} \left( \frac{R_{\rm obs}}{R_{\rm int}} \right), \end{equation} where $k(\lambda)$ is the extinction curve value at the corresponding wavelength. For cases where $R_{\rm obs} < R_{\rm int}$—typically due to measurement uncertainties or residual stellar absorption corrections—we set $E(B-V) = 0$.

The intrinsic H$\alpha$ flux is then \begin{equation} F_{{\rm H}\alpha, \rm int} = F_{{\rm H}\alpha, \rm obs} \times 10^{0.4 E(B-V) k(\lambda_{{\rm H}\alpha})}. \end{equation}

The H$\alpha$ luminosity is calculated using the spectroscopic redshift and adopted cosmology: \begin{equation} L_{{\rm H}\alpha, \rm int} = 4 \pi D_L^2 F_{{\rm H}\alpha, \rm int}. \end{equation}Finally, the dust-corrected star formation rate, SFR(H$\alpha$), is computed using the calibration of \citet{Kennicutt1998}, converted to a Chabrier IMF following \citet{Hao2011} and \citet{Murphy2011}:
\begin{equation} {\rm SFR(H\alpha)} \,[M_\odot \, {\rm yr}^{-1}] = 5.5 \times 10^{-42} \, L_{{\rm H}\alpha, \rm int} \,[{\rm erg \, s}^{-1}]. \end{equation}


This spectroscopically derived, dust-corrected SFR traces recent star formation within the DESI fiber. Throughout this work, we use it for the UVJ-SF Balmer-line-detected sample, including both the SF and operational GV subsets.

\begin{figure*}[htbp]
    \centering
    \includegraphics[width=0.98\textwidth]{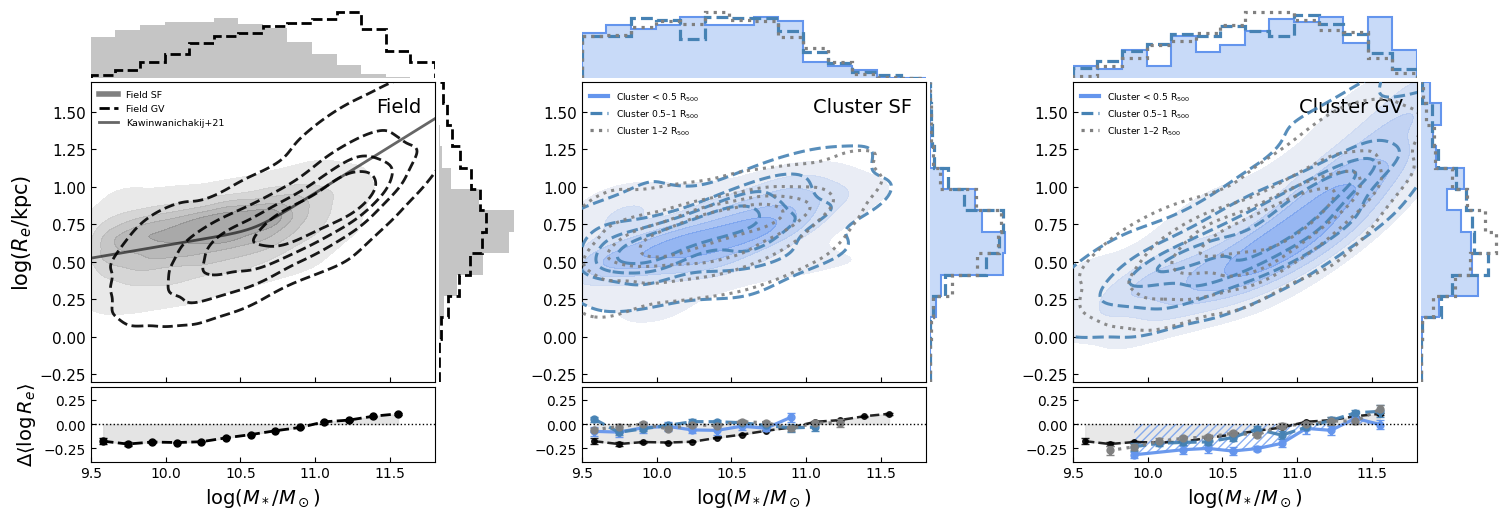}
    \caption{Stellar mass--size relations for the UVJ-SF sample. From left to right, the panels show field SF and H$\alpha$-low UVJ-SF galaxies, cluster SF galaxies, and cluster H$\alpha$-low UVJ-SF galaxies. Cluster medians are split into the three projected radial bins and are compared with the corresponding field relation; residual panels show the cluster-minus-field offset at fixed stellar mass. Plotting the field SF--GV difference alongside the environmental residuals distinguishes an intrinsic stage difference from a cluster-specific offset.} \label{fig:sfg_mass_size}\label{fig:gv_mass_size}
\end{figure*}

\section{Results} \label{sec:results}

We present the empirical results in the order in which the samples are defined: the H$\alpha$ SFR--mass plane, actively star-forming galaxies, the H$\alpha$-low UVJ-SF subset, and UVJ-quiescent galaxies. We compare each cluster subsample with a field reference matched in stellar mass and redshift. The word ``field'' therefore denotes the non-cluster control defined in Section~\ref{subsec:final_sample_env}, not an isolated or void population; it may include galaxies in groups, filaments, and other large-scale structures.

We first use the $M_\ast$--SFR plane only within the UVJ-SF, Balmer-line-detected sample. The split at $\mathrm{sSFR}_{\mathrm{H}\alpha}=10^{-10.5}\,\mathrm{yr}^{-1}$ separates actively star-forming galaxies from the H$\alpha$-low UVJ-SF subset that we call GV for brevity. UVJ-QGs are classified separately and are not defined by this H$\alpha$ threshold. Consequently, SF, GV, and QG are useful analysis categories, but they should not automatically be read as a unique temporal sequence for individual galaxies.

This classification provides an evolutionary framework for the remainder of the analysis. In the following sections, we first investigate environmental trends in star formation activity through the SFR--$M_\ast$ relation and the transition fraction. We then examine how structural properties, including effective radius ($R_{\rm e}$) and bulge-to-total ratio ($B/T$), vary among SFGs, GVs, and QGs as a function of environment. By comparing these populations, we aim to identify when structural transformation occurs relative to the suppression of star formation and to constrain the physical mechanisms responsible for environmentally driven galaxy evolution.

\subsection{SFR--$M_\ast$ Relation and Transition Fraction}
\label{subsec:msfr}

Figure~\ref{fig:mass_sfr_offset_sf_results} presents the H$\alpha$ SFR--$M_\ast$ relation and the fraction below our operational threshold within the UVJ-SF, Balmer-line-detected sample. We emphasize that ``GV'' here means H$\alpha$-low UVJ-SF; it is a compact label for this paper rather than a conventional color selection or proof that each object is caught in transition.

The left panel of Figure~\ref{fig:mass_sfr_offset_sf_results} shows that the star-forming main sequence is present in both cluster and field environments over the full stellar mass range probed by our sample. However, subtle environmental differences are evident. At fixed stellar mass, cluster SFGs exhibit systematically lower SFRs than their field counterparts, particularly within the innermost regions ($R_{\rm proj}<0.5\,R_{500}$). The suppression is modest, typically below 0.2 dex, but becomes increasingly pronounced toward higher stellar masses. Such behavior is consistent with previous studies reporting environmentally driven reductions in star formation activity prior to complete quenching \citep{Peng2010,Wetzel2012,Muzzin2012}.

The residual trends shown in the lower panel further illustrate this environmental dependence. While galaxies beyond $R_{500}$ display SFRs comparable to those of field galaxies, systems located within the cluster core exhibit significantly reduced SFRs, with the strongest suppression occurring at intermediate and high stellar masses. This radial dependence suggests that environmental processes become increasingly effective toward the cluster center.

The two-dimensional distributions in the $M_\ast$--SFR plane provide additional insight into this trend. In the outer cluster regions ($R_{\rm proj}/R_{500}\gtrsim1$), the galaxy distribution closely resembles that of the field population and is dominated by low-mass galaxies with relatively high star formation rates. Toward smaller cluster-centric radii, the peak of the distribution shifts toward higher stellar masses and lower SFRs. This evolution indicates that massive galaxies are preferentially found in dense environments and experience stronger suppression of star formation activity. Such behavior is broadly consistent with environmental quenching mechanisms, including ram-pressure stripping, starvation, and the removal of gas reservoirs, which progressively reduce star formation before galaxies become fully quiescent \citep{Larson1980,Balogh2000,Wetzel2013}.

The right panel quantifies the frequency of the operationally defined H$\alpha$-low subset,

\begin{equation}
f_{\rm GV} = \frac{N_{\rm GV}}{N_{\rm GV}+N_{\rm SF}}.
\end{equation}

Cluster galaxies show a larger $f_{\rm GV}$ than the field over much of the sampled mass range, with the clearest enhancement toward small projected radius. This is consistent with an environmental contribution to suppressed recent star formation, although the Balmer-line selection, fiber aperture, and projection effects prevent a direct conversion of $f_{\rm GV}$ into a quenching timescale. The mass dependence can reflect both mass-related and environmental selection effects \citep{Peng2010,Peng2012}.

\begin{figure*}[htbp]
    \centering
    \includegraphics[width=0.98\textwidth]{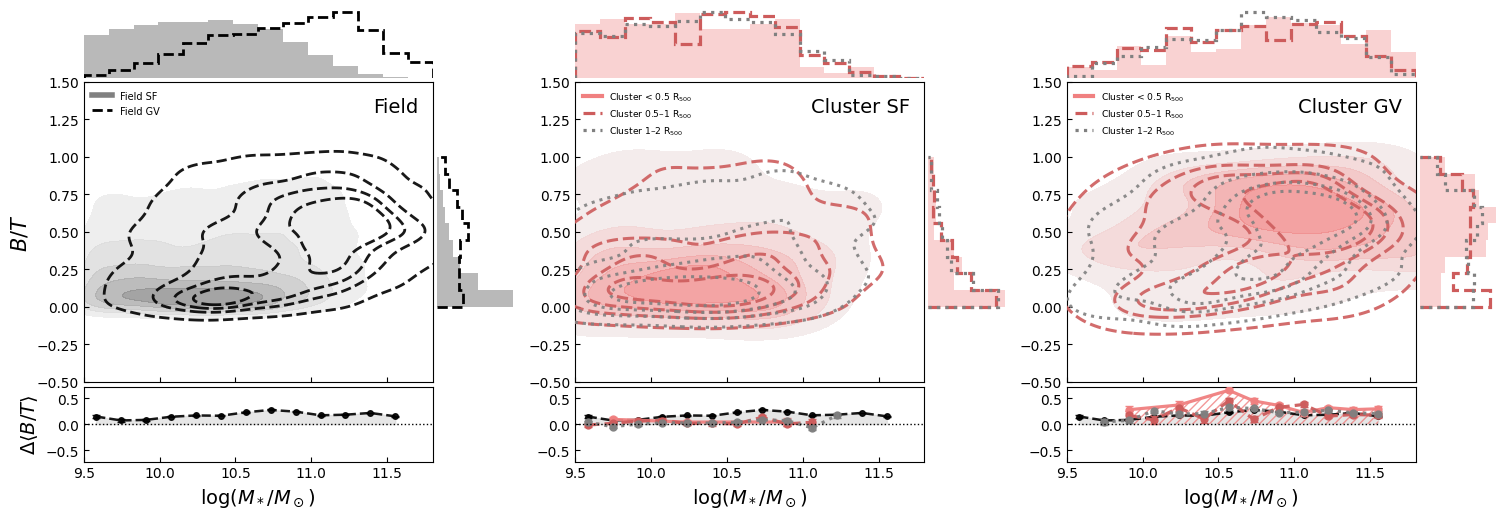}
    \caption{As in Figure~\ref{fig:sfg_mass_size}, but for $B/T$. The common layout shows the field SF--GV difference separately from the cluster-minus-field residuals for cluster SF and cluster H$\alpha$-low UVJ-SF galaxies. Because $B/T$ is obtained from fixed-index two-component fits, offsets smaller than the mock-recovery scatter should be interpreted cautiously.} \label{fig:sfg_mass_bt}\label{fig:gv_mass_bt}
\end{figure*}

\subsection{UVJ Star-forming Population: Active and H$\alpha$-low Subsamples}
\label{subsec:uvjsf}

Figures~\ref{fig:sfg_mass_size}--\ref{fig:sfg_mass_dn4000} use a common three-panel layout. The first panel compares SF and H$\alpha$-low UVJ-SF galaxies within the field control itself; the second isolates the cluster--field difference for active SF galaxies; and the third shows the analogous difference for the H$\alpha$-low subset. This ordering prevents the intrinsic SF--GV contrast in the control sample from being conflated with an environmental residual.

\subsubsection{Actively Star-forming Galaxies}
\label{subsec:sfg}

Having classified galaxies into star-forming (SFG), green-valley (GV), and quiescent (QG) populations based on their locations in the $M_\ast$--SFR plane and UVJ color space, we first examine the structural and stellar population properties of SFGs.

As the least evolved population in our classification scheme, SFGs provide a baseline for assessing the earliest environmental influences on galaxy evolution. Although these galaxies remain actively star-forming, the results of Section~\ref{subsec:msfr} indicate that cluster SFGs already exhibit modestly suppressed star formation activity relative to their field counterparts. An important question is therefore whether this suppression is accompanied by measurable changes in galaxy structure or stellar populations.

To address this issue, we investigate the size--mass relation, $B/T$--mass relation, and $\mathrm{D_n4000}$-mass relation for SFGs in different environments. These diagnostics probe complementary aspects of galaxy evolution, including structural growth, bulge assembly, and stellar population aging. Comparing cluster and field SFGs allows us to determine whether environmental processes first affect star formation activity alone or whether structural transformation has already begun before galaxies enter the transition phase.

\paragraph{Size--mass relation.}
\label{subsubsec:sfg_size}

Figure~\ref{fig:sfg_mass_size} shows the size-mass relation for SFGs in the field and in different cluster-centric radial bins. Field SFGs follow the expected positive correlation between stellar mass and effective radius, consistent with the well-established size--mass relation of star-forming disk galaxies \citep{Shen2003,vanDerWel2014}. Cluster SFGs broadly follow the same relation, indicating that galaxies that remain actively star-forming have not yet experienced a strong global structural transformation in size.

The residual panel shows that the size offsets of cluster SFGs relative to field SFGs are generally small, typically within $\sim0.1$ dex over most of the stellar mass range. This suggests that the modest SFR suppression, described in Section~\ref{subsec:msfr}, is not accompanied by a comparably strong change in galaxy size. The innermost cluster SFGs show a mild enhancement in $R_{\rm e}$ at some masses, but this trend is weak compared to the offset between field GV galaxies and field SFGs, shown by the black dashed curve and further discussed below. Therefore, the size evolution associated with the transition from SFGs to GVs is substantially larger than the environmental size offset among SFGs alone.

These results suggest that environmental processes can begin to suppress star formation in SFGs before producing a strong change in their global size. This behavior is consistent with a scenario in which environmental mechanisms such as starvation or partial gas stripping first reduce the gas supply and star formation activity, while leaving the stellar structure largely intact during the early stages of quenching \citep{Larson1980,Balogh2000,Wetzel2013}.

\paragraph{Bulge-to-total ratio--mass relation.}
\label{subsubsec:sfg_bt}

Figure~\ref{fig:sfg_mass_bt} presents the stellar $B/T$--mass relation for SFGs. Field SFGs are dominated by low $B/T$ systems across most of the stellar mass range, as expected for disk-dominated star-forming galaxies. Cluster SFGs show a broadly similar distribution, with most objects remaining at low $B/T$, indicating that the actively star-forming population in clusters is still predominantly disk dominated.

The residual panel shows that the $B/T$ offsets between cluster and field SFGs are small at most stellar masses and cluster-centric radii. In contrast, the black dashed curve, which represents the offset between field GV galaxies and field SFGs, shows a substantially larger increase in $B/T$. This comparison indicates that the morphological difference between SFGs and GVs is much larger than the environmental difference among SFGs themselves.

The absence of a strong $B/T$ enhancement among cluster SFGs suggests that significant bulge growth is not yet widespread in galaxies that remain actively star-forming. Thus, while environmental effects may already be reducing the SFRs of cluster SFGs, the corresponding morphological transformation appears to occur later, or more gradually, as galaxies move into the transition population. This result is broadly consistent with studies showing that quenching and morphological transformation are related but not necessarily simultaneous processes \citep{Kauffmann2003,Bell2004,Bluck2014}.

\begin{figure*}[htbp]
    \centering
    \includegraphics[width=0.98\textwidth]{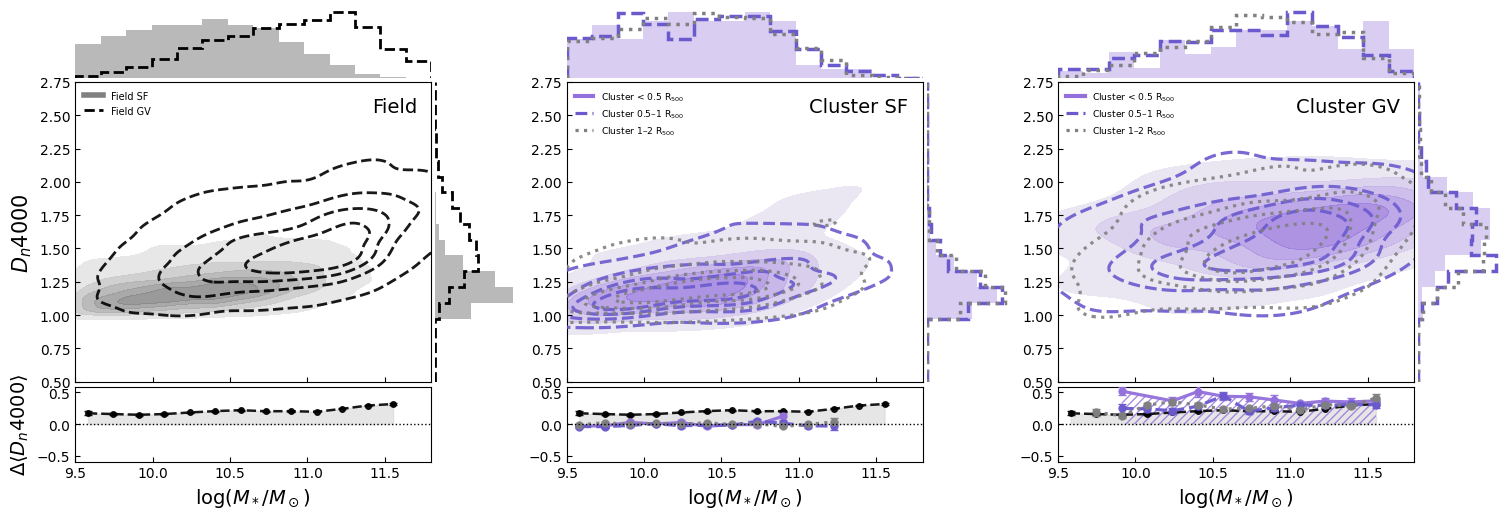}
    \caption{As in Figure~\ref{fig:sfg_mass_size}, but for $\mathrm{D_n}4000$. Larger $\mathrm{D_n}4000$ indicates an older luminosity-weighted stellar population and/or less recent star formation; it is not a one-to-one age measurement without assumptions about metallicity and star-formation history.} \label{fig:sfg_mass_dn4000}\label{fig:gv_mass_dn4000}
\end{figure*}

\paragraph{$\mathrm{D_n4000}$--mass relation.}
\label{subsubsec:sfg_dn4000}

Figure~\ref{fig:sfg_mass_dn4000} shows the $\mathrm{D_n4000}$--mass relation for SFGs. The $\mathrm{D_n4000}$ index traces the strength of the 4000~\AA\ break and is commonly used as an indicator of the luminosity-weighted stellar population age, with larger values corresponding to older stellar populations and/or lower recent star formation activity \citep{Balogh1999,Kauffmann2003}. Field SFGs exhibit low $\mathrm{D_n4000}$ values, consistent with young stellar populations and ongoing star formation.

Cluster SFGs occupy a similar region of the $M_\ast$--$\mathrm{D_n4000}$ plane as field SFGs. The residual panel shows only weak $\mathrm{D_n4000}$ offsets between cluster and field SFGs over most of the stellar mass range. These offsets are much smaller than the difference between field GV galaxies and field SFGs, shown by the black dashed curve. Therefore, although cluster SFGs show modest SFR suppression, their stellar populations remain broadly consistent with those of actively star-forming field galaxies.

This result suggests that the environmental effect on SFGs is still at an early stage. The current star formation activity may already be reduced, but the integrated stellar population has not yet aged enough to produce a strong increase in $\mathrm{D_n4000}$. In this interpretation, SFGs in clusters represent a pre-transition population, in which environmental quenching has begun but has not yet produced the stellar population aging characteristic of GV or QG systems.

\subsubsection{H$\alpha$-low UVJ-SF (``GV'') Galaxies}
\label{subsec:gv}

The active SF results provide the baseline against which we now compare the H$\alpha$-low UVJ-SF subset. These objects occupy an intermediate location in our H$\alpha$ SFR--mass classification, but the cross-sectional selection does not establish that every object will evolve directly from SF through this subset into the present QG sample.

Such H$\alpha$-weak but UVJ-SF systems are plausible candidates for recent or ongoing suppression and therefore provide a useful window on transformation \citep{Martin2007,Schawinski2014}. They can nevertheless include dust- or aperture-affected systems and galaxies with non-monotonic star-formation histories; we therefore report population offsets rather than assigning an evolutionary fate to each galaxy.

To investigate this transition phase, we examine the size--mass relation, $B/T$--mass relation, and $\mathrm{D_n4000}$--mass relation for GVs in different environments. Comparing cluster and field GVs allows us to determine whether environmental quenching is accompanied by structural growth, bulge buildup, and stellar population aging as galaxies evolve toward the quiescent population.

\paragraph{Size--mass relation.}
\label{subsubsec:gv_size}

Figure~\ref{fig:gv_mass_size} shows the stellar size--mass relation for green-valley galaxies. In contrast to the SFG population, cluster GVs exhibit a more pronounced environmental dependence. At fixed stellar mass, cluster GVs are systematically smaller than field GVs, particularly at intermediate stellar masses ($10^{10.2}\lesssim M_\ast/M_\odot \lesssim10^{11}$). The strongest size deficit is observed in the innermost cluster regions, where the median size offset reaches $\sim0.2$--$0.3$ dex.

The comparison with the field SFG relation further illustrates this trend. While field GVs already lie below the field SFG size sequence, cluster GVs occupy an even more compact region of the parameter space. This result suggests that structural evolution becomes increasingly important once galaxies enter the transition phase.

The emergence of smaller sizes among cluster GVs is consistent with scenarios in which environmental quenching preferentially removes star-forming material from the outer disk or suppresses the growth of extended stellar components, thereby increasing the central concentration of stellar mass \citep{Poggianti1999,Koopmann2004,Cibinel2013}. The stronger effect observed toward smaller cluster-centric radii further supports an environmental origin.

\paragraph{Bulge-to-total ratio--mass relation.}
\label{subsubsec:gv_bt}

Figure~\ref{fig:gv_mass_bt} presents the $B/T$--mass relation for green-valley galaxies. Unlike the SFG population, cluster GVs exhibit systematically elevated $B/T$ values across nearly the entire stellar mass range. The enhancement is particularly significant within $R_{\rm proj}<0.5R_{500}$, where the median $B/T$ exceeds the field relation by up to $\sim0.3$--$0.5$.

The increased fitted bulge prominence indicates that structural differences are most evident in this subsample. Compared with field SF galaxies, cluster GVs occupy a more centrally concentrated region of parameter space. This may reflect morphological evolution, preferential quenching of already concentrated galaxies, or both.

This trend is consistent with previous studies showing that galaxies undergoing quenching frequently develop enhanced central mass concentrations and larger bulge fractions \citep{Kauffmann2003,Fang2013,Bluck2014}. The strong environmental dependence observed here implies that cluster-specific processes either accelerate bulge growth or preferentially quench galaxies that already possess more concentrated stellar structures.

\paragraph{$\mathrm{D_n4000}$--mass relation.}
\label{subsubsec:gv_dn4000}

Figure~\ref{fig:gv_mass_dn4000} shows the $\mathrm{D_n4000}$--mass relation for green-valley galaxies. In contrast to the weak environmental dependence observed among SFGs, cluster GVs exhibit systematically larger $\mathrm{D_n4000}$ values than field GVs at nearly all stellar masses. The offset is typically $\sim0.2$--$0.4$, with the strongest enhancement occurring in the innermost cluster regions.

The elevated $\mathrm{D_n4000}$ values indicate older luminosity-weighted stellar populations and reduced recent star formation activity. Together with the enhanced $B/T$ ratios and reduced sizes, these results suggest that cluster GVs represent a more advanced stage of the quenching process than their field counterparts.

The simultaneous population-level offsets in size, fitted $B/T$, and $\mathrm{D_n4000}$ show that the selected H$\alpha$-low cluster sample differs from its field counterpart in several correlated observables \citep{Balogh1999,Kauffmann2003,Schawinski2014}. The data alone do not determine whether structural change precedes, follows, or is selected together with declining star formation; this covariance is quantified explicitly in Section~\ref{subsec:integrated}.

\subsection{Quiescent Galaxies}
\label{subsec:qg}

We finally examine the quiescent galaxy (QG) population, which represents the endpoint of the evolutionary sequence considered in this work. The analyses of SFGs and GVs suggest that environmental effects first manifest as a suppression of star formation and subsequently become associated with significant structural transformation during the transition phase. An important remaining question is whether these environmental signatures persist once galaxies have become fully quenched. Since QGs have largely ceased forming stars, any remaining environmental dependence is expected to reflect differences in their quenching histories, structural assembly, or the timescales over which they entered the quiescent population.

\begin{figure*}[htbp]
    \centering
    \includegraphics[width=0.99\textwidth]{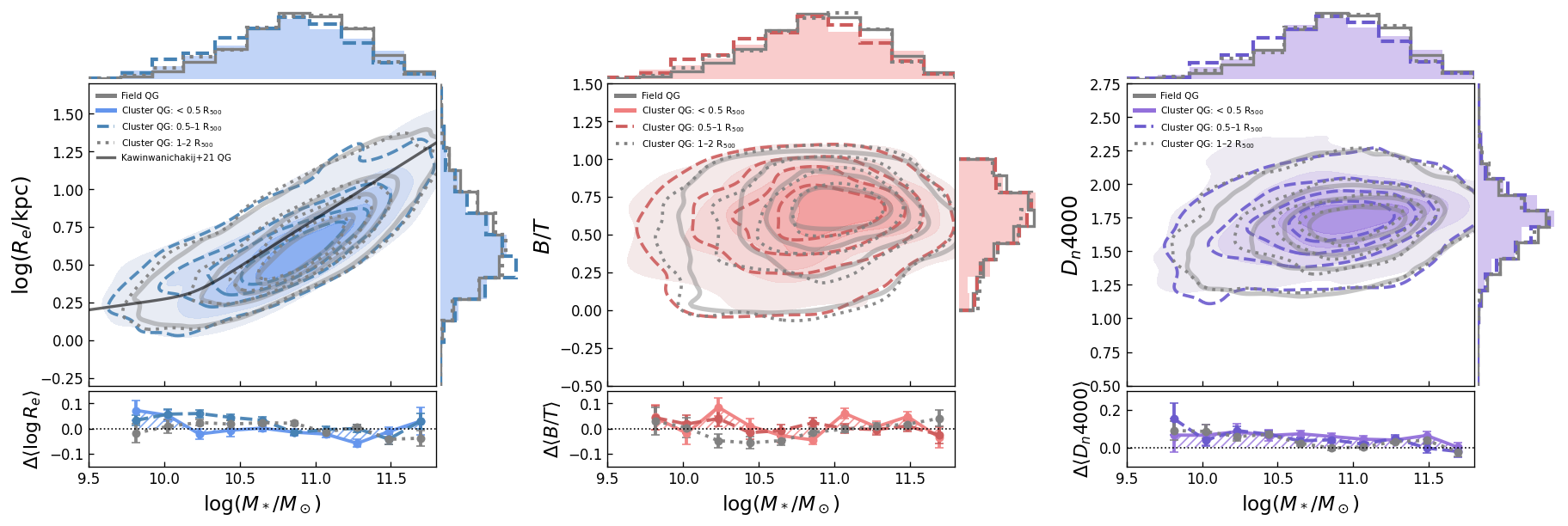}
    \caption{Environmental comparison for UVJ-quiescent galaxies. From left to right: effective radius, fitted $B/T$, and $\mathrm{D_n}4000$ as functions of stellar mass. Each column shows the field distribution and the cluster medians in three projected radial bins, together with cluster-minus-field residuals. The combined presentation makes clear that the structural residuals are generally weaker than the stellar-population residuals and avoids interpreting any one observable in isolation.} \label{fig:qg_mass_size}\label{fig:qg_mass_bt}\label{fig:qg_mass_dn4000}
\end{figure*}

\subsubsection{Size--Mass Relation}
\label{subsubsec:qg_size}

Figure~\ref{fig:qg_mass_size} presents the stellar size--mass relation for quiescent galaxies. In contrast to the clear environmental dependence observed among GVs, cluster and field QGs follow remarkably similar size--mass relations across the full stellar mass range. The residual panel shows that the size offsets remain within $\sim0.05$ dex and are generally consistent with zero within the uncertainties.

This result suggests that the structural differences that emerge during the transition phase do not continue to grow after galaxies become fully quiescent. Instead, the size--mass relation appears to converge toward a common sequence largely independent of environment. The agreement with the relation reported by \citet{Kawinwanichakij2021} further supports the robustness of this result.

The disappearance of the environmental size dependence among QGs implies that the dominant structural transformation likely occurs before or during the transition phase, rather than after quenching is complete.

\subsubsection{Bulge-to-Total Ratio--Mass Relation}
\label{subsubsec:qg_bt}

Figure~\ref{fig:qg_mass_bt} shows the $B/T$--mass relation for quiescent galaxies. Both cluster and field QGs occupy a strongly bulge-dominated region of parameter space, with typical $B/T$ values substantially larger than those observed for SFGs and comparable to those of GVs.

The residual panel indicates only weak environmental differences. Although the lowest-mass cluster QGs exhibit slightly elevated $B/T$ values relative to the field population, the offsets rapidly diminish toward higher stellar masses and remain consistent with zero within the uncertainties over most of the mass range.

The absence of a strong environmental dependence suggests that once galaxies become fully quenched, their bulge-dominated structures are largely established. Therefore, the substantial increase in $B/T$ observed in the GV population likely represents the primary phase of morphological transformation, while subsequent evolution within the quiescent population is comparatively modest.

\subsubsection{$\mathrm{D_n4000}$--Mass Relation}
\label{subsubsec:qg_dn4000}

Figure~\ref{fig:qg_mass_dn4000} presents the $\mathrm{D_n4000}$--mass relation for quiescent galaxies. Unlike the mass--size and $B/T$--mass relations, a clear environmental dependence remains visible in $\mathrm{D_n4000}$. At fixed stellar mass, cluster QGs exhibit systematically larger $\mathrm{D_n4000}$ values than field QGs, with typical offsets of $\sim0.1$--$0.2$.

The enhanced $\mathrm{D_n4000}$ values indicate older luminosity-weighted stellar populations and imply that cluster QGs were quenched earlier than their field counterparts. The strongest offsets are observed at low and intermediate stellar masses, suggesting that environmental quenching is particularly effective in accelerating the cessation of star formation among satellite galaxies in dense environments.

Combined with the weak environmental trends in size and $B/T$, these results indicate that the primary difference between cluster and field QGs lies not in their present-day structures but in their quenching histories. Cluster QGs appear to have reached the quiescent phase earlier, resulting in older stellar populations at fixed stellar mass. This interpretation is consistent with previous studies showing that galaxies in dense environments tend to experience earlier formation and quenching epochs than comparable field galaxies \citep{Thomas2005,Peng2010,Wetzel2013}.

\section{Discussion} \label{sec:discussion}

\subsection{Coupled and Multivariate Environmental Effects}
\label{subsec:integrated}

The environmental signatures in star formation, stellar-population age, and galaxy structure are correlated, and therefore cannot be interpreted independently. We first examine the joint structure of the matched cluster--field residuals directly, and then quantify the total multivariate displacement with a Mahalanobis metric. These two diagnostics address complementary questions: the residual maps show \emph{which combinations of galaxy properties change together}, whereas the Mahalanobis distance measures \emph{the overall amplitude of the multivariate displacement}.

For each cluster satellite, we construct a field reference matched in stellar mass and redshift. In the residual-map analysis, each satellite is compared with the mean of up to five field controls satisfying
\begin{equation}
|\Delta\log M_\ast|<0.05,
\qquad
|\Delta z|<0.01,
\end{equation}
and at least three acceptable controls are required. For an observable $X$, we define
\begin{equation}
\Delta X_i =
X_{i,\mathrm{cl}}
-
\left\langle X_{\mathrm{field}}\right\rangle_{F(i)},
\end{equation}
where $F(i)$ denotes the matched field controls of satellite $i$. Thus, $\Delta X=0$ corresponds to no difference from field galaxies of comparable stellar mass and redshift. Sources with the flagged $B/T=0.5$ value are excluded before matching.

\begin{figure*}[htbp]
    \centering
    \includegraphics[width=0.78\textwidth]{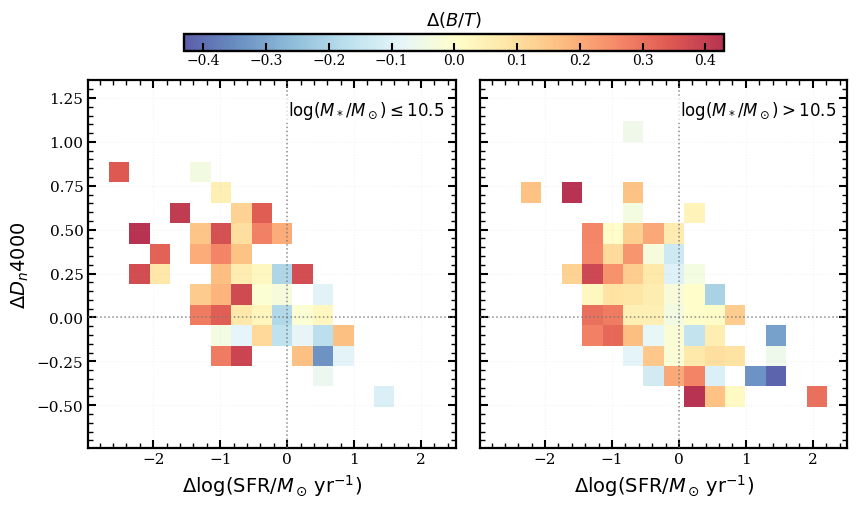}
    \caption{
    Joint cluster--field residuals in star formation, stellar-population age, and galaxy structure.
    Each cluster satellite is compared with stellar-mass- and redshift-matched field controls.
    The horizontal axis shows $\Delta\log{\rm SFR}$, the vertical axis shows $\Delta D_n4000$, and the color gives the mean $\Delta(B/T)$ within each occupied cell.
    The left and right panels show galaxies below and above $\log(M_\ast/M_\odot)=10.5$, respectively.
    Both panels use the same axis limits and color normalization centered on $\Delta(B/T)=0$.
    Only cells containing at least two galaxies are displayed.
    Negative $\Delta\log{\rm SFR}$, positive $\Delta D_n4000$, and positive $\Delta(B/T)$ correspond, respectively, to suppressed star formation, older stellar populations, and enhanced bulge prominence relative to the matched field reference.
    }
    \label{fig:multidimensional_environment}
\end{figure*}

Figure~\ref{fig:multidimensional_environment} visualizes these residuals for the full satellite population, without conditioning on SF, GV, or QG classification. The horizontal and vertical coordinates are $\Delta\log{\rm SFR}$ and $\Delta D_n4000$, respectively, and the color in each occupied cell gives the cluster-weighted mean $\Delta(B/T)$. The sample is divided at $\log(M_\ast/M_\odot)=10.5$, and the two panels use identical coordinate ranges and color normalization. Cells containing fewer than two galaxies are omitted. The figure should therefore be interpreted in terms of coherent trends across multiple neighboring cells rather than individual low-occupancy cells.

The residual maps show that galaxies with more strongly suppressed star formation tend to exhibit more positive $B/T$ residuals. We quantify this trend separately in the two stellar-mass intervals using cluster-weighted linear regressions,
\begin{equation}
\begin{split}
\Delta(B/T) =\;&
\beta_0
+\beta_{\rm SFR}\Delta\log{\rm SFR}
+\beta_{D}\Delta D_n4000 \\
&+\beta_M(\log M_\ast-\widetilde{\log M_\ast})
+\beta_z(z-\widetilde{z}),
\end{split}
\end{equation}
where the tildes denote the median within the corresponding stellar-mass interval. Each host cluster is given equal total weight, and the coefficient uncertainties quoted here are cluster-robust standard errors with host cluster as the clustering unit.

For the low-mass sample, we obtain
\begin{equation}
\beta_{\rm SFR}=-0.105\pm0.032,
\qquad p=0.001,
\end{equation}
whereas the high-mass sample gives
\begin{equation}
\beta_{\rm SFR}=-0.113\pm0.021,
\qquad p<0.001.
\end{equation}
Thus, in both stellar-mass regimes, stronger suppression of star formation relative to the matched field population is associated with a systematically larger $B/T$ residual. In practical terms, a decrease of 1 dex in $\Delta\log{\rm SFR}$ corresponds to an increase of approximately 0.1 in $\Delta(B/T)$ at fixed $\Delta D_n4000$, stellar mass, and redshift.

By contrast, the coefficient of $\Delta D_n4000$ is positive but not independently significant:
\begin{equation}
\beta_D=0.063\pm0.074
\quad (p=0.393)
\end{equation}
at low mass and
\begin{equation}
\beta_D=0.060\pm0.059
\quad (p=0.308)
\end{equation}
at high mass. The color structure in Figure~\ref{fig:multidimensional_environment} therefore does not imply that $D_n4000$ independently predicts the structural residual once the correlated SFR residual is taken into account. Rather, the apparent association between older stellar populations and enhanced bulge prominence is largely embedded in the same population that exhibits suppressed star formation.

Within the high-mass interval, residual trends with stellar mass and redshift remain detectable
($\beta_M=0.128\pm0.057$, $p=0.024$ and
$\beta_z=-0.408\pm0.133$, $p=0.002$),
whereas neither control term is significant in the low-mass interval.
We therefore retain both stellar mass and redshift in the regression rather than assuming that the initial matching removes all within-bin dependence.
These residual trends do not imply a failure of the matching: matching removes systematic differences between each cluster galaxy and its field controls, while the regression tests whether the \emph{environmental residual itself} varies continuously with galaxy mass or redshift.

We next compress the correlated residuals into a single multivariate effect-size statistic. For this purpose, we use the four-dimensional vector
\begin{equation}
\mathbf{y}
=
\left(
\log{\rm SFR},
\log R_{\rm e},
B/T,
D_n4000
\right).
\end{equation}
Stellar mass is not included as a coordinate of the distance because it is already controlled through the matching and is used separately for stratification.

For the Mahalanobis analysis, we use the matched cluster--field residuals from the production multivariate pipeline. The residual vector is
\begin{equation}
\boldsymbol{\delta}_i
=
\mathbf{y}_{i,\mathrm{cl}}
-
\mathbf{y}_{i,\mathrm{field}},
\end{equation}
with the field counterpart selected under the same
$|\Delta\log M_\ast|<0.05$ and $|\Delta z|<0.01$
criteria. A unique field control is assigned within each matched sample so that the same field galaxy does not contribute repeatedly to the group centroid.

The reference covariance,
$\boldsymbol{\Sigma}_{\rm FF}$, is estimated from matched field--field residuals constructed with the same matching procedure. The mean field--field residual is subtracted as a zero-point correction. Importantly, within a given evolutionary-stage and stellar-mass stratum, the same $\boldsymbol{\Sigma}_{\rm FF}$ is used for all three radial bins. Radial changes in the metric therefore reflect changes in the cluster--field centroid displacement rather than changes in the covariance used to measure it.

To prevent rich clusters from dominating the measurement, galaxy residuals are first averaged within each host cluster and the host centroids are subsequently given equal weight. If $K$ host clusters contribute to a bin,
\begin{equation}
\overline{\boldsymbol{\delta}}
=
\frac{1}{K}
\sum_{k=1}^{K}
\left(
\frac{1}{N_k}
\sum_{i\in k}
\boldsymbol{\delta}_i
\right).
\end{equation}
The multivariate environmental offset is then
\begin{equation}
D_{\rm env}
=
\sqrt{
\overline{\boldsymbol{\delta}}^{\mathsf T}
\boldsymbol{\Sigma}_{\rm FF}^{+}
\overline{\boldsymbol{\delta}}
},
\end{equation}
where $\boldsymbol{\Sigma}_{\rm FF}^{+}$ denotes the inverse covariance, evaluated with a Moore--Penrose pseudoinverse when necessary.

$D_{\rm env}\geq0$ is dimensionless and quantifies how far the cluster centroid is displaced from the matched field reference after accounting for the covariance among SFR, size, $B/T$, and $D_n4000$. It is therefore a standardized multivariate effect size, rather than a physical force, evolutionary timescale, or causal ``environmental intensity.''

Uncertainties on $D_{\rm env}$ are estimated from 3000 hierarchical-bootstrap realizations. Host clusters are resampled first, followed by resampling of galaxies within the selected hosts. Statistical significance is assessed with 10,000 cluster-block sign-flip permutations, and the resulting probabilities are corrected with the Benjamini--Hochberg false-discovery-rate procedure across the 18 stage--mass--radius combinations.

\begin{figure*}[htbp]
    \centering
    \includegraphics[width=0.96\textwidth]{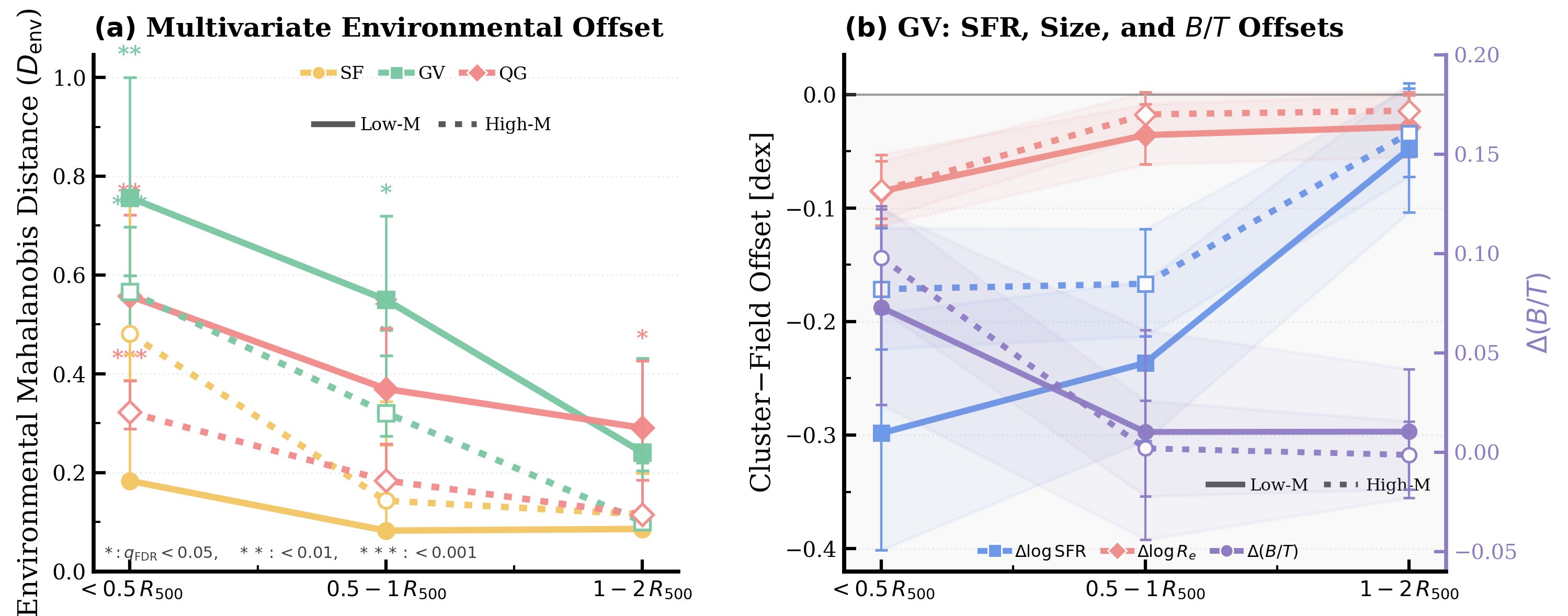}
    \caption{
    Radial dependence of the multivariate environmental offset.
    {\bf (a)} Environmental Mahalanobis distance, $D_{\rm env}$, for SF, GV, and QG satellites in two stellar-mass intervals.
    The covariance metric is estimated from matched field--field residuals and is held fixed across the three radial bins within each stage--mass stratum.
    Error bars show hierarchical-bootstrap uncertainties, and asterisks indicate tests that remain significant after false-discovery-rate correction.
    {\bf (b)} For the GV population, the cluster--field residuals in SFR, $R_{\rm e}$, and $B/T$ are shown separately as a function of projected cluster-centric radius.
    Solid and dotted curves denote the low- and high-mass samples, respectively.
    Larger $D_{\rm env}$ indicates a larger standardized multivariate displacement from the matched field population, but does not by itself identify the physical mechanism or temporal ordering responsible for that displacement.
    }
    \label{fig:environmental_intensity}
\end{figure*}

Figure~\ref{fig:environmental_intensity}(a) shows that the strongest radial variation is associated with the GV population. Low-mass GV satellites exhibit the largest inner-region multivariate displacement and a strong decrease toward larger projected radius. The corresponding high-mass GV population shows a substantially smaller displacement. The QG population exhibits a more moderate radial trend, while the SF population shows a weaker multivariate offset overall.

Because the field--field covariance is defined separately for each evolutionary-stage and stellar-mass stratum, the most direct interpretation of $D_{\rm env}$ is the radial change within a fixed stratum. Differences in the absolute values of $D_{\rm env}$ between distinct populations should be interpreted as differences in standardized effect size relative to the field scatter appropriate to each population, rather than as measurements made with a single universal covariance scale.

The decomposition in Figure~\ref{fig:environmental_intensity}(b) provides physical context for the GV signal. The SFR residual remains negative over a broader range in projected radius than the strongest residuals in $R_{\rm e}$ and $B/T$, whereas the largest structural offsets occur preferentially at smaller $R/R_{500}$. This spatial ordering is consistent with a scenario in which suppression of the gas supply affects star formation over a broader radial range than the strongest changes in the stellar-light distribution \citep{Wetzel2013,Fillingham2015}.

However, projected cluster-centric radius is not an evolutionary clock. Orbital mixing, backsplash, pre-processing, and rejuvenation can place galaxies with substantially different accretion and star-formation histories at the same projected radius. We therefore interpret the radial ordering as evidence that different galaxy properties respond over different spatial scales, rather than as a direct measurement that one physical transformation necessarily precedes another in time.

Taken together, Figures~\ref{fig:multidimensional_environment} and
\ref{fig:environmental_intensity} provide complementary views of the environmental response. Figure~\ref{fig:multidimensional_environment} shows that, after controlling for stellar mass and redshift through field matching, suppressed star formation is systematically associated with enhanced bulge prominence. The regression analysis confirms that this relation remains significant after additionally controlling for $D_n4000$, stellar mass, and redshift. Figure~\ref{fig:environmental_intensity} then shows that the total multivariate displacement is strongest for low-mass GV satellites in the inner cluster regions and declines outward.

These results are consistent with environmental quenching accompanied by structural transformation, but they do not establish a unique evolutionary pathway. The observed covariance can arise from transformation after infall, differences already present among the progenitor population, pre-processing before cluster entry, or preferential survival and selection of structurally distinct galaxies. The similarity of the fitted $\Delta(B/T)$--$\Delta\log{\rm SFR}$ slopes in the two broad mass intervals further suggests that the coupling between star-formation suppression and structural residuals may operate in both regimes, even though the overall magnitude of the environmental displacement is more strongly mass dependent.

\begin{figure*}[htbp]
    \centering
    \includegraphics[width=0.94\textwidth]{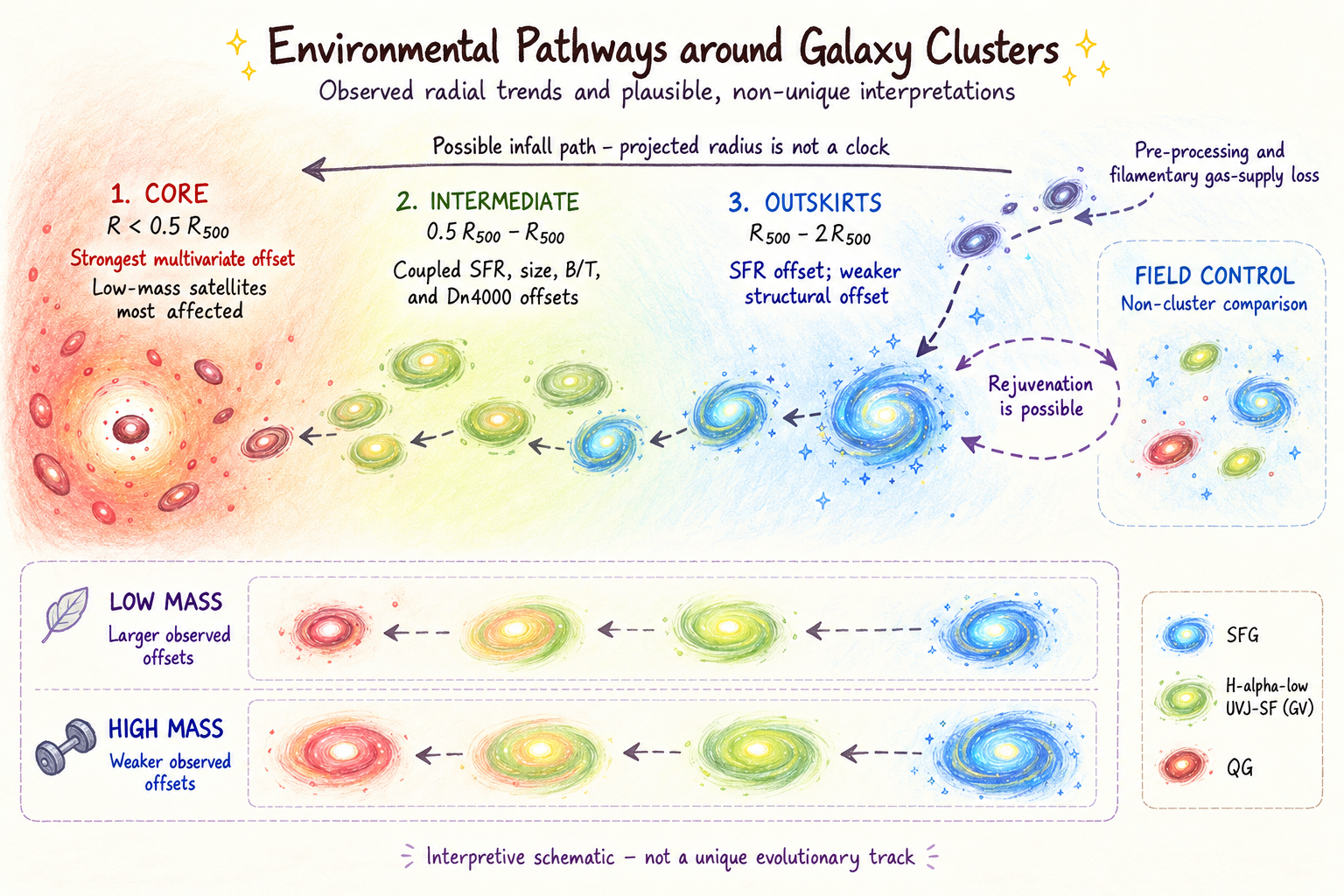}
    \caption{Interpretive summary of environmental pathways around clusters. The upper row sketches the dominant population-level trend suggested by the radial residuals: gas-supply suppression can appear in the outskirts, whereas the strongest coupled stellar-population and structural offsets are found farther inward. The lower row emphasizes the stronger response at lower stellar mass. Pre-processing in groups and filaments, orbital mixing or backsplash, and rejuvenation provide non-exclusive side paths, so the arrows are hypotheses rather than a unique trajectory followed by every galaxy. The separated field control is a matched reference population, not a literal starting location for all satellites.} \label{fig:evolutionary_conveyor}
\end{figure*}

\subsection{A Coherent Picture of Environmentally Driven Galaxy Transformation}
\label{subsec:discussion_sequence}

The results reveal a coherent population-level ordering, but not a unique evolutionary track. Star-formation residuals appear over a broader radial range than the strongest structural residuals; the largest joint offset is found for low-mass H$\alpha$-low UVJ-SF galaxies; and QGs retain a clearer $\mathrm{D_n}4000$ than structural difference. Figure~\ref{fig:evolutionary_conveyor} summarizes one interpretation while explicitly retaining alternative pathways. This distinction is important because projected radius mixes first-infall, virialized, and backsplash populations, and the three galaxy classes are not progenitor-matched.

The earliest environmental signature is seen in the SFR--$M_\ast$ relation. Cluster SFGs exhibit systematically lower SFRs than field SFGs at fixed stellar mass, with the strongest suppression occurring toward smaller cluster-centric radii. However, these galaxies still remain on or near the star-forming sequence, and their sizes, bulge-to-total ratios, and $\mathrm{D_n4000}$ values remain broadly similar to those of field SFGs. This suggests that environmental processes initially act on the gas supply rather than immediately altering the stellar structure. Such behavior is consistent with mechanisms such as starvation, strangulation, or partial gas stripping, in which the gas reservoir available for future star formation is reduced before the existing stellar distribution is significantly modified \citep{Larson1980,Balogh2000,Boselli2006,Wetzel2013}.

The enhanced H$\alpha$-low fraction in clusters is consistent with the environment increasing the incidence of suppressed recent star formation at fixed mass. It is not, by itself, a direct transition rate: the denominator is restricted to UVJ-SF galaxies with measurable Balmer lines, and galaxies can enter or leave the selection through dust, aperture effects, stochasticity, or renewed star formation. We therefore interpret the fraction together with $\mathrm{D_n}4000$ and structure rather than as a standalone clock \citep{Peng2010,Peng2012}.

The H$\alpha$-low UVJ-SF population is nevertheless central to the empirical result because its SFR, size, fitted $B/T$, and $\mathrm{D_n}4000$ residuals are simultaneously displaced. This makes it the clearest observational window on where suppressed recent star formation and structural differences coexist. It does not prove that bulges grew during quenching: preferential quenching of compact or bulge-dominated progenitors, disk fading, and environment-dependent measurement bias remain possible contributors \citep{Yesuf2022,Woo2013,Woo2017}.

The radial ordering is qualitatively consistent with delayed-then-rapid satellite quenching \citep{Wetzel2013,Fillingham2015}, yet cluster entry need not be the beginning of environmental processing. Galaxies can lose gas or be morphologically altered in groups, filaments, and infalling substructure before crossing $R_{500}$---the ``pre-processing'' channel \citep{Sarron2019}. Large-scale gas flows and cosmic-web detachment can also reduce fuel supply outside the nominal halo boundary \citep{AragonCalvo2019,GalarraEspinosa2023}. These channels naturally help explain an SFR signal in the outskirts without requiring the present cluster core to have acted there.

The $\mathrm{D_n}4000$ measurements add a longer-timescale stellar-population constraint. Their persistence among QGs, when size and $B/T$ residuals are weaker, is consistent with earlier average quenching in the cluster population. However, $\mathrm{D_n}4000$ also depends on metallicity and detailed star-formation history; it should not be converted directly into an age difference without spectral modeling.

The covariance-aware $D_{\rm env}$ strengthens the mass-dependent result without double-counting correlated residuals. The larger low-mass offset is physically plausible because shallow potentials are less able to retain gas, whereas high-mass evolution can be more strongly regulated by internal or halo-related processes \citep{Kauffmann2003,Fang2013,Bluck2014,Woo2015}. Yet $D_{\rm env}$ describes a standardized displacement rather than a causal susceptibility; different covariance widths and selection functions across classes are handled within the stated strata but still limit comparisons.

Taken together, the data favor a picture in which suppression of recent star formation can be visible before the strongest structural residuals, with the H$\alpha$-low UVJ-SF population marking where these observables overlap most strongly. Pre-processing, backsplash and orbit mixing, progenitor selection, and rejuvenation broaden that pathway \citep{Mancini2019}. The schematic therefore encodes a set of allowed channels rather than a conveyor belt with a single direction or timescale.

\subsection{Reconciling Previous Environmental Studies}
\label{subsec:discussion_literature}

The environmental trends identified in this work can be placed in the broader context of studies of star formation, stellar populations, and galaxy structure. A broad observational consensus is that galaxies in dense environments, and satellites in particular, are more frequently quenched than comparable field or central galaxies \citep[e.g.,][]{Kauffmann2004,Balogh2004,Hogg2004,Wetzel2012,Wetzel2013,Peng2012,BoselliGavazzi2006,Cortese2021}. The environmental dependence of structural properties is less uniform, however, and depends on stellar mass, evolutionary state, redshift, the definition of environment, rest-frame wavelength, and the adopted structural metric.

Our star-formation results are broadly consistent with this picture. Beyond the well-known enhancement of the quenched fraction, several studies have found reduced star-formation activity among galaxies that have not yet become fully quiescent. HSC measurements, for example, find modestly lower specific star-formation rates for star-forming galaxies in groups and clusters, with stronger environmental quenching toward lower stellar mass \citep{Jian2018}. Studies of nearby clusters likewise favor quenching histories that include a period of reduced gas supply or slow decline followed by more rapid suppression, with pre-processing contributing before final cluster accretion \citep{Wetzel2013,Haines2015}. Our detection of negative SFR residuals among galaxies that remain on the star-forming side of our classification is therefore consistent with environmental processing beginning before complete quiescence.

The abundance of transition or ``green-valley'' galaxies requires more care. Its inferred environmental dependence is sensitive to the transition definition and to the population used for normalization. Some studies identify an enhanced transition population in groups or clusters \citep{Coenda2018,Jian2020}, whereas the SDSS satellite analysis of \citet{Wetzel2012} found no excess in the intermediate-SSFR population despite a strong enhancement of the quenched fraction. At $z\sim1$, \citet{McNab2021} found only a modest cluster excess of transition galaxies, concentrated primarily below $\log(M_*/M_\odot)\simeq10.5$, with little evidence for an excess at higher stellar mass. We therefore interpret the radial behavior of our operational transition class as evidence for a population associated with environmental quenching, rather than assuming that a green-valley fraction is a universal or model-independent measure of the quenching rate.

The stellar-population trends point in the same general direction. Stronger 4000-\AA\ breaks and older luminosity-weighted stellar populations have been reported in denser environments over a broad range of redshift \citep{Balogh1999,Kauffmann2004,PaulinoAfonso2020,Sobral2022}. We likewise find positive $D_n4000$ residuals for the transition and quiescent populations. Because $D_n4000$ depends on recent star-formation history and, particularly for older populations, on metallicity, we do not interpret it as a unique age clock. The persistence of a $D_n4000$ residual when some structural offsets become weak is therefore most conservatively interpreted as evidence that stellar-population differences remain detectable, not as a direct measurement that stellar-population transformation necessarily precedes or outlasts structural evolution.

The environmental dependence of galaxy size is considerably less settled. Several studies report little or no environmental difference in the stellar-mass--size relation once stellar mass, morphology, or color is controlled \citep{Maltby2010,Rettura2010,HuertasCompany2013,Kelkar2015,Morishita2017,Strazzullo2023}. In contrast, larger characteristic sizes in dense environments have been reported for massive, often quiescent, galaxies at intermediate and high redshift \citep{Cooper2012,Papovich2012,Lani2013,Delaye2014,Chan2018,Andreon2018}. Other studies instead find smaller sizes or truncated outer structures in clusters and other overdense regions \citep{Cebrian2014,Matharu2019,Matharu2020,Xu2023,Chamba2024,George2025}. These results are not necessarily mutually exclusive: the studies span different redshifts, stellar-mass ranges, morphological selections, cluster definitions, rest-frame wavelengths, and size estimators. In particular, recent work illustrates that environmental size differences can depend on whether one uses conventional effective radii, low-surface-brightness galaxy edges, or different rest-frame wavelengths \citep{Chamba2024,George2025}.

Our results suggest that evolutionary state provides an additional source of heterogeneity. Within our sample, the size residual is weak for the star-forming and fully quiescent populations but becomes more pronounced for the transition population. Differences in the relative mixture of these populations can therefore dilute or amplify the ensemble environmental size signal. We do not argue that population mixing alone reconciles the literature; redshift evolution, progenitor bias, sample selection, rest-frame wavelength, and the adopted size definition remain important. Rather, separating galaxies by evolutionary state reveals a structural signal that would be substantially less apparent in an undivided population.

The cluster post-starburst results of \citet{Matharu2019,Matharu2020,Matharu2021} provide a particularly relevant comparison. At $z\sim1$, recently quenched cluster galaxies occupy an intermediate locus between the star-forming and quiescent mass--size relations, and models in which the outer stellar populations fade after outside-in suppression can produce both an apparent reduction in light-weighted effective radius and an increase in bulge prominence without requiring an equally large redistribution of stellar mass \citep{Matharu2020}. Spatially resolved H$\alpha$ measurements provide additional evidence for outside-in suppression of star formation in recently quenched cluster systems \citep{Matharu2021}. These results are qualitatively consistent with our finding that the strongest coupled star-formation and structural residuals occur in the transition population, although the redshift ranges, observables, and sample definitions differ and therefore do not require an identical physical mechanism. The compact, evolved star-forming population identified in a $z=2.51$ cluster by \citet{Xu2023} provides a higher-redshift example in which environmental star-formation and structural evolution appear to be coupled.

Our morphology results should be interpreted with the same distinction between association and causation. The classical morphology--density relation establishes a strong population-level connection between galaxy structure and environment \citep{Dressler1980}, but studies controlling for stellar mass and star-formation activity generally find that residual morphology--environment trends are weaker than the corresponding color or star-formation trends \citep{Bamford2009,Skibba2009}. Bulge prominence is itself strongly connected to quiescence at fixed stellar mass, while satellites retain an additional environmental contribution to their quenching and structure \citep{Bluck2016,Bluck2019}. Bulge--disc decompositions also show that environmental effects can act on discs without requiring substantial changes in the bulge component \citep{LacknerGunn2013}. At $z\sim0.8$, \citet{PaulinoAfonso2019} similarly find that morphology depends on both stellar mass and environment, while environmental correlations weaken substantially when star-forming and quiescent populations are considered separately. More recently, \citet{Oxland2024} find evidence that satellite quenching and morphological transformation both proceed with environmental processing, with particularly strong effects at low stellar mass, but need not share the same characteristic timescale.

These results motivate a deliberately conservative interpretation of our $B/T$ measurements. Our cluster-weighted regression shows that $\Delta(B/T)$ is independently associated with $\Delta\log{\rm SFR}$ in both stellar-mass intervals, whereas $\Delta D_n4000$ is not independently significant once SFR, stellar mass, and redshift are controlled. We therefore describe the observed trend as enhanced bulge prominence or structural transformation associated with suppressed star formation, rather than as direct evidence for bulge growth. A larger light-based $B/T$ may arise from genuine growth of the bulge, fading or truncation of the disc, or some combination of these effects.

A key strength of the present work is the combination of spectroscopic galaxy redshifts with X-ray-selected cluster catalogs. The eFEDS cluster catalog is based on extended eROSITA detections and subsequent optical characterization \citep{Liu2022,Klein2022}, while the eRASS1 catalog similarly combines extended X-ray detections with optical confirmation and characterization \citep{Bulbul2024,Kluge2024}. This substantially reduces the projection ambiguity associated with purely photometric overdensity selections, but does not produce a mass-complete or selection-neutral cluster sample. The probability of detecting an extended X-ray source depends on source flux, apparent extent, redshift, exposure, background, and source morphology, and simulations are required to characterize completeness and contamination \citep{Seppi2022}. Our permissive extent-likelihood selection should therefore be viewed as broadening the parent candidate set rather than guaranteeing completeness. The final 348-cluster subset additionally inherits the HSC imaging footprint and DESI spectroscopic targeting selection. Within these limits, the combined HSC--DESI--eROSITA sample enables the environmental trends to be separated by stellar mass, operational evolutionary class, and projected cluster-centric radius. In particular, the strongest multivariate offset is found for the H$\alpha$-low UVJ-SF transition class, which would be mixed into the broader star-forming population in a purely UVJ-based division.

\section{Conclusions}
\label{sec:conclusion}

Using HSC imaging, DESI spectroscopy, and eROSITA-selected clusters, we compared matched cluster satellites and non-cluster controls in SFR, size, fitted $B/T$, and $\mathrm{D_n}4000$. Within the UVJ-SF population we call objects below $\mathrm{sSFR}_{\mathrm{H}\alpha}=10^{-10.5}\,\mathrm{yr}^{-1}$ ``GV'' for brevity; this is an operational H$\alpha$-low definition, distinct from a conventional color-selected green valley. Our main conclusions are as follows.

\begin{enumerate}

\item {\bf Environmental effects are strongest among low-mass galaxies and become progressively weaker toward higher stellar masses.}

Using a covariance-aware Mahalanobis distance in SFR, effective radius, $B/T$, and $\mathrm{D_n}4000$, after matching in mass and redshift, we find larger cluster--field displacements at low mass than at high mass. This is consistent with greater environmental susceptibility of shallow-potential systems, although the metric is an effect size rather than a causal measure.

\item {\bf The cluster environment exerts its strongest influence within the cluster core and weakens with increasing cluster-centric radius.}

The multivariate offset and several individual residuals are largest inside $0.5R_{500}$ and generally decline outward. Because projected radius mixes orbital histories and backsplash populations, this is a radial association rather than a direct measurement of processing time or local three-dimensional density.

\item {\bf Star formation suppression appears to precede the strongest structural transformation.}

Cluster SF galaxies show SFR suppression with weaker structural residuals, whereas the H$\alpha$-low UVJ-SF sample shows joint offsets in size, fitted $B/T$, and $\mathrm{D_n}4000$. This ordering is consistent with star-formation suppression preceding the strongest changes in the stellar-light distribution, but it does not exclude progenitor bias, disk fading, or non-monotonic star-formation histories.

\item {\bf The strongest environmental signatures emerge in the green-valley population.}

Among the three operational classes, low-mass GV galaxies have the largest covariance-normalized cluster--field offset. This population is therefore the most sensitive tracer in our selection of the coexistence of suppressed recent star formation, older luminosity-weighted populations, and structural differences.

\item {\bf Quiescent galaxies retain a stronger memory of environmental history in their stellar populations than in their present-day structures.}

Cluster and field QGs have relatively weak residual differences in size and fitted $B/T$, while the $\mathrm{D_n}4000$ offset is clearer. This is consistent with different average quenching histories even after present-day structural relations have largely converged, subject to the metallicity dependence of $\mathrm{D_n}4000$.

\end{enumerate}

Taken together, these results support---but do not uniquely establish---a pathway in which gas-supply suppression can begin in cluster outskirts or pre-processing environments before the strongest structural differences appear. Orbital mixing, backsplash, rejuvenation, and preferential selection of compact progenitors remain viable. The main empirical conclusion is therefore narrower and more robust: after matching in stellar mass and redshift, the strongest joint cluster--field displacement is associated with low-mass H$\alpha$-low UVJ-SF satellites.


\begin{acknowledgments}
This work is supported by the National Natural
Science Foundation of China (Nos. 12533003, 12192222, 12192220, and 12121003).
This work made extensive use of the following software and analysis tools: \texttt{Astropy} \citep{Astropy2013,Astropy2018,Astropy2022}, \texttt{NumPy} \citep{Harris2020}, \texttt{SciPy} \citep{Virtanen2020}, \texttt{Matplotlib} \citep{Hunter2007}, \texttt{TOPCAT} \citep{Taylor2005}, \texttt{CIGALE} \citep{Boquien2019}, \texttt{EAZY} \citep{Brammer2008}, \texttt{GALIGHT} \citep{Ding2020}, and \texttt{GalSim} \citep{Rowe2015}. The analysis presented in this work was primarily performed using Python and its scientific computing ecosystem.
\end{acknowledgments}

%





\appendix
\setcounter{figure}{0}          
\setcounter{table}{0}           
\renewcommand{\thefigure}{A\arabic{figure}} 
\renewcommand{\thetable}{A\arabic{table}}   

\section{Magnitude-dependent Calibration of the Bulge-to-total Ratio}
\label{app:bt_calibration}

The reliability of a two-component bulge--disk decomposition depends on the signal-to-noise ratio and therefore on apparent magnitude. At progressively fainter magnitudes, the surface-brightness distribution becomes less well constrained and the recovered bulge-to-total light ratio, $B/T$, can become systematically biased. Because $B/T$ is used throughout this work as one of the primary structural diagnostics, we quantify this magnitude dependence with image simulations and derive an empirical calibration that is applied to the science catalog.

The simulated galaxies are analyzed with the same structural-fitting procedure adopted for the observed sample. We divide the simulations into five one-magnitude intervals spanning
\begin{equation}
18\leq i<23,
\end{equation}
and examine the relation between the input (true) and recovered (measured) $B/T$ independently in each interval. Only structural fits satisfying the adopted fit-quality criterion,
\begin{equation}
\chi_\nu^2 \leq 2,
\end{equation}
are used to determine the calibration relation.

\begin{figure*}[htbp]
    \centering
    \includegraphics[width=0.98\textwidth]{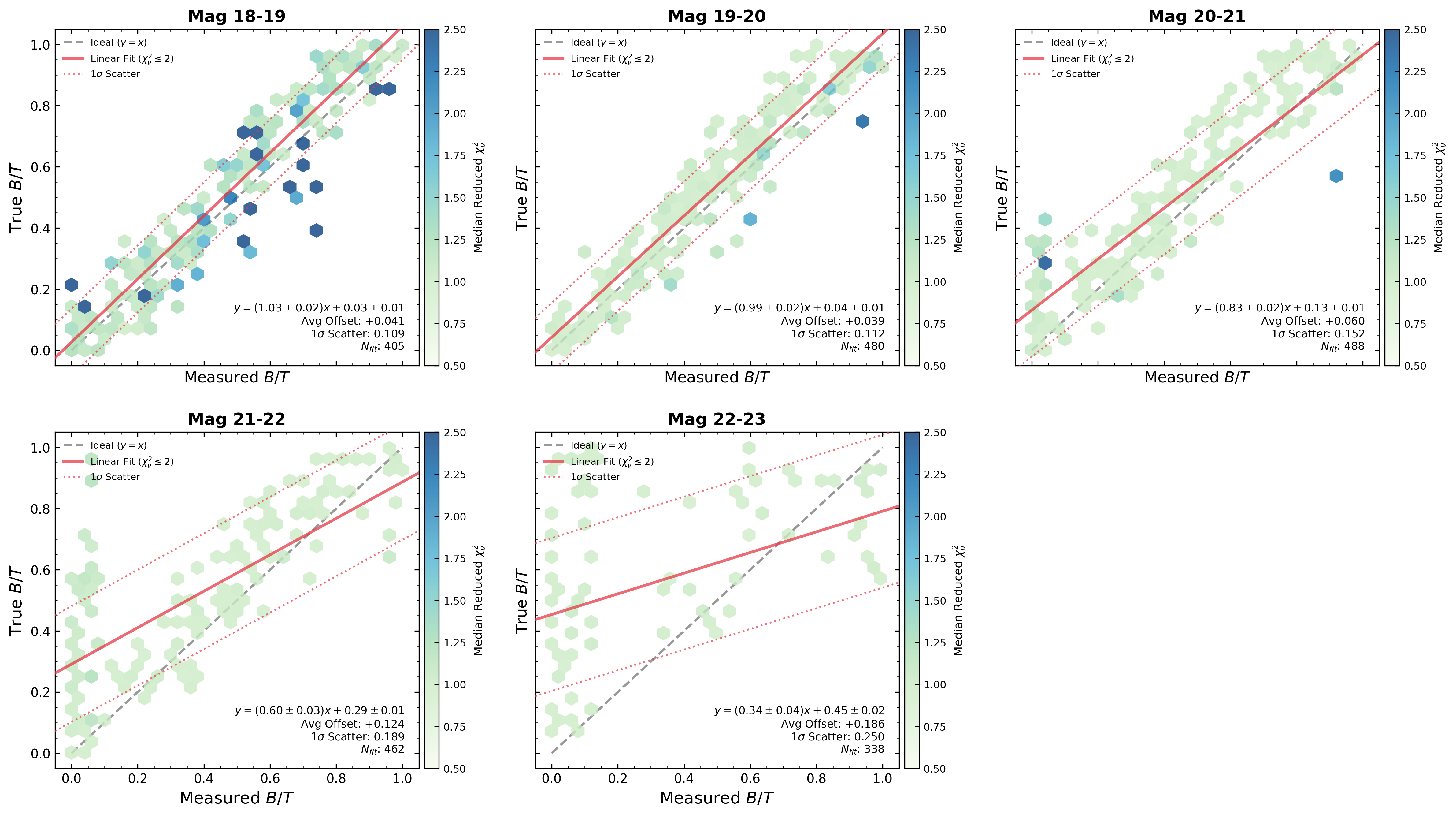}
    \caption{
    Magnitude-dependent recovery of the bulge-to-total ratio in the image simulations.
    The five panels show one-magnitude intervals spanning $18\leq i<23$.
    In each panel, the horizontal axis is the recovered value,
    $(B/T)_{\rm meas}$, and the vertical axis is the known input value,
    $(B/T)_{\rm true}$.
    Hexagonal bins show the simulated galaxies and are colored by the median reduced $\chi^2$ of the structural fits.
    The gray dashed line denotes perfect recovery,
    $(B/T)_{\rm true}=(B/T)_{\rm meas}$.
    The solid red line shows the linear calibration derived from simulations satisfying $\chi_\nu^2\leq2$, and the red dotted lines indicate the corresponding $1\sigma$ scatter.
    The best-fitting relation, mean true-minus-measured offset, scatter, and number of simulations entering the calibration fit are reported in each panel.
    Recovery is close to one-to-one at $i<20$, while both the systematic bias and scatter increase substantially toward the faintest magnitude intervals.
    }
    \label{fig:bt_recovery}
\end{figure*}

Figure~\ref{fig:bt_recovery} shows the resulting recovery tests. The horizontal axis is the measured value returned by the fitting pipeline, $(B/T)_{\rm meas}$, while the vertical axis is the true input value of the simulation, $(B/T)_{\rm true}$. The gray dashed line marks perfect recovery,
\begin{equation}
(B/T)_{\rm true}=(B/T)_{\rm meas}.
\end{equation}
The color coding indicates the median reduced $\chi^2$ of the simulated fits, and the solid red line shows the best-fitting empirical calibration relation for the subset satisfying $\chi_\nu^2\leq2$. The dotted red curves indicate the measured $1\sigma$ scatter about this relation.

Within each magnitude interval $m$, we fit
\begin{equation}
(B/T)_{\rm true}
=
a_m(B/T)_{\rm meas}+b_m.
\label{eq:bt_calibration}
\end{equation}
Because the regression is explicitly performed with true $B/T$ as the dependent variable and measured $B/T$ as the independent variable, Equation~\ref{eq:bt_calibration} can be applied directly to the observed measurements. The calibrated value used in the science analysis is therefore
\begin{equation}
(B/T)_{\rm corr}
=
a_m(B/T)_{\rm meas}+b_m,
\label{eq:bt_corrected}
\end{equation}
where the coefficients corresponding to the galaxy's apparent-magnitude interval are adopted. No inversion of the fitted relation is required.

The best-fitting calibration coefficients are listed in Table~\ref{tab:bt_calibration}. At the bright end, the relation remains close to one-to-one. For $18\leq i<19$, we find
\begin{equation}
(B/T)_{\rm true}
=
(1.03\pm0.02)(B/T)_{\rm meas}
+
(0.03\pm0.01),
\end{equation}
and for $19\leq i<20$,
\begin{equation}
(B/T)_{\rm true}
=
(0.99\pm0.02)(B/T)_{\rm meas}
+
(0.04\pm0.01).
\end{equation}
Thus, for the brightest simulated galaxies, the fitted slope is statistically consistent with unity and the mean recovery bias is small.

The calibration becomes progressively stronger toward fainter magnitudes. In the $20\leq i<21$ interval, the fitted slope decreases to
$0.83\pm0.02$, with an intercept of $0.13\pm0.01$. At $21\leq i<22$, the slope decreases further to $0.60\pm0.03$, while the intercept increases to $0.29\pm0.01$. In the faintest interval, $22\leq i<23$, the recovered $B/T$ values show substantial compression relative to the true dynamic range, with
\begin{equation}
(B/T)_{\rm true}
=
(0.34\pm0.04)(B/T)_{\rm meas}
+
(0.45\pm0.02).
\end{equation}
The increasing deviation from the one-to-one relation demonstrates that a single global correction would not provide an adequate description over the full magnitude range.

The mean recovery offset,
\begin{equation}
\left\langle
(B/T)_{\rm true}-(B/T)_{\rm meas}
\right\rangle,
\end{equation}
increases from approximately $0.04$ at $i<20$ to $0.19$ in the faintest magnitude interval. At the same time, the scatter about the calibration relation increases from $\simeq0.11$ at $i<20$ to $0.16$, $0.19$, and $0.25$ in the progressively fainter bins. This increasing scatter is important: Equation~\ref{eq:bt_corrected} removes the mean magnitude-dependent bias, but it does not remove the object-to-object uncertainty in the recovered structural parameters. Consequently, the calibrated $B/T$ values are most reliable for population-level statistical comparisons rather than for interpreting small differences in individual faint galaxies.

\begin{table}[htbp]
\centering
\caption{
Magnitude-dependent empirical calibration of the bulge-to-total ratio.
The fitted relation is
$(B/T)_{\rm true}=a_m(B/T)_{\rm meas}+b_m$.
The average offset is
$\langle(B/T)_{\rm true}-(B/T)_{\rm meas}\rangle$,
and $\sigma$ denotes the $1\sigma$ scatter about the fitted recovery relation.
Only simulations satisfying $\chi_\nu^2\leq2$ are included in the calibration fit.
}
\label{tab:bt_calibration}
\begin{tabular}{cccccc}
\hline\hline
$i$ range &
$a_m$ &
$b_m$ &
Avg. offset &
$\sigma$ &
$N_{\rm fit}$ \\
\hline
$18\leq i<19$ &
$1.03\pm0.02$ &
$0.03\pm0.01$ &
$+0.041$ &
$0.109$ &
405 \\

$19\leq i<20$ &
$0.99\pm0.02$ &
$0.04\pm0.01$ &
$+0.039$ &
$0.112$ &
480 \\

$20\leq i<21$ &
$0.83\pm0.02$ &
$0.13\pm0.01$ &
$+0.060$ &
$0.162$ &
488 \\

$21\leq i<22$ &
$0.60\pm0.03$ &
$0.29\pm0.01$ &
$+0.124$ &
$0.189$ &
462 \\

$22\leq i<23$ &
$0.34\pm0.04$ &
$0.45\pm0.02$ &
$+0.186$ &
$0.250$ &
338 \\
\hline
\end{tabular}
\end{table}

The behavior in Figure~\ref{fig:bt_recovery} also illustrates why the calibration must be interpreted as a statistical correction rather than an exact recovery of the intrinsic morphology. At bright magnitudes, the fitted slopes are close to unity and the scatter is comparatively small. At faint magnitudes, however, the fitted slope becomes substantially shallower than unity and the range of recoverable $B/T$ values becomes increasingly compressed. The correction therefore restores the mean relation between measured and true $B/T$, but the information content of an individual faint-galaxy decomposition remains limited by the substantially larger scatter.

The calibration also addresses only those biases represented by the simulations. In particular, it captures the magnitude-dependent recovery bias within the adopted parametric model family and under the image conditions represented by the mock sample. It does not explicitly calibrate structural errors caused by bars, spiral arms, tidal features, asymmetric components, or other deviations from the assumed bulge--disk model. Similarly, it does not isolate possible environment-dependent effects associated with crowding, blending, local sky subtraction, or diffuse intracluster light.

These limitations are relevant for the environmental analysis because the most crowded and highest-background regions may preferentially occur toward smaller projected cluster-centric radii. A magnitude-dependent calibration alone therefore cannot demonstrate that the structural measurements are completely independent of environment. For this reason, we interpret the corrected $B/T$ measurements primarily through population-level trends and in conjunction with the independent size, star-formation, and spectroscopic diagnostics. A more direct future assessment would inject mock galaxies into real cluster images over a controlled range of local background surface brightness, crowding, and projected cluster-centric radius.

Finally, Equation~\ref{eq:bt_corrected} should not be interpreted as reducing the random uncertainty quoted in Table~\ref{tab:bt_calibration}. The systematic mean mapping and the object-to-object scatter are distinct components of the recovery performance. The former is corrected empirically in the science catalog, whereas the latter remains part of the uncertainty budget and motivates caution when interpreting small $B/T$ differences, particularly at $i\gtrsim21$.

\section{Central Stellar-Mass Density as an Additional Structural Diagnostic}
\label{app:sigma1}

The main analysis uses the bulge-to-total light ratio, $B/T$, as one measure of structural transformation. Because a two-component bulge--disk decomposition is model dependent, we perform an additional test using the central stellar-mass surface density within 1~kpc,
$\Sigma_{1\,{\rm kpc}}$. This quantity provides a complementary measure of central concentration that does not rely explicitly on separating the galaxy into bulge and disk components. Previous studies have shown that central stellar-mass density is closely related to galaxy compactness and star-formation state and can provide a useful structural diagnostic of galaxies undergoing or having undergone quenching
\citep[e.g.,][]{Fang2013,Barro2017}.

We estimate $\Sigma_{1\,{\rm kpc}}$ from the best-fitting single-S\'ersic structural model. The projected S\'ersic surface-brightness profile is written as
\begin{equation}
I(R)
=
I_{\rm e}
\exp
\left[
-b_n
\left\{
\left(\frac{R}{R_{\rm e}}\right)^{1/n}
-1
\right\}
\right],
\label{eq:sersic_sigma1}
\end{equation}
where $R_{\rm e}$ is the effective radius, $n$ is the S\'ersic index, and $b_n$ is defined such that $R_{\rm e}$ encloses one half of the total S\'ersic luminosity
\citep{Sersic1968,Ciotti1991,GrahamDriver2005}. Specifically,
\begin{equation}
\Gamma(2n) = 2\,\gamma(2n,b_n),
\label{eq:bn_definition}
\end{equation}
where $\Gamma$ and $\gamma$ are the complete and lower incomplete gamma functions, respectively
\citep{Ciotti1991,CiottiBertin1999,GrahamDriver2005}.

For a S\'ersic profile, the fraction of the projected luminosity enclosed within radius $R$ is
\begin{equation}
f(<R)
=
\frac{
\gamma
\left[
2n,\,
b_n
\left(
\frac{R}{R_{\rm e}}
\right)^{1/n}
\right]
}{
\Gamma(2n)
}.
\label{eq:sersic_enclosed_fraction}
\end{equation}
Assuming that stellar mass follows the fitted light profile with a spatially constant mass-to-light ratio for an individual galaxy, the stellar mass enclosed within 1~kpc is therefore
\begin{equation}
M_\ast(<1\,{\rm kpc})
=
M_\ast
\frac{
\gamma
\left[
2n,\,
b_n
\left(
\frac{1\,{\rm kpc}}{R_{\rm e}}
\right)^{1/n}
\right]
}{
\Gamma(2n)
},
\label{eq:mstar_1kpc}
\end{equation}
and the corresponding projected central stellar-mass surface density is
\begin{equation}
\Sigma_{1\,{\rm kpc}}
=
\frac{
M_\ast(<1\,{\rm kpc})
}{
\pi(1\,{\rm kpc})^2
}.
\label{eq:sigma1_definition}
\end{equation}
We use $\log\Sigma_{1\,{\rm kpc}}$ in units of
$M_\odot\,{\rm kpc}^{-2}$ throughout the analysis.
This definition follows the commonly adopted central-density measure used in studies of galaxy structure and quenching
\citep[e.g.,][]{Fang2013,Barro2017}.

\begin{figure*}[htbp]
    \centering
    \includegraphics[width=0.98\textwidth]{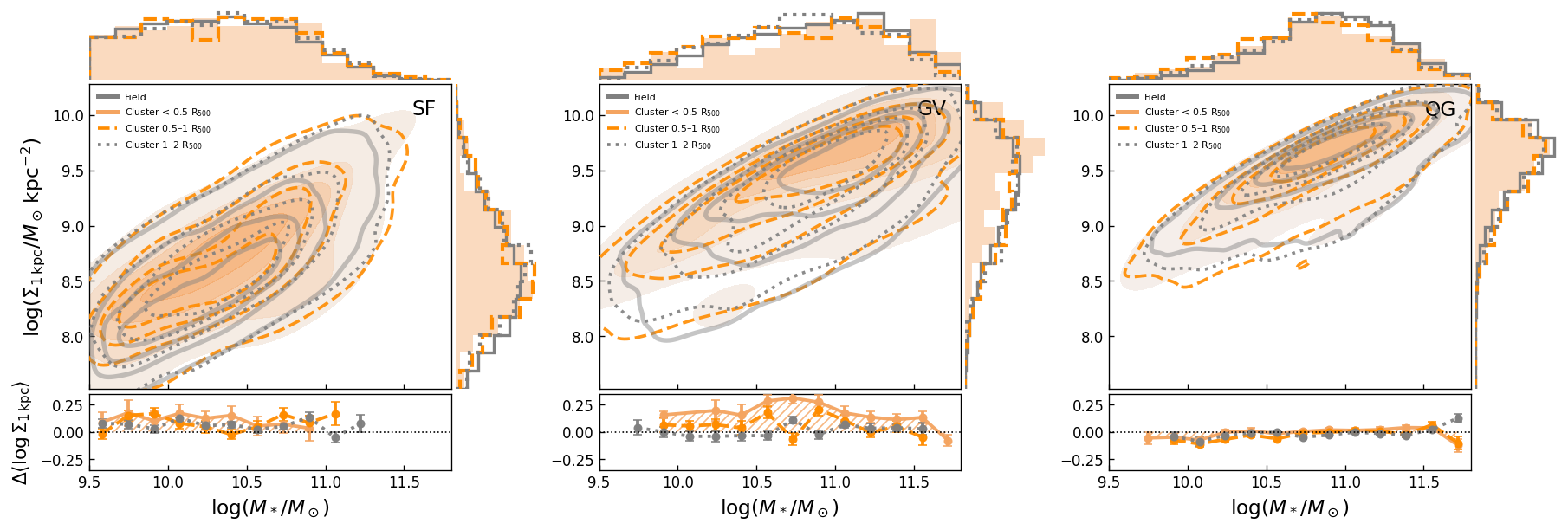}
    \caption{
    Central stellar-mass surface density as an additional structural diagnostic.
    From left to right, the panels show the SF, transition (GV), and quiescent populations in the
    $\log\Sigma_{1\,{\rm kpc}}$--$\log M_\ast$ plane.
    Gray contours show the matched field population, while the cluster samples are separated into
    $R<0.5R_{500}$,
    $0.5<R/R_{500}<1$, and
    $1<R/R_{500}<2$.
    The upper and right-hand marginal distributions show the corresponding stellar-mass and
    $\Sigma_{1\,{\rm kpc}}$ distributions.
    The lower panels show the cluster-minus-field residuals in
    $\log\Sigma_{1\,{\rm kpc}}$ after stellar-mass and redshift matching; the horizontal dotted line marks zero residual.
    Positive values indicate a larger projected stellar mass density within the central 1~kpc than in the matched field population.
    }
    \label{fig:sigma1_appendix_diagnostics}
\end{figure*}

Figure~\ref{fig:sigma1_appendix_diagnostics} presents the
$\Sigma_{1\,{\rm kpc}}$--stellar-mass relation separately for the SF, transition (GV), and quiescent populations.
The gray distributions show the matched field reference, while the cluster samples are divided into the same three projected radial intervals used in the main analysis:
$R<0.5R_{500}$,
$0.5<R/R_{500}<1$, and
$1<R/R_{500}<2$.
The marginal distributions above and to the right of the main panels illustrate the corresponding stellar-mass and central-density distributions, while the lower panels show the cluster-minus-field residual,
\begin{equation}
\Delta\log\Sigma_{1\,{\rm kpc}}
=
\log\Sigma_{1\,{\rm kpc},{\rm cl}}
-
\log\Sigma_{1\,{\rm kpc},{\rm field}},
\end{equation}
after the same stellar-mass and redshift matching adopted in the main analysis.

The central-density trends provide a useful cross-check on the $B/T$ results presented in the main text.
The transition population shows the clearest positive
$\Delta\log\Sigma_{1\,{\rm kpc}}$ signal, particularly at small and intermediate projected cluster-centric radii.
The magnitude of the offset is weaker and less coherent for the SF population, while the quiescent population is generally much closer to the matched field relation.
Thus, the population for which we find the strongest environmental enhancement in $B/T$ also tends to exhibit the largest enhancement in central stellar-mass density.

This qualitative agreement is important because $B/T$ and
$\Sigma_{1\,{\rm kpc}}$ probe structure in different ways.
The former is derived from an explicit bulge--disk decomposition and measures the relative contribution of the fitted bulge component to the total galaxy light.
In contrast, $\Sigma_{1\,{\rm kpc}}$ measures the total projected stellar mass inferred to lie within a fixed physical aperture and does not require assigning the central light to a separate bulge component.
The consistent behavior of the two diagnostics therefore argues against the environmental $B/T$ trend being produced solely by a peculiarity of the two-component decomposition.

At the same time, the two measurements are not statistically or physically independent.
Both ultimately depend on the galaxy surface-brightness distribution, and
$\Sigma_{1\,{\rm kpc}}$ additionally relies on the assumption that the stellar mass-to-light ratio does not vary strongly with radius.
Strong stellar-population or dust gradients could therefore cause the true stellar-mass profile to differ from the fitted light profile.
Moreover, because $\Sigma_{1\,{\rm kpc}}$ is inferred from the best-fitting S\'ersic model, the measurement can be affected by uncertainties in $n$ and $R_{\rm e}$, particularly when the physical 1-kpc aperture approaches or falls below the effective spatial resolution.

We therefore regard $\Sigma_{1\,{\rm kpc}}$ as a complementary structural consistency check rather than as a fully independent measurement of central stellar mass.
Within these limitations, the fact that the strongest positive central-density residuals occur in the same transition population that exhibits the strongest $B/T$ offsets supports the interpretation that the environmental structural signal is associated with an enhanced central concentration of the stellar-light distribution.
It does not, however, by itself demonstrate physical bulge mass growth: an increase in central concentration can also arise through changes in the relative prominence of the disk and central component, including disk fading or truncation.



\setcounter{table}{0}
\renewcommand{\thetable}{B\arabic{table}}

\section{Comparison with Previous Environmental Studies}
\label{app:literature_comparison}

The observational literature on environmental galaxy evolution is heterogeneous by construction: studies use different definitions of environment (e.g., cluster membership, host-halo mass, projected density, or cosmic-web class), different population selections, and different structural estimators and rest-frame wavelengths. The tables below are therefore intended as a structured comparison of representative results rather than as a formal meta-analysis. In particular, apparently different size results need not be mutually inconsistent when the samples differ in redshift, stellar mass, star-formation state, morphology, or size definition. This point is especially important for the present work, where the environmental signal changes substantially when galaxies are separated by evolutionary state and stellar mass.

\subsection{Environmental dependence of galaxy size}
\label{app:size_literature}

Table~\ref{tab:size_env_lit} summarizes representative measurements of the environmental dependence of galaxy size. We group the studies by their reported result, but retain the principal qualification in the final column. This avoids treating conditional results---for example, a trend confined to low-mass disks or to very massive early-type galaxies---as universal statements about all galaxies.

\begin{table*}[htbp]
\centering
\scriptsize
\caption{Representative observational studies of the environmental dependence of galaxy size. The categories describe the principal result reported by each study for the population stated in the table; they should not be interpreted as measurements made with identical environment definitions, selections, or size estimators.}
\label{tab:size_env_lit}
\resizebox{\textwidth}{!}{%
\begin{tabular}{lllllll}
\toprule
Reference & $z$ & Environment & Population & Size measure & Reported environmental trend & Important qualification \\
\midrule
\multicolumn{7}{l}{\textit{No significant or weak dependence for the quoted population}}\\
\citet{Maltby2010} & $\sim0.17$ & Cluster/field & E, S0, spirals & $R_{\rm e}$ & No trend for E/S0 and high-$M_\ast$ spirals & Low-$M_\ast$ spirals show a possible cluster-size reduction \\
\citet{HuertasCompany2013Local} & $\sim0$ & Halo mass; central/satellite & ETGs & $R_{\rm e}$ & No significant dependence at fixed $M_\ast$ & Rules out large ($\gtrsim30$--40\%) offsets for massive ETGs \\
\citet{Shankar2014} & $<0.3$ & Host-halo mass & ETGs & $R_{\rm e}$ & Observed SDSS sizes show no dependence at fixed $M_\ast$ & Used primarily to test hierarchical-model predictions \\
\citet{Rettura2010} & $\sim1.2$ & Cluster/field & Massive ETGs & $R_{\rm e}$ & Similar mass--size relations & Cluster and field stellar populations also compared \\
\citet{Kelkar2015} & $0.4$--$0.8$ & Cluster/field & Morphology- or color-matched & $R_{\rm e}$ & No significant difference & Average differences $\gtrsim10$--20\% are disfavored \\
\citet{Newman2014} & $1.80$ & Cluster/field & QGs & $R_{\rm e}$ & No significant difference after morphology mix is considered & JKCS~041; mean QG ages also similar to field \\
\citet{Saracco2017} & $1.2$--$1.4$ & Cluster/field & Ellipticals & $R_{\rm e}$, $\Sigma$ & Same structural relations at fixed $M_\ast$ & Field lacks the most massive/large ellipticals \\
\citet{Strazzullo2023} & $1.4$--$1.7$ & Cluster/field comparison & Massive cluster galaxies & $R_{\rm e}$, S\'ersic structure & No compelling cluster-specific size offset at fixed population & Very massive SPT-SZ clusters; central regions \\
\addlinespace[2pt]
\multicolumn{7}{l}{\textit{Weak or conditional dependence}}\\
\citet{ZhangYang2019} & $\lesssim0.2$ & Halo mass; central/satellite; cosmic web & Mixed & $R_{\rm e}$ & Weak halo/central--satellite dependence & Satellites slightly smaller at $9\lesssim\log M_\ast\lesssim10.5$; cosmic-web trend nearly negligible \\
\addlinespace[2pt]
\multicolumn{7}{l}{\textit{Larger sizes in denser environments for the quoted population}}\\
\citet{Yoon2017} & $0.10$--$0.15$ & Large-scale density & ETGs & Non-parametric radius & Dense-region ETGs larger & Trend appears for $M_\ast\gtrsim10^{11.2}M_\odot$; amplitude $\sim20$--40\% \\
\citet{Cooper2012} & $0.4$--$1.2$ & Local overdensity & Massive ETGs & $R_{\rm e}$ & Larger in denser regions & Interpreted in connection with earlier assembly \\
\citet{Lani2013} & $0.5$--$2.0$ & Local density & QGs and SFGs & $R_{\rm e}$ & Positive size--density relation mainly for QGs & Strength depends on redshift and population \\
\citet{Papovich2012} & $1.62$ & Cluster/field & Quiescent galaxies & $R_{\rm e}$ & Cluster QGs larger on average & High-redshift single-cluster comparison \\
\citet{Delaye2014} & $0.8$--$1.5$ & Cluster/field & Massive Q/ETGs & $R_{\rm e}$ & Mean cluster sizes larger & Driven partly by a high-size tail; median difference is weaker \\
\citet{Strazzullo2013} & $\sim2$ & Cluster/field & Passive ETGs & $R_{\rm e}$ & Cluster passive ETGs larger than field analogs & Cluster galaxies remain much smaller than local ETGs \\
\citet{Chan2018} & $1.39$--$1.61$ & Cluster/field & Massive passive galaxies & $R_{\rm e}$ & Passive galaxies in evolved clusters are larger & Significant cluster-to-cluster variation \\
\citet{Andreon2018} & $\lesssim2$ & Cluster/field & Red-sequence ETGs & $R_{\rm e}$ & Faster early size growth in clusters & Environmental difference decreases toward low $z$ \\
\citet{Noordeh2021} & $1.98$ & Cluster/field & Bulge-like QGs & $R_{\rm e}$ & Cluster QGs larger & $63^{+31}_{-24}$\% larger at fixed mass in XLSSC~122 \\
\citet{Afanasiev2023} & $1.4$--$2.8$ & Cluster/field & Passive ETGs & $R_{\rm e}$ & Massive cluster ETGs larger & Main signal at $\log(M_\ast/M_\odot)>10.5$ \\
\citet{Ghosh2024} & $0.3$--$0.7$ & Projected density & Mixed; morphology/SF split & Radius & Denser-region galaxies larger & Up to $\sim25$\%; weakens with redshift and depends on mass/morphology \\
\addlinespace[2pt]
\multicolumn{7}{l}{\textit{Smaller sizes in denser environments for the quoted population}}\\
\citet{Cebrian2014} & $<0.12$ & Local density & Early- and late-type & $R_{\rm e}$ & Dense-environment galaxies slightly smaller & Stronger for low-mass late types \\
\citet{Matharu2019} & $0.86$--$1.34$ & Cluster/field & SFGs and QGs & $R_{\rm e}$ & Cluster galaxies smaller & Offsets of order $\sim0.07$--0.08 dex for the quoted populations \\
\citet{Xu2023} & $2.51$ & Cluster/field comparison & Massive cluster galaxies & $R_{\rm e}$, morphology & Cluster galaxies more compact & High-$z$ starbursting cluster; different evolutionary regime \\
\citet{Chamba2024} & $\lesssim0.02$ & Fornax/field & Broad mass range, including dwarfs & Galaxy-edge radius & Fornax galaxies smaller & Uses an edge-based size metric rather than conventional $R_{\rm e}$ \\
\citet{George2025} & $0.1$--$0.85$ & Cluster/field & QGs & UV/optical $R_{\rm e}$ & Cluster QGs smaller & Difference is stronger in rest-UV than rest-optical \\
\bottomrule
\end{tabular}}
\end{table*}

The diversity in Table~\ref{tab:size_env_lit} motivates caution when reducing the size literature to a single ``cluster versus field'' sign. For example, \citet{Maltby2010} found no measurable environmental dependence for ellipticals, S0s, and high-mass spirals, while finding tentative evidence that low-mass cluster spirals are smaller. Conversely, the high-mass ETG signal reported by \citet{Yoon2017} appears only above $M_\ast\sim10^{11.2}M_\odot$. At high redshift, several studies report larger passive galaxies in dense environments \citep[e.g.,][]{Papovich2012,Delaye2014,Strazzullo2013,Noordeh2021,Afanasiev2023}, whereas other work finds no significant difference \citep[e.g.,][]{Newman2014,Saracco2017} or smaller systems in clusters for different populations and redshift ranges \citep[e.g.,][]{Matharu2019,Xu2023}. More recent work further shows that the inferred environmental signal can depend on the adopted size definition and rest-frame wavelength \citep{Chamba2024,George2025}. Our evolutionary-stage split should therefore be viewed as one additional axis that helps organize this heterogeneous literature, not as a unique explanation for all previous discrepancies.

\subsection{Star formation, quenching, and transition populations}
\label{app:sfr_literature}

The environmental dependence of star formation is considerably more robust than that of size, but even here it is important to distinguish a change in the \emph{fraction} of star-forming galaxies from a shift in the SFR distribution \emph{within} the star-forming population. This distinction is directly relevant to our detection of residual SFR suppression among galaxies that have not yet entered the fully quiescent class. Reviews of gas-removal and satellite-quenching processes are given by \citet{BoselliGavazzi2006} and \citet{Cortese2021}.

\begin{table*}[htbp]
\centering
\scriptsize
\caption{Representative environmental studies of star formation, quenching, and transition populations. The final column highlights the aspect most relevant to the interpretation in this work.}
\label{tab:sfr_env_lit}
\resizebox{\textwidth}{!}{%
\begin{tabular}{llllll}
\toprule
Reference & $z$ & Environment/sample & Principal SF result & Population-level implication & Relevance/caveat \\
\midrule
\citet{Kauffmann2004} & Low $z$ & SDSS; local density & SFH indicators vary strongly with environment at fixed $M_\ast$ & Dense regions host more evolved populations & Structural indicators show weaker residual trends \\
\citet{Balogh2004Ecology} & $0.05$--$0.095$ & SDSS+2dFGRS; density/groups & SF fraction declines with density & Strong population-mix effect & H$\alpha$ EW distribution of the SF population is nearly environment-independent \\
\citet{Peng2010} & $0$--$1$ & SDSS+zCOSMOS; density & Mass- and environment-quenching are statistically separable & Environment preferentially raises passive fraction & Framework rather than a cluster-specific mechanism \\
\citet{Peng2012} & Low $z$ & SDSS; centrals/satellites & Satellite quenching drives much of the environmental signal & Satellites are the environmentally affected population & Useful for interpreting our satellite-only analysis \\
\citet{Wetzel2012} & Low $z$ & SDSS groups/clusters & Satellite quenched fraction is enhanced & Bimodal SF distribution persists & Active satellites need not show a large continuous SFR shift \\
\citet{Wetzel2013} & Low $z$ & SDSS satellites & Delayed-then-rapid quenching histories & Environmental effects depend on time since infall & Projected radius is not itself a clock \\
\citet{Haines2015} & $\sim0.15$--$0.30$ & LoCuSS clusters & Slow suppression plus pre-processing is required & Quenching begins beyond the core for some systems & Supports a broad radial response \\
\citet{Jian2018} & $0.2$--$1.1$ & HSC groups/clusters & Environment enhances quenching relative to field & Strong mass dependence & Directly relevant HSC-based comparison \\
\citet{Coenda2018} & Low $z$ & Field/groups/X-ray clusters & GV properties and abundance depend on environment & Transition populations carry environmental information & GV definition is selection-dependent \\
\citet{Jian2020} & $0.2$--$1.1$ & HSC environments & GV evolution depends on environment and redshift & Transition statistics depend on normalization/selection & Avoids treating GV fraction as a universal quenching-rate proxy \\
\citet{McNab2021} & $\sim1$ & GOGREEN clusters/field & Transition excess is strongest at lower stellar mass & Environmental transition population is mass dependent & High-mass excess can be weak/absent \\
\citet{PaulinoAfonso2020} & $\sim0.8$ & VIS$^3$COS density field & Spectral SFH indicators vary with environment & Denser environments host more evolved SFHs & Connects emission lines and $D_n4000$ \\
\citet{Oman2021} & Low $z$ & SDSS+ALFALFA groups/clusters & Gas stripping precedes observable SF quenching & Different phases respond on different timescales & Supports caution in interpreting radial ordering temporally \\
\citet{Goubert2024} & Low $z$ & SDSS + hydrodynamical simulations & Environment is most predictive for low-mass satellite quenching & Intrinsic predictors dominate centrals/high-mass satellites & Closely related to our mass-dependent environmental response \\
\bottomrule
\end{tabular}}
\end{table*}

An important nuance is provided by \citet{Balogh2004Ecology}: the fraction of star-forming systems changes strongly with density, while the H$\alpha$ equivalent-width distribution among galaxies that remain star forming is approximately independent of environment in their low-redshift sample. The literature therefore does not require a universal continuous SFR suppression within every star-forming sample. Our residual result should instead be stated as an additional observational constraint for our particular mass, redshift, environment, and SFR selection. Likewise, transition-galaxy fractions are definition dependent \citep{Coenda2018,Jian2020,McNab2021}; we therefore interpret our operational H$\alpha$-low UVJ-SF class as a useful transition tracer rather than a unique or model-independent quenching stage.

\subsection{Stellar-population indicators}
\label{app:stellar_population_literature}

\begin{table*}[htbp]
\centering
\scriptsize
\caption{Representative studies using spectroscopic stellar-population or recent-SFH indicators in different environments.}
\label{tab:dn4000_env_lit}
\resizebox{\textwidth}{!}{%
\begin{tabular}{llllll}
\toprule
Reference & $z$ & Indicator(s) & Environment/sample & Reported result & Interpretation relevant here \\
\midrule
\citet{Balogh1999} & $0.18$--$0.55$ & D4000, H$\delta$, [O\,II] & CNOC1 clusters/field & Spectral populations vary with cluster radius/environment & Direct evidence for environmentally varying recent SFHs \\
\citet{Kauffmann2004} & Low $z$ & D4000, sSFR and structure & SDSS local density & SFH indicators show strong residual environment dependence & Spectral response can exceed structural response \\
\citet{PaulinoAfonso2020} & $\sim0.8$ & $D_n4000$, H$\delta$, [O\,II] & VIS$^3$COS density field & Spectral indices vary systematically with environment & Directly comparable to our $D_n4000$ analysis \\
\citet{Sobral2022} & $\sim0.6$--$1.0$ & $D_n4000$, H$\delta$ & LEGA-C environments; QGs & Environment-dependent stellar-population assembly histories & Demonstrates that different age-sensitive indices need not respond identically \\
\citet{Newman2014} & $1.80$ & Composite absorption-line spectra/ages & JKCS~041 cluster/field QGs & Mean QG ages are similar to field comparison & Useful high-$z$ null result; environment can change fractions without shifting mean QG age \\
\bottomrule
\end{tabular}}
\end{table*}

These studies support using $D_n4000$ as a tracer of differences in recent star-formation history, but not as a one-parameter age clock. In particular, $D_n4000$ is sensitive to both stellar age and metallicity, and its interpretation becomes increasingly degenerate for older populations. We therefore describe positive cluster--field $D_n4000$ residuals as evidence for systematically different, typically more evolved recent SFHs, rather than converting them directly into a unique age offset or quenching timescale.

\subsection{Morphology, bulge prominence, and structural transformation}
\label{app:morph_literature}

\begin{table*}[htbp]
\centering
\scriptsize
\caption{Representative environmental studies of morphology, bulge/disc structure, and quenching. ``Structure'' is used broadly because the studies employ different diagnostics (visual morphology, concentration, S\'ersic index, or bulge--disc decomposition).}
\label{tab:bt_morph_env_lit}
\resizebox{\textwidth}{!}{%
\begin{tabular}{llllll}
\toprule
Reference & $z$ & Structural diagnostic & Environment/sample & Main result & Relevance/caveat \\
\midrule
\citet{Dressler1980} & Low $z$ & Visual morphology & Rich-cluster projected density & Early-type fraction rises with density & Classical morphology--density relation \\
\citet{Kauffmann2004} & Low $z$ & Concentration/structure & SDSS local density & Structural residuals are weaker than SFH residuals at fixed $M_\ast$ & Environment acts strongly through population/SFH changes \\
\citet{Bamford2009} & Low $z$ & Galaxy Zoo morphology + color & Density/group-centric distance & Both color and morphology depend on environment & Color--density relation remains stronger at fixed mass \\
\citet{Skibba2009} & Low $z$ & Galaxy Zoo morphology/color & Marked correlations & Residual morphology--environment relation weakens strongly at fixed color & Highlights covariance between SF state and morphology \\
\citet{LacknerGunn2013} & Low $z$ & Bulge--disc decomposition & SDSS environment & Environment affects disc/bulge populations differently & A change in light-based $B/T$ need not imply pure bulge mass growth \\
\citet{Bluck2016} & Low $z$ & Central structure + quenching diagnostics & SDSS centrals/satellites & Internal and environmental predictors differ for centrals/satellites & Motivates separating satellite-specific environmental effects \\
\citet{Bluck2019} & Low $z$ & Stellar $B/T$ and other structure & SDSS/MaNGA; multiple environments & $\Delta$SFR and $M_\ast$ are the strongest predictors of $B/T$; environment is secondary but stronger for satellites & Direct analogue of our $\Delta$SFR--$\Delta(B/T)$ result \\
\citet{PaulinoAfonso2019} & $\sim0.8$ & Size, S\'ersic index, morphology & VIS$^3$COS density field & Structure depends on both nature and nurture, with strong mass/SF covariance & Relevant intermediate-$z$ comparison \\
\citet{Noordeh2021} & $1.98$ & S\'ersic index + size & XLSSC~122 cluster/field & Excess bulge-like QGs and larger QG sizes in the cluster & Demonstrates coupled population/structural differences at high $z$ \\
\citet{Oxland2024} & Low $z$ & Morphology + sSFR after infall & Groups/clusters & Quenching and morphological transformation both evolve after infall & Their characteristic responses need not be synchronous; strongest environmental effects at low mass \\
\bottomrule
\end{tabular}}
\end{table*}

The terminology is important for the interpretation of our $B/T$ result. A larger light-based $B/T$ can reflect genuine growth of the bulge, fading or truncation of the disk, or both. We therefore use ``enhanced bulge prominence'' or ``structural transformation'' rather than treating $\Delta(B/T)>0$ as direct evidence for bulge stellar-mass growth. This conservative interpretation is consistent with studies showing that star-formation state, stellar mass, and morphology are strongly covariant and that residual environmental effects can remain for satellites \citep{Bamford2009,Skibba2009,Bluck2019,Oxland2024}.

\subsection{Studies jointly connecting star formation and structure}
\label{app:cross_parameter_literature}

Table~\ref{tab:cross_parameter_lit} collects a smaller set of studies that are particularly useful for interpreting our multivariate analysis because they explicitly connect more than one of star formation, stellar populations, and structure. We intentionally omit non-environmental age--size studies from this table; such work can inform physical interpretation but is not a direct environmental comparison.

\begin{table*}[htbp]
\centering
\scriptsize
\caption{Selected studies that jointly connect star-formation state and galaxy structure in an environmental context. The final column states the specific connection to the present analysis rather than assigning a subjective ``agreement'' score.}
\label{tab:cross_parameter_lit}
\resizebox{\textwidth}{!}{%
\begin{tabular}{lllllll}
\toprule
Reference & $z$ & Environment/sample & Star-formation/stellar-population result & Structural result & Main interpretation & Connection to this work \\
\midrule
\citet{Kauffmann2004} & Low $z$ & SDSS local density & Strong environmental differences in SFH indicators & Weaker structural residuals at fixed $M_\ast$ & SFH is highly environment sensitive & Supports treating SFH and structure as correlated but non-equivalent responses \\
\citet{Bamford2009} & Low $z$ & Galaxy Zoo + SDSS & Color strongly environment dependent & Morphology also environment dependent & Color and morphology respond differently & Motivates avoiding a single-property environmental interpretation \\
\citet{Bluck2019} & Low $z$ & SDSS/MaNGA; centrals/satellites & $\Delta$SFR is a dominant predictor & $B/T$ tightly coupled to $\Delta$SFR and $M_\ast$ & Structure--quenching coupling exceeds most direct environment predictors & Closely parallels our significant $\Delta$SFR--$\Delta(B/T)$ relation \\
\citet{PaulinoAfonso2019} & $\sim0.8$ & VIS$^3$COS density field & SF state covaries with environment & Size/S\'ersic/morphology studied versus density & Nature and nurture both contribute & Structural companion to the VIS$^3$COS spectral analysis \\
\citet{PaulinoAfonso2020} & $\sim0.8$ & VIS$^3$COS density field & [O\,II], H$\delta$, and $D_n4000$ vary with environment & Interpreted together with morphology work & Environment modifies recent SFHs & Direct comparison for our spectroscopic residuals \\
\citet{Matharu2019} & $0.86$--$1.34$ & GCLASS cluster/field & Galaxies separated by SF state & Cluster SFGs and QGs are smaller than field analogs & Environment-dependent size evolution is population dependent & Relevant to our stage-dependent $R_{\rm e}$ residuals \\
\citet{Matharu2020} & $0.86$--$1.34$ & Cluster post-starbursts & PSBs identify recently quenched systems & PSB sizes are intermediate between SFGs and QGs & Fading/quenching can alter light-weighted structure & Relevant to the strong structural signal in our transition population \\
\citet{Matharu2021} & $\sim1$ & Cluster/field; resolved H$\alpha$ & Recently quenched systems show rapid environmental suppression & H$\alpha$ spatial extent is truncated relative to stellar continuum & Supports outside-in quenching of star formation & Physical analogue for spatially ordered SF suppression, not a direct continuum-size measurement \\
\citet{Noordeh2021} & $1.98$ & XLSSC~122 cluster/field & Strongly enhanced QG fraction & Excess bulge-like QGs; larger QG sizes & Population mix and structure are jointly environment dependent & Demonstrates that structure and quenching can covary at high $z$ with a different sign in size \\
\citet{Oxland2024} & Low $z$ & Groups/clusters; infall histories & sSFR decreases after infall & Morphological transformation also proceeds after infall & Quenching and morphology have related but non-identical evolution & Directly supports our cautious ``accompanied by'' language \\
\citet{Xu2023} & $2.51$ & High-$z$ cluster/field comparison & Active cluster population in a starbursting environment & Accelerated structural evolution/compactness & Environment can couple SF state and structure already at high $z$ & Provides a high-$z$ comparison, not a direct analogue of the low-$z$ pathway \\
\bottomrule
\end{tabular}}
\end{table*}

Taken together, the literature supports three conservative conclusions that are directly relevant to our analysis. First, environmental quenching is robust at the population level, but the amount of SFR suppression within galaxies that remain formally star forming is sample dependent \citep{Balogh2004Ecology,Wetzel2012,Jian2018}. Second, the sign and amplitude of the environmental size residual depend on population selection, mass, redshift, and measurement definition \citep{Maltby2010,Kelkar2015,Delaye2014,Matharu2019,Afanasiev2023,Ghosh2024,Chamba2024,George2025}. Third, morphology and star-formation state are tightly covariant, so an environmental $B/T$ residual should not be interpreted independently of the accompanying SFR and stellar-population residuals \citep{Bamford2009,Skibba2009,Bluck2019,Oxland2024}. These considerations motivate the matched, stage-resolved, and multivariate analysis adopted in the main text.

\bibliography{references}{}

@article{Bond1996,
  author = {{Bond}, J.~R. and {Kofman}, L. and {Pogosyan}, D.},
  title = {{How filaments of galaxies are woven into the cosmic web}},
  journal = {Nature},
  year = {1996},
  month = {apr},
  volume = {380},
  number = {6575},
  pages = {603--606},
  doi = {10.1038/380603a0},
  url = {https://doi.org/10.1038/380603a0},
  adsurl = {https://ui.adsabs.harvard.edu/abs/1996Natur.380..603B}
}

@article{Dressler1980,
  author = {{Dressler}, Alan},
  title = {{Galaxy morphology in rich clusters: implications for the formation and evolution of galaxies}},
  journal = {ApJ},
  year = {1980},
  month = {mar},
  volume = {236},
  pages = {351},
  doi = {10.1086/157753},
  url = {https://doi.org/10.1086/157753},
  adsurl = {https://ui.adsabs.harvard.edu/abs/1980ApJ...236..351D}
}

@article{BlantonMoustakas2009,
  author = {{Blanton}, M.~R. and {Moustakas}, J.},
  title = {{Physical Properties and Environments of Nearby Galaxies}},
  journal = {ARA\&A},
  year = {2009},
  month = {sep},
  volume = {47},
  number = {1},
  pages = {159--210},
  doi = {10.1146/annurev-astro-082708-101734},
  url = {https://doi.org/10.1146/annurev-astro-082708-101734},
  adsurl = {https://ui.adsabs.harvard.edu/abs/2009ARA%26A..47..159B}
}

@article{Strateva2001,
  author = {{Strateva}, Iskra and {Ivezi\'c}, \v{Z}eljko and {Knapp}, Gillian R. and {Narayanan}, Vijay K. and {Strauss}, Michael A. and {Gunn}, James E. and {Lupton}, Robert H. and {Schlegel}, David and {Bahcall}, Neta A. and {Brinkmann}, Jon and {Brunner}, Robert J. and {Budav\'ari}, Tam\'as and {Csabai}, Istv\'an and {Castander}, Francisco Javier and {Doi}, Mamoru and {Fukugita}, Masataka and {Gy\H{o}ry}, Zsuzsanna and {Hamabe}, Masaru and {Hennessy}, Greg and {Ichikawa}, Takashi and {Kunszt}, Peter Z. and {Lamb}, Don Q. and {McKay}, Timothy A. and {Okamura}, Sadanori and {Racusin}, Judith and {Sekiguchi}, Maki and {Schneider}, Donald P. and {Shimasaku}, Kazuhiro and {York}, Donald},
  title = {{Color Separation of Galaxy Types in the Sloan Digital Sky Survey Imaging Data}},
  journal = {AJ},
  year = {2001},
  month = {oct},
  volume = {122},
  number = {4},
  pages = {1861--1874},
  doi = {10.1086/323301},
  url = {https://doi.org/10.1086/323301},
  adsurl = {https://ui.adsabs.harvard.edu/abs/2001AJ....122.1861S}
}

@article{Thomas2005,
  author = {{Thomas}, Daniel and {Maraston}, Claudia and {Bender}, Ralf and {Mendes de Oliveira}, Claudia},
  title = {{The Epochs of Early-Type Galaxy Formation as a Function of Environment}},
  journal = {ApJ},
  year = {2005},
  month = {mar},
  volume = {621},
  number = {2},
  pages = {673--694},
  doi = {10.1086/426932},
  url = {https://doi.org/10.1086/426932},
  archivePrefix = {arXiv},
  eprint = {astro-ph/0410209},
  primaryClass = {astro-ph},
  adsurl = {https://ui.adsabs.harvard.edu/abs/2005ApJ...621..673T}
}

@article{Noordeh2021,
  author = {{Noordeh}, E. and {Canning}, R.~E.~A. and {Willis}, J.~P. and {Allen}, S.~W. and {Mantz}, A. and {Stanford}, S.~A. and {Brammer}, G.},
  title = {{Quiescent galaxies in a virialized cluster at redshift 2: evidence for accelerated size growth}},
  journal = {MNRAS},
  year = {2021},
  month = {sep},
  volume = {507},
  number = {4},
  pages = {5272--5280},
  doi = {10.1093/mnras/stab2459},
  url = {https://doi.org/10.1093/mnras/stab2459},
  archivePrefix = {arXiv},
  eprint = {2109.02200},
  primaryClass = {astro-ph.GA},
  adsurl = {https://ui.adsabs.harvard.edu/abs/2021MNRAS.507.5272N}
}

@article{George2025,
  author = {{George}, Angelo and {Damjanov}, Ivana and {Sawicki}, Marcin and {Williams}, Devin J. and {Chen}, Lingjian and {Desprez}, Guillaume and {Annunziatella}, Marianna and {Arnouts}, St{\'e}phane and {Gwyn}, Stephen and {Marchesini}, Danilo and {Moutard}, Thibaud and {Sajina}, Anna},
  title = {{Effects of Environment on the Size Evolution of Quiescent Galaxies: Comparing Galaxies in Clusters and in the Field at Two Rest-frame Wavelengths}},
  journal = {ApJ},
  year = {2025},
  month = {jul},
  volume = {987},
  number = {1},
  eid = {45},
  pages = {45},
  doi = {10.3847/1538-4357/addc6b},
  url = {https://doi.org/10.3847/1538-4357/addc6b},
  archivePrefix = {arXiv},
  eprint = {2503.22788},
  primaryClass = {astro-ph.GA},
  adsurl = {https://ui.adsabs.harvard.edu/abs/2025ApJ...987...45G}
}

@article{Baldry2004,
  author = {{Baldry}, Ivan K. and {Glazebrook}, Karl and {Brinkmann}, Jon and {Ivezi\'c}, \v{Z}eljko and {Lupton}, Robert H. and {Nichol}, Robert C. and {Szalay}, Alexander S.},
  title = {{Quantifying the Bimodal Color-Magnitude Distribution of Galaxies}},
  journal = {ApJ},
  year = {2004},
  month = {jan},
  volume = {600},
  number = {2},
  pages = {681--694},
  doi = {10.1086/380092},
  url = {https://doi.org/10.1086/380092},
  adsurl = {https://ui.adsabs.harvard.edu/abs/2004ApJ...600..681B}
}

@article{Wetzel2012,
  author = {{Wetzel}, Andrew R. and {Tinker}, Jeremy L. and {Conroy}, Charlie},
  title = {{Galaxy evolution in groups and clusters: star formation rates, red sequence fractions and the persistent bimodality}},
  journal = {MNRAS},
  year = {2012},
  month = {jul},
  volume = {424},
  number = {1},
  pages = {232--243},
  doi = {10.1111/j.1365-2966.2012.21188.x},
  url = {https://doi.org/10.1111/j.1365-2966.2012.21188.x},
  adsurl = {https://ui.adsabs.harvard.edu/abs/2012MNRAS.424..232W}
}

@article{vanDerBurg2018,
  author = {{van der Burg}, Remco F. J. and {McGee}, Sean and {Aussel}, Herv\'e and {Dahle}, H\r{a}kon and {Arnaud}, Monique and {Pratt}, Gabriel W. and {Muzzin}, Adam},
  title = {{The stellar mass function of galaxies in Planck-selected clusters at $0.5 < z < 0.7$: new constraints on the timescale and location of satellite quenching}},
  journal = {Astronomy \& Astrophysics},
  year = {2018},
  month = {oct},
  volume = {618},
  pages = {A140},
  doi = {10.1051/0004-6361/201833572},
  url = {https://doi.org/10.1051/0004-6361/201833572},
  adsurl = {https://ui.adsabs.harvard.edu/abs/2018A%26A...618A.140V}
}

@article{GunnGott1972,
  author = {{Gunn}, J.~E. and {Gott}, J.~Richard III},
  title = {{On the Infall of Matter Into Clusters of Galaxies and Some Effects on Their Evolution}},
  journal = {ApJ},
  year = {1972},
  month = {aug},
  volume = {176},
  pages = {1},
  doi = {10.1086/151605},
  url = {https://doi.org/10.1086/151605},
  adsurl = {https://ui.adsabs.harvard.edu/abs/1972ApJ...176....1G}
}

@article{BoselliGavazzi2006,
  author = {{Boselli}, Alessandro and {Gavazzi}, Giuseppe},
  title = {{Environmental Effects on Late-Type Galaxies in Nearby Clusters}},
  journal = {PASP},
  year = {2006},
  month = {apr},
  volume = {118},
  number = {842},
  pages = {517--559},
  doi = {10.1086/500691},
  url = {https://doi.org/10.1086/500691},
  adsurl = {https://ui.adsabs.harvard.edu/abs/2006PASP..118..517B}
}

@article{Moore1996,
  author = {{Moore}, B. and {Katz}, N. and {Lake}, G. and {Dressler}, A. and {Oemler}, A.},
  title = {{Galaxy harassment and the evolution of clusters of galaxies}},
  journal = {Nature},
  year = {1996},
  month = {feb},
  volume = {379},
  number = {6566},
  pages = {613--616},
  doi = {10.1038/379613a0},
  url = {https://doi.org/10.1038/379613a0},
  adsurl = {https://ui.adsabs.harvard.edu/abs/1996Natur.379..613M}
}

@article{Merritt1983,
  author = {{Merritt}, D.},
  title = {{Relaxation and tidal stripping in rich clusters of galaxies. I. Evolution of the mass distribution}},
  journal = {ApJ},
  year = {1983},
  month = {jan},
  volume = {264},
  pages = {24--48},
  doi = {10.1086/160571},
  url = {https://doi.org/10.1086/160571},
  adsurl = {https://ui.adsabs.harvard.edu/abs/1983ApJ...264...24M}
}

@article{Boselli2022,
  author = {{Boselli}, A. and {Fossati}, M. and {Sun}, M.},
  title = {{Ram pressure stripping in high-density environments}},
  journal = {The Astronomy and Astrophysics Review},
  year = {2022},
  month = {dec},
  volume = {30},
  number = {1},
  pages = {3},
  doi = {10.1007/s00159-022-00140-3},
  url = {https://doi.org/10.1007/s00159-022-00140-3},
  adsurl = {https://ui.adsabs.harvard.edu/abs/2022A%26ARv..30....3B}
}

@article{Jaffe2015,
  author = {{Jaff\'e}, Yara L. and {Smith}, Rory and {Candlish}, Graeme N. and {Poggianti}, Bianca M. and {Sheen}, Yun-Kyeong and {Verheijen}, Marc A. W.},
  title = {{BUDHIES II: a phase-space view of H I gas stripping and star formation quenching in cluster galaxies}},
  journal = {MNRAS},
  year = {2015},
  month = {apr},
  volume = {448},
  number = {2},
  pages = {1715--1728},
  doi = {10.1093/mnras/stv100},
  url = {https://doi.org/10.1093/mnras/stv100},
  adsurl = {https://ui.adsabs.harvard.edu/abs/2015MNRAS.448.1715J}
}

@article{OmanHudson2016,
  author = {{Oman}, K.~A. and {Hudson}, M.~J.},
  title = {{Satellite quenching time-scales in clusters from projected phase space measurements matched to simulated orbits}},
  journal = {MNRAS},
  year = {2016},
  month = {dec},
  volume = {463},
  number = {3},
  pages = {3083--3095},
  doi = {10.1093/mnras/stw2195},
  url = {https://doi.org/10.1093/mnras/stw2195},
  adsurl = {https://ui.adsabs.harvard.edu/abs/2016MNRAS.463.3083O}
}

@article{Daddi2005,
  author = {{Daddi}, E. and {Renzini}, A. and {Pirzkal}, N. and {Cimatti}, A. and {Malhotra}, S. and {Stiavelli}, M. and {Xu}, C. and {Pasquali}, A. and {Rhoads}, J. E. and {Brusa}, M. and {di Serego Alighieri}, S. and {Ferguson}, H. C. and {Koekemoer}, A. M. and {Moustakas}, L. A. and {Panagia}, N. and {Windhorst}, R. A.},
  title = {{Passively Evolving Early-Type Galaxies at 1.4 < z < 2.5 in the Hubble Ultra Deep Field}},
  journal = {ApJ},
  year = {2005},
  month = {jun},
  volume = {626},
  number = {2},
  pages = {680--697},
  doi = {10.1086/430104},
  url = {https://doi.org/10.1086/430104},
  adsurl = {https://ui.adsabs.harvard.edu/abs/2005ApJ...626..680D}
}

@article{Trujillo2006,
  author = {{Trujillo}, Ignacio and {Forster Schreiber}, Natascha M. and {Rudnick}, Gregory and {Barden}, Marco and {Franx}, Marijn and {Rix}, HansWalter and {Caldwell}, J. A. R. and {McIntosh}, Daniel H. and {Toft}, Sune and {Haussler}, Boris and {Zirm}, Andrew and {van Dokkum}, Pieter G. and {Labbe}, Ivo and {Moorwood}, Alan and {Rottgering}, Huub and {van der Wel}, Arjen and {van der Werf}, Paul and {van Starkenburg}, Lottie},
  title = {{The Size Evolution of Galaxies since $z \sim 3$: Combining SDSS, GEMS, and FIRES}},
  journal = {ApJ},
  year = {2006},
  month = {oct},
  volume = {650},
  number = {1},
  pages = {18--41},
  doi = {10.1086/506464},
  url = {https://doi.org/10.1086/506464},
  adsurl = {https://ui.adsabs.harvard.edu/abs/2006ApJ...650...18T}
}

@article{Shen2003,
  author = {{Shen}, S. and {Mo}, H. J. and {White}, S. D. M. and {Blanton}, M. R. and {Kauffmann}, G. and {Voges}, W. and {Brinkmann}, J. and {Csabai}, I.},
  title = {{The size distribution of galaxies in the Sloan Digital Sky Survey}},
  journal = {MNRAS},
  year = {2003},
  month = {aug},
  volume = {343},
  number = {3},
  pages = {978--994},
  doi = {10.1046/j.1365-8711.2003.06740.x},
  url = {https://doi.org/10.1046/j.1365-8711.2003.06740.x},
  adsurl = {https://ui.adsabs.harvard.edu/abs/2003MNRAS.343..978S}
}

@article{vanDerWel2014,
  author = {{van der Wel}, A. and {Franx}, M. and {van Dokkum}, P. G. and {Skelton}, R. E. and {Momcheva}, I. G. and {Whitaker}, K. E. and {Brammer}, G. B. and {Bell}, E. F. and {Rix}, H.-W. and {Wuyts}, S. and {Ferguson}, H. C. and {Holden}, B. P. and {Barro}, G. and {Koekemoer}, A. M. and {Chang}, Yu-Yen and {McGrath}, E. J. and {H\"aussler}, B. and {Dekel}, A. and {Behroozi}, P. and {Fumagalli}, M. and {Leja}, J. and {Lundgren}, B. F. and {Maseda}, M. V. and {Nelson}, E. J. and {Wake}, D. A. and {Patel}, S. G. and {Labb\'e}, I. and {Faber}, S. M. and {Grogin}, N. A. and {Kocevski}, D. D.},
  title = {{3D-HST+CANDELS: The Evolution of the Galaxy Size-Mass Distribution since z = 3}},
  journal = {ApJ},
  year = {2014},
  month = {may},
  volume = {788},
  number = {1},
  eid = {28},
  pages = {28},
  doi = {10.1088/0004-637X/788/1/28},
  url = {https://doi.org/10.1088/0004-637X/788/1/28},
  adsurl = {https://ui.adsabs.harvard.edu/abs/2014ApJ...788...28V}
}

@article{Lange2015,
  author = {{Lange}, Rebecca and {Driver}, Simon P. and {Robotham}, Aaron S. G. and {Kelvin}, Lee S. and {Graham}, Alister W. and {Alpaslan}, Mehmet and {Andrews}, Stephen K. and {Baldry}, Ivan K. and {Bamford}, Steven and {Bland-Hawthorn}, Joss and {Brough}, Sarah and {Cluver}, Michelle E. and {Conselice}, Christopher J. and {Davies}, Luke J. M. and {Haeussler}, Boris and {Konstantopoulos}, Iraklis S. and {Loveday}, Jon and {Moffett}, Amanda J. and {Norberg}, Peder and {Phillipps}, Steven and {Taylor}, Edward N. and {L\'opez-S\'anchez}, \'Angel R. and {Wilkins}, Stephen M.},
  title = {{Galaxy And Mass Assembly (GAMA): mass-size relations of z < 0.1 galaxies subdivided by S{\'e}rsic index, colour and morphology}},
  journal = {MNRAS},
  year = {2015},
  month = {mar},
  volume = {447},
  number = {3},
  pages = {2603--2630},
  doi = {10.1093/mnras/stu2467},
  url = {https://doi.org/10.1093/mnras/stu2467},
  adsurl = {https://ui.adsabs.harvard.edu/abs/2015MNRAS.447.2603L}
}

@article{Kelkar2015,
  author = {{Kelkar}, Kshitija and {Arag{\'o}n-Salamanca}, Alfonso and {Gray}, Meghan E. and {Maltby}, David and {Vulcani}, Benedetta and {De Lucia}, Gabriella and {Poggianti}, Bianca M. and {Zaritsky}, Dennis},
  title = {{Galaxy sizes as a function of environment at intermediate redshift from the ESO Distant Cluster Survey}},
  journal = {MNRAS},
  year = {2015},
  month = {jun},
  volume = {450},
  number = {2},
  pages = {1246--1255},
  doi = {10.1093/mnras/stv670},
  url = {https://doi.org/10.1093/mnras/stv670},
  adsurl = {https://ui.adsabs.harvard.edu/abs/2015MNRAS.450.1246K}
}

@article{Kuchner2017,
  author = {{Kuchner}, U. and {Ziegler}, B. and {Verdugo}, M. and {Bamford}, S. and {H\"au{\ss}ler}, B.},
  title = {{The effects of the cluster environment on the galaxy mass-size relation in MACS J1206.2-0847}},
  journal = {Astronomy \& Astrophysics},
  year = {2017},
  month = {aug},
  volume = {604},
  pages = {A54},
  doi = {10.1051/0004-6361/201630252},
  url = {https://doi.org/10.1051/0004-6361/201630252},
  adsurl = {https://ui.adsabs.harvard.edu/abs/2017A%26A...604A..54K}
}

@article{Valentinuzzi2010,
  author = {{Valentinuzzi}, T. and {Fritz}, J. and {Poggianti}, B. M. and {Cava}, A. and {Bettoni}, D. and {Fasano}, G. and {D'Onofrio}, M. and {Couch}, W. J. and {Dressler}, A. and {Moles}, M. and {Moretti}, A. and {Omizzolo}, A. and {Kj{\ae}rgaard}, P. and {Vanzella}, E. and {Varela}, J.},
  title = {{Superdense massive galaxies in WINGS local clusters}},
  journal = {ApJ},
  year = {2010},
  month = {mar},
  volume = {712},
  number = {1},
  pages = {226--237},
  doi = {10.1088/0004-637X/712/1/226},
  url = {https://doi.org/10.1088/0004-637X/712/1/226},
  adsurl = {https://ui.adsabs.harvard.edu/abs/2010ApJ...712..226V}
}

@article{Poggianti2013,
  author = {{Poggianti}, B. M. and {Calvi}, R. and {Bindoni}, D. and {D'Onofrio}, M. and {Moretti}, A. and {Valentinuzzi}, T. and {Fasano}, G. and {Fritz}, J. and {De Lucia}, G. and {Vulcani}, B. and {Bettoni}, D. and {Gullieuszik}, M. and {Omizzolo}, A.},
  title = {{Superdense Galaxies and the Mass-Size Relation at Low Redshift}},
  journal = {ApJ},
  year = {2013},
  month = {jan},
  volume = {762},
  number = {2},
  eid = {77},
  pages = {77},
  doi = {10.1088/0004-637X/762/2/77},
  url = {https://doi.org/10.1088/0004-637X/762/2/77},
  adsurl = {https://ui.adsabs.harvard.edu/abs/2013ApJ...762...77P}
}

@article{Raichoor2012,
  author = {{Raichoor}, A. and {Mei}, S. and {Stanford}, S. A. and {Holden}, B. P. and {Nakata}, F. and {Rosati}, P. and {Shankar}, F. and {Tanaka}, M. and {Ford}, H. and {Huertas-Company}, M. and {Illingworth}, G. and {Kodama}, T. and {Postman}, M. and {Rettura}, A. and {Blakeslee}, J. P. and {Demarco}, R. and {Jee}, M. J. and {White}, R. L.},
  title = {{Early-type Galaxies at $z \sim 1.3$. IV. Scaling Relations in Different Environments}},
  journal = {ApJ},
  year = {2012},
  month = {feb},
  volume = {745},
  number = {2},
  eid = {130},
  pages = {130},
  doi = {10.1088/0004-637X/745/2/130},
  url = {https://doi.org/10.1088/0004-637X/745/2/130},
  adsurl = {https://ui.adsabs.harvard.edu/abs/2012ApJ...745..130R}
}

@article{Delaye2014,
  author = {{Delaye}, L. and {Huertas-Company}, M. and {Mei}, S. and {Lidman}, C. and {Licitra}, R. and {Newman}, A. B. and {Raichoor}, A. and {Shankar}, F. and {Barrientos}, F. and {Bernardi}, M. and {Cerulo}, P. and {Couch}, W. and {Demarco}, R. and {Mu{\~n}oz}, R. and {S{\'a}nchez-Janssen}, R. and {Tanaka}, M.},
  title = {{Larger sizes of massive quiescent early-type galaxies in clusters than in the field at 0.8 < z < 1.5}},
  journal = {MNRAS},
  year = {2014},
  month = {jun},
  volume = {441},
  number = {1},
  pages = {203--223},
  doi = {10.1093/mnras/stu496},
  url = {https://doi.org/10.1093/mnras/stu496},
  adsurl = {https://ui.adsabs.harvard.edu/abs/2014MNRAS.441..203D}
}

@article{Papovich2012,
  author = {{Papovich}, C. and {Bassett}, R. and {Lotz}, J. M. and {van der Wel}, A. and {Tran}, K.-V. and {Finkelstein}, S. L. and {Bell}, E. F. and {Conselice}, C. J. and {Dekel}, A. and {Dunlop}, J. S. and {Guo}, Yicheng and {Faber}, S. M. and {Farrah}, D. and {Ferguson}, H. C. and {Finkelstein}, K. D. and {H{\"a}u{\ss}ler}, B. and {Kocevski}, D. D. and {Koekemoer}, A. M. and {Koo}, D. C. and {McGrath}, E. J. and {McLure}, R. J. and {McIntosh}, D. H. and {Momcheva}, I. and {Newman}, J. A. and {Rudnick}, G. and {Weiner}, B. and {Willmer}, C. N. A. and {Wuyts}, S.},
  title = {{CANDELS Observations of the Structural Properties of Cluster Galaxies at z=1.62}},
  journal = {ApJ},
  year = {2012},
  month = {may},
  volume = {750},
  number = {2},
  eid = {93},
  pages = {93},
  doi = {10.1088/0004-637X/750/2/93},
  url = {https://doi.org/10.1088/0004-637X/750/2/93},
  adsurl = {https://ui.adsabs.harvard.edu/abs/2012ApJ...750...93P}
}

@article{Lani2013,
  author = {{Lani}, Caterina and {Almaini}, Omar and {Hartley}, William G. and {Mortlock}, Alice and {H{\"a}u{\ss}ler}, Boris and {Chuter}, Robert W. and {Simpson}, Chris and {van der Wel}, Arjen and {Gr{\"u}tzbauch}, Ruth and {Conselice}, Christopher J. and {Bradshaw}, Emma J. and {Cooper}, Michael C. and {Faber}, Sandra M. and {Grogin}, Norman A. and {Kocevski}, Dale D. and {Koekemoer}, Anton M. and {Lai}, Kamson},
  title = {{Evidence for a correlation between the sizes of quiescent galaxies and local environment to $z \sim 2$}},
  journal = {MNRAS},
  year = {2013},
  month = {oct},
  volume = {435},
  number = {1},
  pages = {207--221},
  doi = {10.1093/mnras/stt1275},
  url = {https://doi.org/10.1093/mnras/stt1275},
  adsurl = {https://ui.adsabs.harvard.edu/abs/2013MNRAS.435..207L}
}

@article{CebrianTrujillo2014,
  author = {{Cebri{\'a}n}, M. and {Trujillo}, I.},
  title = {{The effect of the environment on the stellar mass-size relationship for present-day galaxies}},
  journal = {MNRAS},
  year = {2014},
  month = {oct},
  volume = {444},
  number = {1},
  pages = {682--699},
  doi = {10.1093/mnras/stu1375},
  url = {https://doi.org/10.1093/mnras/stu1375},
  adsurl = {https://ui.adsabs.harvard.edu/abs/2014MNRAS.444..682C}
}

@article{Sweet2017,
  author = {{Sweet}, Sarah M. and {Sharp}, Robert and {Glazebrook}, Karl and {Rigaut}, Francois and {Carrasco}, Eleazar R. and {Brodwin}, Mark and {Bayliss}, Matthew and {Stalder}, Brian and {Abraham}, Roberto and {McGregor}, Peter},
  title = {{The stellar mass{\textendash}size relation for cluster galaxies at z = 1 with high angular resolution from the Gemini/GeMS multiconjugate adaptive optics system}},
  journal = {MNRAS},
  year = {2017},
  month = {jan},
  volume = {464},
  number = {3},
  pages = {2910--2929},
  doi = {10.1093/mnras/stw2411},
  url = {https://doi.org/10.1093/mnras/stw2411},
  adsurl = {https://ui.adsabs.harvard.edu/abs/2017MNRAS.464.2910S}
}

@article{Vulcani2011,
  author = {{Vulcani}, Benedetta and {Poggianti}, Bianca M. and {Arag\'on-Salamanca}, Alfonso and {Fasano}, Giovanni and {Rudnick}, Gregory and {Valentinuzzi}, Tiziano and {Dressler}, Alan and {Bettoni}, Daniela and {Cava}, Antonio and {D'Onofrio}, Mauro and {Fritz}, Jacopo and {Moretti}, Alessia and {Omizzolo}, Alessandro and {Varela}, Jes\'us},
  title = {{Galaxy stellar mass functions of different morphological types in clusters, and their evolution between $z = 0.8$ and 0}},
  journal = {MNRAS},
  year = {2011},
  month = {mar},
  volume = {412},
  number = {1},
  pages = {246--268},
  doi = {10.1111/j.1365-2966.2010.17904.x},
  url = {https://doi.org/10.1111/j.1365-2966.2010.17904.x},
  adsurl = {https://ui.adsabs.harvard.edu/abs/2011MNRAS.412..246V}
}

@article{Bluck2014,
  author = {{Bluck}, Asa F. L. and {Mendel}, J. Trevor and {Ellison}, Sara L. and {Moreno}, Jorge and {Simard}, Luc and {Patton}, David R. and {Starkenburg}, Else},
  title = {{Bulge mass is king: the dominant role of the bulge in determining the fraction of passive galaxies in the Sloan Digital Sky Survey}},
  journal = {MNRAS},
  year = {2014},
  month = {jun},
  volume = {441},
  number = {1},
  pages = {599--629},
  doi = {10.1093/mnras/stu594},
  url = {https://doi.org/10.1093/mnras/stu594},
  adsurl = {https://ui.adsabs.harvard.edu/abs/2014MNRAS.441..599B}
}

@article{Kauffmann2003,
  author = {{Kauffmann}, Guinevere and {Heckman}, Timothy M. and {White}, S.~D.~M. and {Charlot}, St\'ephane and {Tremonti}, Christy and {Brinchmann}, Jarle and {Bruzual}, Gustavo and {Peng}, Eric W. and {Seibert}, Mark and {Bernardi}, Mariangela and {Blanton}, Michael and {Brinkmann}, Jon and {Castander}, Francisco and {Cs\'abai}, Istvan and {Fukugita}, Masataka and {Ivezic}, Zeljko and {Munn}, Jeffrey A. and {Nichol}, Robert C. and {Padmanabhan}, Nikhil and {Thakar}, Aniruddha R. and {Weinberg}, David H. and {York}, Donald},
  title = {{Stellar masses and star formation histories for $10^5$ galaxies from the Sloan Digital Sky Survey}},
  journal = {MNRAS},
  year = {2003},
  month = {may},
  volume = {341},
  number = {1},
  pages = {33--53},
  doi = {10.1046/j.1365-8711.2003.06291.x},
  url = {https://doi.org/10.1046/j.1365-8711.2003.06291.x},
  adsurl = {https://ui.adsabs.harvard.edu/abs/2003MNRAS.341...33K}
}

@article{Muzzin2012,
  author = {{Muzzin}, Adam and {Wilson}, Gillian and {Yee}, H. K. C. and {Gilbank}, David and {Hoekstra}, Henk and {Demarco}, Ricardo and {Balogh}, Michael and {van Dokkum}, Pieter and {Franx}, Marijn and {Ellingson}, Erica and {Hicks}, Amalia and {Nantais}, Julie and {Noble}, Allison and {Lacy}, Mark and {Lidman}, Chris and {Rettura}, Alessandro and {Surace}, Jason and {Webb}, Tracy},
  title = {{The Gemini Cluster Astrophysics Spectroscopic Survey (GCLASS): The Role of Environment and Self-regulation in Galaxy Evolution at $z \sim 1$}},
  journal = {ApJ},
  year = {2012},
  month = {feb},
  volume = {746},
  number = {2},
  eid = {188},
  pages = {188},
  doi = {10.1088/0004-637X/746/2/188},
  url = {https://doi.org/10.1088/0004-637X/746/2/188},
  adsurl = {https://ui.adsabs.harvard.edu/abs/2012ApJ...746..188M}
}

@article{DESI2023,
  author = {{DESI Collaboration} and {Adame}, A. G. and {Aguilar}, J. and {Ahlen}, S. and {Alam}, S. and {Aldering}, G. and {Alexander}, D. M. and {Alfarsy}, R. and {Allende Prieto}, C. and {Alvarez}, M. and {Alves}, O. and {Anand}, A. and {Andrade-Oliveira}, F. and {Armengaud}, E. and {Asorey}, J. and {Avila}, S. and {Aviles}, A. and {Bailey}, S. and {Balaguera-Antol{\'\i}nez}, A. and {Ballester}, O. and {Baltay}, C. and {Bault}, A. and {Bautista}, J. and {Behera}, J. and {Beltran}, S. F. and {BenZvi}, S. and {Beraldo e Silva}, L. and {Bermejo-Climent}, J. R. and {Berti}, A. and {Besuner}, R. and {Beutler}, F. and {Bianchi}, D. and {Blake}, C. and {Blum}, R. and {Bolton}, A. S. and {Brieden}, S. and {Brodzeller}, A. and {Brooks}, D. and {Brown}, Z. and {Buckley-Geer}, E. and {Burtin}, E. and {Cabayol-Garcia}, L. and {Cai}, Z. and {Canning}, R. and {Cardiel-Sas}, L. and {Carnero Rosell}, A. and {Castander}, F. J. and {Cervantes-Cota}, J. L. and {Chabanier}, S. and {Chaussidon}, E. and {Chaves-Montero}, J. and {Chen}, S. and {Chen}, X. and {Chuang}, C. and {Claybaugh}, T. and {Cole}, S. and {Cooper}, A. P. and {Cuceu}, A. and {Davis}, T. M. and {Dawson}, K. and {de Belsunce}, R. and {de la Cruz}, R. and {de la Macorra}, A. and {Della Costa}, J. and {de Mattia}, A. and {Demina}, R. and {Demirbozan}, U. and {DeRose}, J. and {Dey}, A. and {Dey}, B. and {Dhungana}, G. and {Ding}, J. and {Ding}, Z. and {Doel}, P. and {Doshi}, R. and {Douglass}, K. and {Edge}, A. and {Eftekharzadeh}, S. and {Eisenstein}, D. J. and {Elliott}, A. and {Ereza}, J. and {Escoffier}, S. and {Fagrelius}, P. and {Fan}, X. and {Fanning}, K. and {Fawcett}, V. A. and {Ferraro}, S. and {Flaugher}, B. and {Font-Ribera}, A. and {Forero-Romero}, J. E. and {Forero-S\'anchez}, D. and {Frenk}, C. S. and {G\"ansicke}, B. T. and {Garc{\'\i}a}, L. \'A. and {Garc{\'\i}a-Bellido}, J. and {Garcia-Quintero}, C. and {Garrison}, L. H. and {Gil-Mar{\'\i}n}, H. and {Golden-Marx}, J. and {Gontcho A Gontcho}, S. and {Gonzalez-Morales}, A. X. and {Gonzalez-Perez}, V. and {Gordon}, C. and {Graur}, O. and {Green}, D. and {Gruen}, D. and {Guy}, J. and {Hadzhiyska}, B. and {Hahn}, C. and {Han}, J. J. and {Hanif}, M. M. S. and {Herrera-Alcantar}, H. K. and {Honscheid}, K. and {Hou}, J. and {Howlett}, C. and {Huterer}, D. and {Ir\v{s}i\v{c}}, V. and {Ishak}, M. and {Jacques}, A. and {Jana}, A. and {Jiang}, L. and {Jimenez}, J. and {Jing}, Y. P. and {Joudaki}, S. and {Joyce}, R. and {Jullo}, E. and {Juneau}, S. and {Kara\c{c}ayl{\i}}, N. G. and {Karim}, T. and {Kehoe}, R. and {Kent}, S. and {Khederlarian}, A. and {Kim}, S. and {Kirkby}, D. and {Kisner}, T. and {Kitaura}, F. and {Kizhuprakkat}, N. and {Kneib}, J. and {Koposov}, S. E. and {Kov\'acs}, A. and {Kremin}, A. and {Krolewski}, A. and {LHuillier}, B. and {Lahav}, O. and {Lambert}, A. and {Lamman}, C. and {Lan}, T.-W. and {Landriau}, M. and {Lang}, D. and {Lange}, J. U. and {Lasker}, J. and {Leauthaud}, A. and {Le Guillou}, L. and {Levi}, M. E. and {Li}, T. S. and {Linder}, E. and {Lyons}, A. and {Magneville}, C. and {Manera}, M. and {Manser}, C. J. and {Margala}, D. and {Martini}, P. and {McDonald}, P. and {Medina}, G. E. and {Medina-Varela}, L. and {Meisner}, A. and {Mena-Fern\'andez}, J. and {Meneses-Rizo}, J. and {Mezcua}, M. and {Miquel}, R. and {Montero-Camacho}, P. and {Moon}, J. and {Moore}, S. and {Moustakas}, J. and {Mueller}, E. and {Mundet}, J. and {Mu\~noz-Guti\'errez}, A. and {Myers}, A. D. and {Nadathur}, S. and {Napolitano}, L. and {Neveux}, R. and {Newman}, J. A. and {Nie}, J. and {Nikutta}, R. and {Niz}, G. and {Norberg}, P. and {Noriega}, H. E. and {Paillas}, E. and {Palanque-Delabrouille}, N. and {Palmese}, A. and {Pan}, Z. and {Parkinson}, D. and {Penmetsa}, S. and {Percival}, W. J. and {P\'erez-Fern\'andez}, A. and {P\'erez-R\`afols}, I. and {Pieri}, M. and {Poppett}, C. and {Porredon}, A. and {Pothier}, S. and {Prada}, F. and {Pucha}, R. and {Raichoor}, A. and {Ram{\'\i}rez-P\'erez}, C. and {Ramirez-Solano}, S. and {Rashkovetskyi}, M. and {Ravoux}, C. and {Rocher}, A. and {Rockosi}, C. and {Ross}, A. J. and {Rossi}, G. and {Ruggeri}, R. and {Ruhlmann-Kleider}, V. and {Sabiu}, C. G. and {Said}, K. and {Saintonge}, A. and {Samushia}, L. and {Sanchez}, E. and {Saulder}, C. and {Schaan}, E. and {Schlafly}, E. F. and {Schlegel}, D. and {Scholte}, D. and {Schubnell}, M. and {Seo}, H. and {Shafieloo}, A. and {Sharples}, R. and {Sheu}, W. and {Silber}, J. and {Sinigaglia}, F. and {Siudek}, M. and {Slepian}, Z. and {Smith}, A. and {Soumagnac}, M. T. and {Sprayberry}, D. and {Stephey}, L. and {Su\'arez-P\'erez}, J. and {Sun}, Z. and {Tan}, T. and {Tarl\'e}, G. and {Tojeiro}, R. and {Ure\~na-L\'opez}, L. A. and {Vaisakh}, R. and {Valcin}, D. and {Valdes}, F. and {Valluri}, M. and {Vargas-Maga\~na}, M. and {Variu}, A. and {Verde}, L. and {Walther}, M. and {Wang}, B. and {Wang}, M. S. and {Weaver}, B. A. and {Weaverdyck}, N. and {Wechsler}, R. H. and {White}, M. and {Xie}, Y. and {Yang}, J. and {Y\`eche}, C. and {Yu}, J. and {Yuan}, S. and {Zhang}, H. and {Zhang}, Z. and {Zhao}, C. and {Zheng}, Z. and {Zhou}, R. and {Zhou}, Z. and {Zou}, H. and {Zou}, S. and {Zu}, Y.},
  title = {{The Early Data Release of the Dark Energy Spectroscopic Instrument}},
  journal = {AJ},
  year = {2024},
  month = {aug},
  volume = {168},
  number = {2},
  eid = {58},
  pages = {58},
  doi = {10.3847/1538-3881/ad3217},
  url = {https://doi.org/10.3847/1538-3881/ad3217},
  archivePrefix = {arXiv},
  eprint = {2306.06308},
  primaryClass = {astro-ph.IM},
  adsurl = {https://ui.adsabs.harvard.edu/abs/2024AJ....168...58D}
}

@article{Liu2022,
  author = {{Liu}, A. and {Bulbul}, E. and {Ghirardini}, V. and {Liu}, T. and {Klein}, M. and {Clerc}, N. and {{\"O}zsoy}, Y. and {Ramos-Ceja}, M. E. and {Pacaud}, F. and {Comparat}, J. and {Okabe}, N. and {Bahar}, Y. E. and {Biffi}, V. and {Brunner}, H. and {Br{\"u}ggen}, M. and {Buchner}, J. and {Ider Chitham}, J. and {Chiu}, I. and {Dolag}, K. and {Gatuzz}, E. and {Gonzalez}, J. and {Hoang}, D. N. and {Lamer}, G. and {Merloni}, A. and {Nandra}, K. and {Oguri}, M. and {Ota}, N. and {Predehl}, P. and {Reiprich}, T. H. and {Salvato}, M. and {Schrabback}, T. and {Sanders}, J. S. and {Seppi}, R. and {Thibaud}, Q.},
  title = {{The eROSITA Final Equatorial-Depth Survey (eFEDS): Catalog of galaxy clusters and groups}},
  journal = {Astronomy \& Astrophysics},
  year = {2022},
  month = {may},
  volume = {661},
  eid = {A2},
  pages = {A2},
  doi = {10.1051/0004-6361/202141120},
  url = {https://doi.org/10.1051/0004-6361/202141120},
  archivePrefix = {arXiv},
  eprint = {2106.14518},
  primaryClass = {astro-ph.CO},
  adsurl = {https://ui.adsabs.harvard.edu/abs/2022A%26A...661A...2L}
}

@ARTICLE{Sunyaev2021,
       author = {{Sunyaev}, R. and {Arefiev}, V. and {Babyshkin}, V. and {Bogomolov}, A. and {Borisov}, K. and {Buntov}, M. and {Brunner}, H. and {Burenin}, R. and {Churazov}, E. and {Coutinho}, D. and {Eder}, J. and {Eismont}, N. and {Freyberg}, M. and {Gilfanov}, M. and {Gureyev}, P. and {Hasinger}, G. and {Khabibullin}, I. and {Kolmykov}, V. and {Komovkin}, S. and {Krivonos}, R. and {Lapshov}, I. and {Levin}, V. and {Lomakin}, I. and {Lutovinov}, A. and {Medvedev}, P. and {Merloni}, A. and {Mernik}, T. and {Mikhailov}, E. and {Molodtsov}, V. and {Mzhelsky}, P. and {M{\"u}ller}, S. and {Nandra}, K. and {Nazarov}, V. and {Pavlinsky}, M. and {Poghodin}, A. and {Predehl}, P. and {Robrade}, J. and {Sazonov}, S. and {Scheuerle}, H. and {Shirshakov}, A. and {Tkachenko}, A. and {Voron}, V.},
        title = "{SRG X-ray orbital observatory. Its telescopes and first scientific results}",
      journal = {\aap},
         year = 2021,
        month = dec,
       volume = {656},
          eid = {A132},
        pages = {A132},
          doi = {10.1051/0004-6361/202141179},
archivePrefix = {arXiv},
       eprint = {2104.13267},
 primaryClass = {astro-ph.HE},
       adsurl = {https://ui.adsabs.harvard.edu/abs/2021A&A...656A.132S}
}

@article{Predehl2021,
  author = {{Predehl}, P. and {Andritschke}, R. and {Arefiev}, V. and {Babyshkin}, V. and {Batanov}, O. and {Becker}, W. and {B\"ohringer}, H. and {Bogomolov}, A. and {Boller}, T. and {Borm}, K. and {Bornemann}, W. and {Br\"auninger}, H. and {Br\"uggen}, M. and {Brunner}, H. and {Brusa}, M. and {Bulbul}, E. and {Buntov}, M. and {Burwitz}, V. and {Burkert}, W. and {Clerc}, N. and {Churazov}, E. and {Coutinho}, D. and {Dauser}, T. and {Dennerl}, K. and {Doroshenko}, V. and {Eder}, J. and {Emberger}, V. and {Eraerds}, T. and {Finoguenov}, A. and {Freyberg}, M. and {Friedrich}, P. and {Friedrich}, S. and {F\"urmetz}, M. and {Georgakakis}, A. and {Gilfanov}, M. and {Granato}, S. and {Grossberger}, C. and {Gueguen}, A. and {Gureev}, P. and {Haberl}, F. and {H\"alker}, O. and {Hartner}, G. and {Hasinger}, G. and {Huber}, H. and {Ji}, L. and {Kienlin}, A. v. and {Kink}, W. and {Korotkov}, F. and {Kreykenbohm}, I. and {Lamer}, G. and {Lomakin}, I. and {Lapshov}, I. and {Liu}, T. and {Maitra}, C. and {Meidinger}, N. and {Menz}, B. and {Merloni}, A. and {Mernik}, T. and {Mican}, B. and {Mohr}, J. and {M\"uller}, S. and {Nandra}, K. and {Nazarov}, V. and {Pacaud}, F. and {Pavlinsky}, M. and {Perinati}, E. and {Pfeffermann}, E. and {Pietschner}, D. and {Ramos-Ceja}, M. E. and {Rau}, A. and {Reiffers}, J. and {Reiprich}, T. H. and {Robrade}, J. and {Salvato}, M. and {Sanders}, J. and {Santangelo}, A. and {Sasaki}, M. and {Scheuerle}, H. and {Schmid}, C. and {Schmitt}, J. and {Schwope}, A. and {Shirshakov}, A. and {Steinmetz}, M. and {Stewart}, I. and {Str\"uder}, L. and {Sunyaev}, R. and {Tenzer}, C. and {Tiedemann}, L. and {Tr\"umper}, J. and {Voron}, V. and {Weber}, P. and {Wilms}, J. and {Yaroshenko}, V.},
  title = {{The eROSITA X-ray telescope on SRG}},
  journal = {Astronomy \& Astrophysics},
  year = {2021},
  month = {mar},
  volume = {647},
  pages = {A1},
  doi = {10.1051/0004-6361/202039313},
  url = {https://doi.org/10.1051/0004-6361/202039313},
  adsurl = {https://ui.adsabs.harvard.edu/search/q=doi%3A%2210.1051%2F0004-6361%2F202039313%22}
}

@article{Brunner2022,
  author = {{Brunner}, H. and {Liu}, T. and {Lamer}, G. and {Georgakakis}, A. and {Merloni}, A. and {Brusa}, M. and {Bulbul}, E. and {Dennerl}, K. and {Friedrich}, S. and {Liu}, A. and {Maitra}, C. and {Nandra}, K. and {Ramos-Ceja}, M. E. and {Sanders}, J. S. and {Stewart}, I. M. and {Boller}, T. and {Buchner}, J. and {Clerc}, N. and {Comparat}, J. and {Dwelly}, T. and {Eckert}, D. and {Finoguenov}, A. and {Freyberg}, M. and {Ghirardini}, V. and {Gueguen}, A. and {Haberl}, F. and {Kreykenbohm}, I. and {Krumpe}, M. and {Osterhage}, S. and {Pacaud}, F. and {Predehl}, P. and {Reiprich}, T. H. and {Robrade}, J. and {Salvato}, M. and {Santangelo}, A. and {Schrabback}, T. and {Schwope}, A. and {Wilms}, J.},
  title = {{The eROSITA Final Equatorial-Depth Survey (eFEDS): The X-ray catalog}},
  journal = {Astronomy \& Astrophysics},
  year = {2022},
  month = {may},
  volume = {661},
  pages = {A1},
  doi = {10.1051/0004-6361/202141266},
  url = {https://doi.org/10.1051/0004-6361/202141266},
  adsurl = {https://ui.adsabs.harvard.edu/search/q=doi%3A%2210.1051%2F0004-6361%2F202141266%22}
}

@article{Liu2022eFEDS,
  author = {{Liu}, A. and {Bulbul}, E. and {Ghirardini}, V. and {Liu}, T. and {Klein}, M. and {Clerc}, N. and {\"Ozsoy}, Y. and {Ramos-Ceja}, M. E. and {Pacaud}, F. and {Comparat}, J. and {Okabe}, N. and {Bahar}, Y. E. and {Biffi}, V. and {Brunner}, H. and {Br\"uggen}, M. and {Buchner}, J. and {Ider Chitham}, J. and {Chiu}, I. and {Dolag}, K. and {Gatuzz}, E. and {Gonzalez}, J. and {Hoang}, D. N. and {Lamer}, G. and {Merloni}, A. and {Nandra}, K. and {Oguri}, M. and {Ota}, N. and {Predehl}, P. and {Reiprich}, T. H. and {Salvato}, M. and {Schrabback}, T. and {Sanders}, J. S. and {Seppi}, R. and {Thibaud}, Q.},
  title = {{The eROSITA Final Equatorial-Depth Survey (eFEDS): Catalog of galaxy clusters and groups}},
  journal = {Astronomy \& Astrophysics},
  year = {2022},
  month = {may},
  volume = {661},
  pages = {A2},
  doi = {10.1051/0004-6361/202141120},
  url = {https://doi.org/10.1051/0004-6361/202141120},
  adsurl = {https://ui.adsabs.harvard.edu/search/q=doi%3A%2210.1051%2F0004-6361%2F202141120%22}
}

@article{Klein2022,
  author = {{Klein}, M. and {Oguri}, M. and {Mohr}, J. J. and {Grandis}, S. and {Ghirardini}, V. and {Liu}, T. and {Liu}, A. and {Bulbul}, E. and {Wolf}, J. and {Comparat}, J. and {Ramos-Ceja}, M. E. and {Buchner}, J. and {Chiu}, I. and {Clerc}, N. and {Merloni}, A. and {Miyatake}, H. and {Miyazaki}, S. and {Okabe}, N. and {Ota}, N. and {Pacaud}, F. and {Salvato}, M. and {Driver}, S. P.},
  title = {{The eROSITA Final Equatorial-Depth Survey (eFEDS). Optical confirmation, redshifts, and properties of the cluster and group catalog}},
  journal = {Astronomy \& Astrophysics},
  year = {2022},
  month = {may},
  volume = {661},
  eid = {A4},
  pages = {A4},
  doi = {10.1051/0004-6361/202141123},
  url = {https://doi.org/10.1051/0004-6361/202141123},
  adsurl = {https://ui.adsabs.harvard.edu/abs/2022A&A...661A...4K}
}

@article{Chiu2022,
  author = {{Chiu}, I-Non and {Ghirardini}, Vittorio and {Liu}, Ang and {Grandis}, Sebastian and {Bulbul}, Esra and {Emre Bahar}, Y. and {Comparat}, Johan and {Bocquet}, Sebastian and {Clerc}, Nicolas and {Klein}, Matthias and {Liu}, Teng and {Li}, Xiangchong and {Miyatake}, Hironao and {Mohr}, Joseph and {More}, Surhud and {Oguri}, Masamune and {Okabe}, Nobuhiro and {Pacaud}, Florian and {Ramos-Ceja}, Miriam E. and {Reiprich}, Thomas H. and {Schrabback}, Tim and {Umetsu}, Keiichi},
  title = {{The eROSITA Final Equatorial-Depth Survey (eFEDS): X-ray observable-to-mass-and-redshift relations of galaxy clusters and groups with weak-lensing mass calibration from the Hyper Suprime-Cam Subaru Strategic Program survey}},
  journal = {Astronomy \& Astrophysics},
  year = {2022},
  month = {may},
  volume = {661},
  pages = {A11},
  doi = {10.1051/0004-6361/202141755},
  url = {https://doi.org/10.1051/0004-6361/202141755},
  adsurl = {https://ui.adsabs.harvard.edu/search/q=doi%3A%2210.1051%2F0004-6361%2F202141755%22}
}

@article{Kawinwanichakij2021,
  author = {{Kawinwanichakij}, Lalitwadee and {Silverman}, John D. and {Ding}, Xuheng and {George}, Angelo and {Damjanov}, Ivana and {Sawicki}, Marcin and {Tanaka}, Masayuki and {Taranu}, Dan S. and {Birrer}, Simon and {Huang}, Song and {Li}, Junyao and {Onodera}, Masato and {Shibuya}, Takatoshi and {Yasuda}, Naoki},
  title = {{Hyper Suprime-Cam Subaru Strategic Program: A Mass-dependent Slope of the Galaxy Size-Mass Relation at $z < 1$}},
  journal = {ApJ},
  year = {2021},
  month = {nov},
  volume = {921},
  number = {1},
  eid = {38},
  pages = {38},
  doi = {10.3847/1538-4357/ac1f21},
  url = {https://doi.org/10.3847/1538-4357/ac1f21},
  adsurl = {https://ui.adsabs.harvard.edu/search/q=doi%3A%2210.3847%2F1538-4357%2Fac1f21%22}
}

@article{Bulbul2024,
  author = {{Bulbul}, E. and {Liu}, A. and {Kluge}, M. and {Zhang}, X. and {Sanders}, J. S. and {Bahar}, Y. E. and {Ghirardini}, V. and {Artis}, E. and {Seppi}, R. and {Garrel}, C. and {Ramos-Ceja}, M. E. and {Comparat}, J. and {Balzer}, F. and {Boeckmann}, K. and {Br{\"u}ggen}, M. and {Clerc}, N. and {Dennerl}, K. and {Dolag}, K. and {Freyberg}, M. and {Grandis}, S. and {Gruen}, D. and {Kleinebreil}, F. and {Krippendorf}, S. and {Lamer}, G. and {Merloni}, A. and {Migkas}, K. and {Nandra}, K. and {Pacaud}, F. and {Predehl}, P. and {Reiprich}, T. H. and {Schrabback}, T. and {Veronica}, A. and {Weller}, J. and {Zelmer}, S.},
  title = {{The SRG/eROSITA All-Sky Survey. The first catalog of galaxy clusters and groups in the Western Galactic Hemisphere}},
  journal = {Astronomy \& Astrophysics},
  year = {2024},
  month = {may},
  volume = {685},
  eid = {A106},
  pages = {A106},
  doi = {10.1051/0004-6361/202348264},
  url = {https://doi.org/10.1051/0004-6361/202348264},
  adsurl = {https://ui.adsabs.harvard.edu/abs/2024A&A...685A.106B}
}

@article{Dey2019,
  author = {{Dey}, Arjun and {Schlegel}, David J. and {Lang}, Dustin and {Blum}, Robert and {Burleigh}, Kaylan and {Fan}, Xiaohui and {Findlay}, Joseph R. and {Finkbeiner}, Doug and {Herrera}, David and {Juneau}, St\'ephanie and {Landriau}, Martin and {Levi}, Michael and {McGreer}, Ian and {Meisner}, Aaron and {Myers}, Adam D. and {Moustakas}, John and {Nugent}, Peter and {Patej}, Anna and {Schlafly}, Edward F. and {Walker}, Alistair R. and {Valdes}, Francisco and {Weaver}, Benjamin A. and {Y\`eche}, Christophe and {Zou}, Hu and {Zhou}, Xu and {Abareshi}, Behzad and {Abbott}, T. M. C. and {Abolfathi}, Bela and {Aguilera}, C. and {Alam}, Shadab and {Allen}, Lori and {Alvarez}, A. and {Annis}, James and {Ansarinejad}, Behzad and {Aubert}, Marie and {Beechert}, Jacqueline and {Bell}, Eric F. and {BenZvi}, Segev Y. and {Beutler}, Florian and {Bielby}, Richard M. and {Bolton}, Adam S. and {Brice\~no}, C\'esar and {Buckley-Geer}, Elizabeth J. and {Butler}, Karen and {Calamida}, Annalisa and {Carlberg}, Raymond G. and {Carter}, Paul and {Casas}, Ricard and {Castander}, Francisco J. and {Choi}, Yumi and {Comparat}, Johan and {Cukanovaite}, Elena and {Delubac}, Timoth\'ee and {DeVries}, Kaitlin and {Dey}, Sharmila and {Dhungana}, Govinda and {Dickinson}, Mark and {Ding}, Zhejie and {Donaldson}, John B. and {Duan}, Yutong and {Duckworth}, Christopher J. and {Eftekharzadeh}, Sarah and {Eisenstein}, Daniel J. and {Etourneau}, Thomas and {Fagrelius}, Parker A. and {Farihi}, Jay and {Fitzpatrick}, Mike and {Font-Ribera}, Andreu and {Fulmer}, Leah and {G\"ansicke}, Boris T. and {Gaztanaga}, Enrique and {George}, Koshy and {Gerdes}, David W. and {A Gontcho}, Satya Gontcho and {Gorgoni}, Claudio and {Green}, Gregory and {Guy}, Julien and {Harmer}, Diane and {Hernandez}, M. and {Honscheid}, Klaus and {Huang}, Lijuan (Wendy) and {James}, David J. and {Jannuzi}, Buell T. and {Jiang}, Linhua and {Joyce}, Richard and {Karcher}, Armin and {Karkar}, Sonia and {Kehoe}, Robert and {Kneib}, Jean-Paul and {Kueter-Young}, Andrea and {Lan}, Ting-Wen and {Lauer}, Tod R. and {Guillou}, Laurent Le and {Van Suu}, Auguste Le and {Lee}, Jae Hyeon and {Lesser}, Michael and {Levasseur}, Laurence Perreault and {Li}, Ting S. and {Mann}, Justin L. and {Marshall}, Robert and {Mart{\'\i}nez-V\'azquez}, C. E. and {Martini}, Paul and {du Mas des Bourboux}, H\'elion and {McManus}, Sean and {Meier}, Tobias Gabriel and {M\'enard}, Brice and {Metcalfe}, Nigel and {Mu\~noz-Guti\'errez}, Andrea and {Najita}, Joan and {Napier}, Kevin and {Narayan}, Gautham and {Newman}, Jeffrey A. and {Nie}, Jundan and {Nord}, Brian and {Norman}, Dara J. and {Olsen}, Knut A. G. and {Paat}, Anthony and {Palanque-Delabrouille}, Nathalie and {Peng}, Xiyan and {Poppett}, Claire L. and {Poremba}, Megan R. and {Prakash}, Abhishek and {Rabinowitz}, David and {Raichoor}, Anand and {Rezaie}, Mehdi and {Robertson}, A. N. and {Roe}, Natalie A. and {Ross}, Ashley J. and {Ross}, Nicholas P. and {Rudnick}, Gregory and {Gaines}, Sasha and {Saha}, Abhijit and {S\'anchez}, F. Javier and {Savary}, Elodie and {Schweiker}, Heidi and {Scott}, Adam and {Seo}, Hee-Jong and {Shan}, Huanyuan and {Silva}, David R. and {Slepian}, Zachary and {Soto}, Christian and {Sprayberry}, David and {Staten}, Ryan and {Stillman}, Coley M. and {Stupak}, Robert J. and {Summers}, David L. and {Tie}, Suk Sien and {Tirado}, H. and {Vargas-Maga\~na}, Mariana and {Vivas}, A. Katherina and {Wechsler}, Risa H. and {Williams}, Doug and {Yang}, Jinyi and {Yang}, Qian and {Yapici}, Tolga and {Zaritsky}, Dennis and {Zenteno}, A. and {Zhang}, Kai and {Zhang}, Tianmeng and {Zhou}, Rongpu and {Zhou}, Zhimin},
  title = {{Overview of the DESI Legacy Imaging Surveys}},
  journal = {AJ},
  year = {2019},
  month = {may},
  volume = {157},
  number = {5},
  eid = {168},
  pages = {168},
  doi = {10.3847/1538-3881/ab089d},
  url = {https://doi.org/10.3847/1538-3881/ab089d},
  adsurl = {https://ui.adsabs.harvard.edu/search/q=doi%3A%2210.3847%2F1538-3881%2Fab089d%22}
}

@article{Kluge2024,
  author = {{Kluge}, M. and {Comparat}, J. and {Liu}, A. and {Balzer}, F. and {Bulbul}, E. and {Ider Chitham}, J. and {Ghirardini}, V. and {Garrel}, C. and {Bahar}, Y. E. and {Artis}, E. and {Bender}, R. and {Clerc}, N. and {Dwelly}, T. and {Fabricius}, M. H. and {Grandis}, S. and {Hern{\'a}ndez-Lang}, D. and {Hill}, G. J. and {Joshi}, G. and {Lamer}, G. and {Merloni}, A. and {Nandra}, K. and {Pacaud}, F. and {Predehl}, P. and {Ramos-Ceja}, M. E. and {Reiprich}, T. H. and {Salvato}, M. and {Sanders}, J. S. and {Schrabback}, T. and {Seppi}, R. and {Zelmer}, S. and {Zenteno}, A. and {Zhang}, X.},
  title = {{The SRG/eROSITA All-Sky Survey. Optical identification and properties of galaxy clusters and groups in the Western Galactic Hemisphere}},
  journal = {Astronomy \& Astrophysics},
  year = {2024},
  month = {aug},
  volume = {688},
  eid = {A210},
  pages = {A210},
  doi = {10.1051/0004-6361/202349031},
  url = {https://doi.org/10.1051/0004-6361/202349031},
  adsurl = {https://ui.adsabs.harvard.edu/abs/2024A&A...688A.210K}
}

@article{Aihara2022,
  author = {{Aihara}, Hiroaki and {AlSayyad}, Yusra and {Ando}, Makoto and {Armstrong}, Robert and {Bosch}, James and {Egami}, Eiichi and {Furusawa}, Hisanori and {Furusawa}, Junko and {Harasawa}, Sumiko and {Harikane}, Yuichi and {Hsieh}, Bau-Ching and {Ikeda}, Hiroyuki and {Ito}, Kei and {Iwata}, Ikuru and {Kodama}, Tadayuki and {Koike}, Michitaro and {Kokubo}, Mitsuru and {Komiyama}, Yutaka and {Li}, Xiangchong and {Liang}, Yongming and {Lin}, Yen-Ting and {Lupton}, Robert H and {Lust}, Nate B and {MacArthur}, Lauren A and {Mawatari}, Ken and {Mineo}, Sogo and {Miyatake}, Hironao and {Miyazaki}, Satoshi and {More}, Surhud and {Morishima}, Takahiro and {Murayama}, Hitoshi and {Nakajima}, Kimihiko and {Nakata}, Fumiaki and {Nishizawa}, Atsushi J and {Oguri}, Masamune and {Okabe}, Nobuhiro and {Okura}, Yuki and {Ono}, Yoshiaki and {Osato}, Ken and {Ouchi}, Masami and {Pan}, Yen-Chen and {Plazas Malag\'on}, Andr\'es A and {Price}, Paul A and {Reed}, Sophie L and {Rykoff}, Eli S and {Shibuya}, Takatoshi and {Simunovic}, Mirko and {Strauss}, Michael A and {Sugimori}, Kanako and {Suto}, Yasushi and {Suzuki}, Nao and {Takada}, Masahiro and {Takagi}, Yuhei and {Takata}, Tadafumi and {Takita}, Satoshi and {Tanaka}, Masayuki and {Tang}, Shenli and {Taranu}, Dan S and {Terai}, Tsuyoshi and {Toba}, Yoshiki and {Turner}, Edwin L and {Uchiyama}, Hisakazu and {Vijarnwannaluk}, Bovornpratch and {Waters}, Christopher Z and {Yamada}, Yoshihiko and {Yamamoto}, Naoaki and {Yamashita}, Takuji},
  title = {{Third data release of the Hyper Suprime-Cam Subaru Strategic Program}},
  journal = {PASJ},
  year = {2022},
  month = {apr},
  volume = {74},
  number = {2},
  pages = {247--272},
  doi = {10.1093/pasj/psab122},
  url = {https://doi.org/10.1093/pasj/psab122},
  adsurl = {https://ui.adsabs.harvard.edu/search/q=doi%3A%2210.1093%2Fpasj%2Fpsab122%22}
}

@article{Miyazaki2018,
  author = {{Miyazaki}, Satoshi and {Komiyama}, Yutaka and {Kawanomoto}, Satoshi and {Doi}, Yoshiyuki and {Furusawa}, Hisanori and {Hamana}, Takashi and {Hayashi}, Yusuke and {Ikeda}, Hiroyuki and {Kamata}, Yukiko and {Karoji}, Hiroshi and {Koike}, Michitaro and {Kurakami}, Tomio and {Miyama}, Shoken and {Morokuma}, Tomoki and {Nakata}, Fumiaki and {Namikawa}, Kazuhito and {Nakaya}, Hidehiko and {Nariai}, Kyoji and {Obuchi}, Yoshiyuki and {Oishi}, Yukie and {Okada}, Norio and {Okura}, Yuki and {Tait}, Philip and {Takata}, Tadafumi and {Tanaka}, Yoko and {Tanaka}, Masayuki and {Terai}, Tsuyoshi and {Tomono}, Daigo and {Uraguchi}, Fumihiro and {Usuda}, Tomonori and {Utsumi}, Yousuke and {Yamada}, Yoshihiko and {Yamanoi}, Hitomi and {Aihara}, Hiroaki and {Fujimori}, Hiroki and {Mineo}, Sogo and {Miyatake}, Hironao and {Oguri}, Masamune and {Uchida}, Tomohisa and {Tanaka}, Manobu M and {Yasuda}, Naoki and {Takada}, Masahiro and {Murayama}, Hitoshi and {Nishizawa}, Atsushi J and {Sugiyama}, Naoshi and {Chiba}, Masashi and {Futamase}, Toshifumi and {Wang}, Shiang-Yu and {Chen}, Hsin-Yo and {Ho}, Paul T P and {Liaw}, Eric J Y and {Chiu}, Chi-Fang and {Ho}, Cheng-Lin and {Lai}, Tsang-Chih and {Lee}, Yao-Cheng and {Jeng}, Dun-Zen and {Iwamura}, Satoru and {Armstrong}, Robert and {Bickerton}, Steve and {Bosch}, James and {Gunn}, James E and {Lupton}, Robert H and {Loomis}, Craig and {Price}, Paul and {Smith}, Steward and {Strauss}, Michael A and {Turner}, Edwin L and {Suzuki}, Hisanori and {Miyazaki}, Yasuhito and {Muramatsu}, Masaharu and {Yamamoto}, Koei and {Endo}, Makoto and {Ezaki}, Yutaka and {Ito}, Noboru and {Kawaguchi}, Noboru and {Sofuku}, Satoshi and {Taniike}, Tomoaki and {Akutsu}, Kotaro and {Dojo}, Naoto and {Kasumi}, Kazuyuki and {Matsuda}, Toru and {Imoto}, Kohei and {Miwa}, Yoshinori and {Suzuki}, Masayuki and {Takeshi}, Kunio and {Yokota}, Hideo},
  title = {{Hyper Suprime-Cam: System design and verification of image quality}},
  journal = {PASJ},
  year = {2018},
  month = {jan},
  volume = {70},
  number = {SP1},
  eid = {S1},
  pages = {S1},
  doi = {10.1093/pasj/psx063},
  url = {https://doi.org/10.1093/pasj/psx063},
  adsurl = {https://ui.adsabs.harvard.edu/search/q=doi%3A%2210.1093%2Fpasj%2Fpsx063%22}
}

@article{Kawanomoto2018,
  author = {{Kawanomoto}, Satoshi and {Uraguchi}, Fumihiro and {Komiyama}, Yutaka and {Miyazaki}, Satoshi and {Furusawa}, Hisanori and {Finet}, Fran\c{c}ois and {Hattori}, Takashi and {Wang}, Shiang-Yu and {Yasuda}, Naoki and {Suzuki}, Naotaka},
  title = {{Hyper Suprime-Cam: Filters}},
  journal = {PASJ},
  year = {2018},
  month = {aug},
  volume = {70},
  number = {4},
  eid = {66},
  pages = {66},
  doi = {10.1093/pasj/psy056},
  url = {https://doi.org/10.1093/pasj/psy056},
  adsurl = {https://ui.adsabs.harvard.edu/search/q=doi%3A%2210.1093%2Fpasj%2Fpsy056%22}
}

@article{Bosch2018,
  author = {{Bosch}, James and {Armstrong}, Robert and {Bickerton}, Steven and {Furusawa}, Hisanori and {Ikeda}, Hiroyuki and {Koike}, Michitaro and {Lupton}, Robert and {Mineo}, Sogo and {Price}, Paul and {Takata}, Tadafumi and {Tanaka}, Masayuki and {Yasuda}, Naoki and {AlSayyad}, Yusra and {Becker}, Andrew C and {Coulton}, William and {Coupon}, Jean and {Garmilla}, Jose and {Huang}, Song and {Krughoff}, K Simon and {Lang}, Dustin and {Leauthaud}, Alexie and {Lim}, Kian-Tat and {Lust}, Nate B and {MacArthur}, Lauren A and {Mandelbaum}, Rachel and {Miyatake}, Hironao and {Miyazaki}, Satoshi and {Murata}, Ryoma and {More}, Surhud and {Okura}, Yuki and {Owen}, Russell and {Swinbank}, John D and {Strauss}, Michael A and {Yamada}, Yoshihiko and {Yamanoi}, Hitomi},
  title = {{The Hyper Suprime-Cam software pipeline}},
  journal = {PASJ},
  year = {2018},
  month = {jan},
  volume = {70},
  number = {SP1},
  eid = {S5},
  pages = {S5},
  doi = {10.1093/pasj/psx080},
  url = {https://doi.org/10.1093/pasj/psx080},
  adsurl = {https://ui.adsabs.harvard.edu/search/q=doi%3A%2210.1093%2Fpasj%2Fpsx080%22}
}

@article{Schlafly2012,
  author = {{Schlafly}, E. F. and {Finkbeiner}, D. P. and {Juri\'c}, M. and {Magnier}, E. A. and {Burgett}, W. S. and {Chambers}, K. C. and {Grav}, T. and {Hodapp}, K. W. and {Kaiser}, N. and {Kudritzki}, R.-P. and {Martin}, N. F. and {Morgan}, J. S. and {Price}, P. A. and {Rix}, H.-W. and {Stubbs}, C. W. and {Tonry}, J. L. and {Wainscoat}, R. J.},
  title = {{Photometric calibration of the first 1.5 years of the Pan-STARRS1 survey}},
  journal = {ApJ},
  year = {2012},
  month = {sep},
  volume = {756},
  number = {2},
  eid = {158},
  pages = {158},
  doi = {10.1088/0004-637X/756/2/158},
  url = {https://doi.org/10.1088/0004-637X/756/2/158},
  adsurl = {https://ui.adsabs.harvard.edu/search/q=doi%3A%2210.1088%2F0004-637X%2F756%2F2%2F158%22}
}

@article{Tonry2012,
  author = {{Tonry}, J. L. and {Stubbs}, C. W. and {Lykke}, K. R. and {Doherty}, P. and {Shivvers}, I. S. and {Burgett}, W. S. and {Chambers}, K. C. and {Hodapp}, K. W. and {Kaiser}, N. and {Kudritzki}, R.-P. and {Magnier}, E. A. and {Morgan}, J. S. and {Price}, P. A. and {Wainscoat}, R. J.},
  title = {{The Pan-STARRS1 photometric system}},
  journal = {ApJ},
  year = {2012},
  month = {may},
  volume = {750},
  number = {2},
  eid = {99},
  pages = {99},
  doi = {10.1088/0004-637X/750/2/99},
  url = {https://doi.org/10.1088/0004-637X/750/2/99},
  adsurl = {https://ui.adsabs.harvard.edu/search/q=doi%3A%2210.1088%2F0004-637X%2F750%2F2%2F99%22}
}

@article{DESIDR1_2025,
  author = {{DESI Collaboration} and {Abdul Karim}, M. and {Adame}, A. G. and {Aguado}, D. and {Aguilar}, J. and {Ahlen}, S. and {Alam}, S. and {Aldering}, G. and {Alexander}, D. M. and {Alfarsy}, R. and {Allen}, L. and {Allende Prieto}, C. and {Alves}, O. and {Anand}, A. and {Andrade}, U. and {Armengaud}, E. and {Avila}, S. and {Aviles}, A. and {Awan}, H. and {Bailey}, S. and {Baleato Lizancos}, A. and {Ballester}, O. and {Bault}, A. and {Bautista}, J. and {Bean}, R. and {Behera}, J. and {BenZvi}, S. and {Beraldo e Silva}, L. and {Bermejo-Climent}, J. R. and {Beutler}, F. and {Bianchi}, D. and {Blake}, C. and {Blum}, R. and {Bolton}, A. S. and {Bonici}, M. and {Brieden}, S. and {Brodzeller}, A. and {Brooks}, D. and {Buckley-Geer}, E. and {Burtin}, E. and {Bystr\"om}, A. and {Canning}, R. and {Carnero Rosell}, A. and {Carr}, A. and {Carrilho}, P. and {Casas}, L. and {Castander}, F. J. and {Cereskaite}, R. and {Cervantes-Cota}, J. L. and {Chaussidon}, E. and {Chaves-Montero}, J. and {Chen}, S. and {Chen}, X. and {Circosta}, C. and {Claybaugh}, T. and {Cole}, S. and {Cooper}, A. P. and {Cousinou}, M.-C. and {Cuceu}, A. and {Davis}, T. M. and {Dawson}, K. S. and {de Belsunce}, R. and {de la Cruz}, R. and {de la Macorra}, A. and {de Mattia}, A. and {Deiosso}, N. and {Della Costa}, J. and {Demina}, R. and {Demirbozan}, U. and {DeRose}, J. and {Dey}, A. and {Dey}, B. and {Ding}, J. and {Ding}, Z. and {Doel}, P. and {Douglass}, K. and {Dowicz}, M. and {Ebina}, H. and {Edelstein}, J. and {Eisenstein}, D. J. and {Elbers}, W. and {Emas}, N. and {Escoffier}, S. and {Fagrelius}, P. and {Fan}, X. and {Fanning}, K. and {Favole}, G. and {Fawcett}, V. A. and {Fern\'andez-Garc{\'\i}a}, E. and {Ferraro}, S. and {Findlay}, N. and {Font-Ribera}, A. and {Forero-Romero}, J. E. and {Forero-S\'anchez}, D. and {Frenk}, C. S. and {G\"ansicke}, B. T. and {Galbany}, L. and {Garc{\'\i}a-Bellido}, J. and {Garcia-Quintero}, C. and {Garrison}, L. H. and {Gazta\~naga}, E. and {Gil-Mar{\'\i}n}, H. and {Gloudemans}, A. and {Gnedin}, O. Y. and {Gontcho A Gontcho}, S. and {Gonzalez}, D. and {Gonzalez-Morales}, A. X. and {Gonzalez-Perez}, V. and {Gordon}, C. and {Graur}, O. and {Green}, D. and {Gruen}, D. and {Gsponer}, R. and {Guandalin}, C. and {Gutierrez}, G. and {Guy}, J. and {Hahn}, C. and {Han}, J. J. and {Han}, J. and {He}, S. and {Herrera-Alcantar}, H. K. and {Heydenreich}, S. and {Honscheid}, K. and {Hou}, J. and {Howlett}, C. and {Huterer}, D. and {Ir\v{s}i\v{c}}, V. and {Ishak}, M. and {Jacques}, A. and {Jiang}, L. and {Jimenez}, J. and {Jing}, Y. P. and {Joachimi}, B. and {Joudaki}, S. and {Joyce}, R. and {Jullo}, E. and {Juneau}, S. and {Kara\c{c}ayl{\i}}, N. G. and {Karim}, T. and {Kehoe}, R. and {Kent}, S. and {Khederlarian}, A. and {Kirkby}, D. and {Kisner}, T. and {Kitaura}, F.-S. and {Kizhuprakkat}, N. and {Kong}, H. and {Koposov}, S. E. and {Kremin}, A. and {Krolewski}, A. and {Lahav}, O. and {Lai}, Y. and {Lamman}, C. and {Lan}, T.-W. and {Landriau}, M. and {Lang}, D. and {Lange}, J. U. and {Lasker}, J. and {Le Goff}, J.M. and {Le Guillou}, L. and {Leauthaud}, A. and {Levi}, M. E. and {Li}, S. and {Li}, T. S. and {Liu}, W. and {Lodha}, K. and {Lokken}, M. and {Luo}, Y. and {Luo}, Y. and {Magneville}, C. and {Manera}, M. and {Manser}, C. J. and {Margala}, D. and {Martini}, P. and {Maus}, M. and {McCullough}, J. and {McDonald}, P. and {Medina}, G. E. and {Medina-Varela}, L. and {Meisner}, A. and {Mena-Fern\'andez}, J. and {Menegas}, A. and {Meneses-Rizo}, J. and {Mezcua}, M. and {Miquel}, R. and {Montero-Camacho}, P. and {Moon}, J. and {Moustakas}, J. and {Mu\~noz-Guti\'errez}, A. and {Mu noz-Santos}, D. and {Myers}, A. D. and {Myles}, J. and {Nadathur}, S. and {Najita}, J. and {Napolitano}, L. and {Newman}, J. A. and {Nikakhtar}, F. and {Nikutta}, R. and {Niz}, G. and {Noriega}, H. E. and {Nugent}, P. and {Padmanabhan}, N. and {Paillas}, E. and {Palanque-Delabrouille}, N. and {Palmese}, A. and {Pan}, J. and {Pan}, Z. and {Parkinson}, D. and {Peacock}, J. A. and {Ibanez}, M. P. and {Percival}, W. J. and {P\'erez-Fern\'andez}, A. and {P\'erez-R\`afols}, I. and {Peterson}, P. and {Piat}, J. and {Pieri}, M. M. and {Pinon}, M. and {Poppett}, C. and {Porredon}, A. and {Prada}, F. and {Pucha}, R. and {Qin}, F. and {Rabinowitz}, D. and {Raichoor}, A. and {Ram{\'\i}rez-P\'erez}, C. and {Ramirez-Solano}, S. and {Rashkovetskyi}, M. and {Ravoux}, C. and {Ried Guachalla}, B. and {Riley}, A. H. and {Rocher}, A. and {Rockosi}, C. and {Rohlf}, J. and {Rosado-Mar{\'\i}n}, A. J. and {Ross}, A. J. and {Ross}, C. and {Rossi}, G. and {Ruggeri}, R. and {Ruhlmann-Kleider}, V. and {Sabiu}, C. G. and {Said}, K. and {Sailer}, N. and {Saintonge}, A. and {Salcedo Hernandez}, Y. and {Samushia}, L. and {Sanchez}, E. and {Sanders}, N. and {Sandford}, N. and {Satyavolu}, S. and {Saulder}, C. and {Saydjari}, A. K. and {Schlafly}, E. F. and {Schlegel}, D. and {Scholte}, D. and {Schubnell}, M. and {Semenaite}, A. and {Seo}, H. and {Shafieloo}, A. and {Sharples}, R. and {Silber}, J. and {Sinigaglia}, F. and {Siudek}, M. and {Slepian}, Z. and {Smith}, A. and {Soumagnac}, M. and {Sprayberry}, D. and {Su\'arez-P\'erez}, J. and {Swanson}, J. and {Tan}, T. and {Tarl\'e}, G. and {Taylor}, P. and {Thomas}, G. and {Tojeiro}, R. and {Turner}, R. J. and {Turner}, W. and {Ure\~na-L\'opez}, L. A. and {Vaisakh}, R. and {Valluri}, M. and {Valogiannis}, G. and {Vargas-Maga\~na}, M. and {Verde}, L. and {Vielzeuf}, P. and {Walther}, M. and {Wang}, B. and {Wang}, M. S. and {Wang}, W. and {Weaver}, B. A. and {Weaverdyck}, N. and {Wechsler}, R. H. and {Weinberg}, D. H. and {White}, M. and {Whitford}, A. and {Wolfson}, M. and {Yang}, J. and {Y\`eche}, C. and {Youles}, S. and {Yu}, J. and {Yuan}, S. and {Zaborowski}, E. A. and {Zarrouk}, P. and {Zhang}, H. and {Zhao}, C. and {Zhao}, R. and {Zheng}, Z. and {Zhou}, C. and {Zhou}, R. and {Zhou}, Y. and {Zou}, H. and {Zou}, S. and {Zu}, Y.},
  title = {{Data Release 1 of the Dark Energy Spectroscopic Instrument}},
  journal = {AJ},
  year = {2026},
  month = {may},
  volume = {171},
  number = {5},
  eid = {285},
  pages = {285},
  doi = {10.3847/1538-3881/ae4c43},
  url = {https://doi.org/10.3847/1538-3881/ae4c43},
  archivePrefix = {arXiv},
  eprint = {2503.14745},
  primaryClass = {astro-ph.CO},
  adsurl = {https://ui.adsabs.harvard.edu/search/q=doi%3A%2210.3847%2F1538-3881%2Fae4c43%22}
}

@article{DESI2022,
  author = {{DESI Collaboration} and {Aghamousa}, Amir and {Aguilar}, Jessica and {Ahlen}, Steve and {Alam}, Shadab and {Allen}, Lori E. and {Prieto}, Carlos Allende and {Annis}, James and {Bailey}, Stephen and {Balland}, Christophe and {Ballester}, Otger and {Baltay}, Charles and {Beaufore}, Lucas and {Bebek}, Chris and {Beers}, Timothy C. and {Bell}, Eric F. and {Bernal}, Jos\'e Luis and {Besuner}, Robert and {Beutler}, Florian and {Blake}, Chris and {Bleuler}, Hannes and {Blomqvist}, Michael and {Blum}, Robert and {Bolton}, Adam S. and {Briceno}, Cesar and {Brooks}, David and {Brownstein}, Joel R. and {Buckley-Geer}, Elizabeth and {Burden}, Angela and {Burtin}, Etienne and {Busca}, Nicolas G. and {Cahn}, Robert N. and {Cai}, Yan-Chuan and {Cardiel-Sas}, Laia and {Carlberg}, Raymond G. and {Carton}, Pierre-Henri and {Casas}, Ricard and {Castander}, Francisco J. and {Cervantes-Cota}, Jorge L. and {Claybaugh}, Todd M. and {Close}, Madeline and {Coker}, Carl T. and {Cole}, Shaun and {Comparat}, Johan and {Cooper}, Andrew P. and {Cousinou}, M. -C. and {Crocce}, Martin and {Cuby}, Jean-Gabriel and {Cunningham}, Daniel P. and {Davis}, Tamara M. and {Dawson}, Kyle S. and {de la Macorra}, Axel and {De Vicente}, Juan and {Delubac}, Timoth\'ee and {Derwent}, Mark and {Dey}, Arjun and {Dhungana}, Govinda and {Ding}, Zhejie and {Doel}, Peter and {Duan}, Yutong T. and {Ealet}, Anne and {Edelstein}, Jerry and {Eftekharzadeh}, Sarah and {Eisenstein}, Daniel J. and {Elliott}, Ann and {Escoffier}, St\'ephanie and {Evatt}, Matthew and {Fagrelius}, Parker and {Fan}, Xiaohui and {Fanning}, Kevin and {Farahi}, Arya and {Farihi}, Jay and {Favole}, Ginevra and {Feng}, Yu and {Fernandez}, Enrique and {Findlay}, Joseph R. and {Finkbeiner}, Douglas P. and {Fitzpatrick}, Michael J. and {Flaugher}, Brenna and {Flender}, Samuel and {Font-Ribera}, Andreu and {Forero-Romero}, Jaime E. and {Fosalba}, Pablo and {Frenk}, Carlos S. and {Fumagalli}, Michele and {Gaensicke}, Boris T. and {Gallo}, Giuseppe and {Garcia-Bellido}, Juan and {Gaztanaga}, Enrique and {Fusillo}, Nicola Pietro Gentile and {Gerard}, Terry and {Gershkovich}, Irena and {Giannantonio}, Tommaso and {Gillet}, Denis and {Gonzalez-de-Rivera}, Guillermo and {Gonzalez-Perez}, Violeta and {Gott}, Shelby and {Graur}, Or and {Gutierrez}, Gaston and {Guy}, Julien and {Habib}, Salman and {Heetderks}, Henry and {Heetderks}, Ian and {Heitmann}, Katrin and {Hellwing}, Wojciech A. and {Herrera}, David A. and {Ho}, Shirley and {Holland}, Stephen and {Honscheid}, Klaus and {Huff}, Eric and {Hutchinson}, Timothy A. and {Huterer}, Dragan and {Hwang}, Ho Seong and {Laguna}, Joseph Maria Illa and {Ishikawa}, Yuzo and {Jacobs}, Dianna and {Jeffrey}, Niall and {Jelinsky}, Patrick and {Jennings}, Elise and {Jiang}, Linhua and {Jimenez}, Jorge and {Johnson}, Jennifer and {Joyce}, Richard and {Jullo}, Eric and {Juneau}, St\'ephanie and {Kama}, Sami and {Karcher}, Armin and {Karkar}, Sonia and {Kehoe}, Robert and {Kennamer}, Noble and {Kent}, Stephen and {Kilbinger}, Martin and {Kim}, Alex G. and {Kirkby}, David and {Kisner}, Theodore and {Kitanidis}, Ellie and {Kneib}, Jean-Paul and {Koposov}, Sergey and {Kovacs}, Eve and {Koyama}, Kazuya and {Kremin}, Anthony and {Kron}, Richard and {Kronig}, Luzius and {Kueter-Young}, Andrea and {Lacey}, Cedric G. and {Lafever}, Robin and {Lahav}, Ofer and {Lambert}, Andrew and {Lampton}, Michael and {Landriau}, Martin and {Lang}, Dustin and {Lauer}, Tod R. and {Goff}, Jean-Marc Le and {Guillou}, Laurent Le and {Van Suu}, Auguste Le and {Lee}, Jae Hyeon and {Lee}, Su-Jeong and {Leitner}, Daniela and {Lesser}, Michael and {Levi}, Michael E. and {L'Huillier}, Benjamin and {Li}, Baojiu and {Liang}, Ming and {Lin}, Huan and {Linder}, Eric and {Loebman}, Sarah R. and {Luki\'c}, Zarija and {Ma}, Jun and {MacCrann}, Niall and {Magneville}, Christophe and {Makarem}, Laleh and {Manera}, Marc and {Manser}, Christopher J. and {Marshall}, Robert and {Martini}, Paul and {Massey}, Richard and {Matheson}, Thomas and {McCauley}, Jeremy and {McDonald}, Patrick and {McGreer}, Ian D. and {Meisner}, Aaron and {Metcalfe}, Nigel and {Miller}, Timothy N. and {Miquel}, Ramon and {Moustakas}, John and {Myers}, Adam and {Naik}, Milind and {Newman}, Jeffrey A. and {Nichol}, Robert C. and {Nicola}, Andrina and {da Costa}, Luiz Nicolati and {Nie}, Jundan and {Niz}, Gustavo and {Norberg}, Peder and {Nord}, Brian and {Norman}, Dara and {Nugent}, Peter and {O'Brien}, Thomas and {Oh}, Minji and {Olsen}, Knut A. G. and {Padilla}, Cristobal and {Padmanabhan}, Hamsa and {Padmanabhan}, Nikhil and {Palanque-Delabrouille}, Nathalie and {Palmese}, Antonella and {Pappalardo}, Daniel and {P\^aris}, Isabelle and {Park}, Changbom and {Patej}, Anna and {Peacock}, John A. and {Peiris}, Hiranya V. and {Peng}, Xiyan and {Percival}, Will J. and {Perruchot}, Sandrine and {Pieri}, Matthew M. and {Pogge}, Richard and {Pollack}, Jennifer E. and {Poppett}, Claire and {Prada}, Francisco and {Prakash}, Abhishek and {Probst}, Ronald G. and {Rabinowitz}, David and {Raichoor}, Anand and {Ree}, Chang Hee and {Refregier}, Alexandre and {Regal}, Xavier and {Reid}, Beth and {Reil}, Kevin and {Rezaie}, Mehdi and {Rockosi}, Constance M. and {Roe}, Natalie and {Ronayette}, Samuel and {Roodman}, Aaron and {Ross}, Ashley J. and {Ross}, Nicholas P. and {Rossi}, Graziano and {Rozo}, Eduardo and {Ruhlmann-Kleider}, Vanina and {Rykoff}, Eli S. and {Sabiu}, Cristiano and {Samushia}, Lado and {Sanchez}, Eusebio and {Sanchez}, Javier and {Schlegel}, David J. and {Schneider}, Michael and {Schubnell}, Michael and {Secroun}, Aur\'elia and {Seljak}, Uros and {Seo}, Hee-Jong and {Serrano}, Santiago and {Shafieloo}, Arman and {Shan}, Huanyuan and {Sharples}, Ray and {Sholl}, Michael J. and {Shourt}, William V. and {Silber}, Joseph H. and {Silva}, David R. and {Sirk}, Martin M. and {Slosar}, Anze and {Smith}, Alex and {Smoot}, George F. and {Som}, Debopam and {Song}, Yong-Seon and {Sprayberry}, David and {Staten}, Ryan and {Stefanik}, Andy and {Tarle}, Gregory and {Tie}, Suk Sien and {Tinker}, Jeremy L. and {Tojeiro}, Rita and {Valdes}, Francisco and {Valenzuela}, Octavio and {Valluri}, Monica and {Vargas-Magana}, Mariana and {Verde}, Licia and {Walker}, Alistair R. and {Wang}, Jiali and {Wang}, Yuting and {Weaver}, Benjamin A. and {Weaverdyck}, Curtis and {Wechsler}, Risa H. and {Weinberg}, David H. and {White}, Martin and {Yang}, Qian and {Yeche}, Christophe and {Zhang}, Tianmeng and {Zhao}, Gong-Bo and {Zheng}, Yi and {Zhou}, Xu and {Zhou}, Zhimin and {Zhu}, Yaling and {Zou}, Hu and {Zu}, Ying},
  title = {{The DESI Experiment Part I: Science, Targeting, and Survey Design}},
  journal = {arXiv e-prints},
  year = {2016},
  eid = {arXiv:1611.00036},
  pages = {arXiv:1611.00036},
  doi = {10.48550/arXiv.1611.00036},
  url = {https://doi.org/10.48550/arXiv.1611.00036},
  archivePrefix = {arXiv},
  eprint = {1611.00036},
  primaryClass = {astro-ph.IM},
  adsurl = {https://ui.adsabs.harvard.edu/search/q=doi%3A%2210.48550%2FarXiv.1611.00036%22}
}

@article{Silber2023,
  author = {{Silber}, Joseph Harry and {Fagrelius}, Parker and {Fanning}, Kevin and {Schubnell}, Michael and {Aguilar}, Jessica Nicole and {Ahlen}, Steven and {Ameel}, Jon and {Ballester}, Otger and {Baltay}, Charles and {Bebek}, Chris and {Beard}, Dominic Benton and {Besuner}, Robert and {Cardiel-Sas}, Laia and {Casas}, Ricard and {Castander}, Francisco Javier and {Claybaugh}, Todd and {Dobson}, Carl and {Duan}, Yutong and {Dunlop}, Patrick and {Edelstein}, Jerry and {Emmet}, William T. and {Elliott}, Ann and {Evatt}, Matthew and {Gershkovich}, Irena and {Guy}, Julien and {Harris}, Stu and {Heetderks}, Henry and {Heetderks}, Ian and {Honscheid}, Klaus and {Illa}, Jose Maria and {Jelinsky}, Patrick and {Jelinsky}, Sharon R. and {Jimenez}, Jorge and {Karcher}, Armin and {Kent}, Stephen and {Kirkby}, David and {Kneib}, Jean-Paul and {Lambert}, Andrew and {Lampton}, Mike and {Leitner}, Daniela and {Levi}, Michael and {McCauley}, Jeremy and {Meisner}, Aaron and {Miller}, Timothy N. and {Miquel}, Ramon and {Mundet}, Juli\'a and {Poppett}, Claire and {Rabinowitz}, David and {Reil}, Kevin and {Roman}, David and {Schlegel}, David and {Serrano}, Santiago and {Van Shourt}, William and {Sprayberry}, David and {Tarl\'e}, Gregory and {Tie}, Suk Sien and {Weaverdyck}, Curtis and {Zhang}, Kai and {Azzaro}, Marco and {Bailey}, Stephen and {Becerril}, Santiago and {Blackwell}, Tami and {Bouri}, Mohamed and {Brooks}, David and {Buckley-Geer}, Elizabeth and {Castro}, Jose Pe\~nate and {Derwent}, Mark and {Dey}, Arjun and {Dhungana}, Govinda and {Doel}, Peter and {Eisenstein}, Daniel J. and {Fahim}, Nasib and {Garcia-Bellido}, Juan and {Gazta\~naga}, Enrique and {A Gontcho}, Satya Gontcho and {Gutierrez}, Gaston and {H\"orler}, Philipp and {Kehoe}, Robert and {Kisner}, Theodore and {Kremin}, Anthony and {Kronig}, Luzius and {Landriau}, Martin and {Le Guillou}, Laurent and {Martini}, Paul and {Moustakas}, John and {Palanque-Delabrouille}, Nathalie and {Peng}, Xiyan and {Percival}, Will and {Prada}, Francisco and {Prieto}, Carlos Allende and {de Rivera}, Guillermo Gonzalez and {Sanchez}, Eusebio and {Sanchez}, Justo and {Sharples}, Ray and {Soares-Santos}, Marcelle and {Schlafly}, Edward and {Weaver}, Benjamin Alan and {Zhou}, Zhimin and {Zhu}, Yaling and {Zou}, Hu and {(DESI Collaboration)}},
  title = {{The Robotic Multiobject Focal Plane System of the Dark Energy Spectroscopic Instrument (DESI)}},
  journal = {AJ},
  year = {2023},
  month = {jan},
  volume = {165},
  number = {1},
  eid = {9},
  pages = {9},
  doi = {10.3847/1538-3881/ac9ab1},
  url = {https://doi.org/10.3847/1538-3881/ac9ab1},
  adsurl = {https://ui.adsabs.harvard.edu/search/q=doi%3A%2210.3847%2F1538-3881%2Fac9ab1%22}
}

@article{Guy2023,
  author = {{Guy}, J. and {Bailey}, S. and {Kremin}, A. and {Alam}, Shadab and {Alexander}, D. M. and {Allende Prieto}, C. and {BenZvi}, S. and {Bolton}, A. S. and {Brooks}, D. and {Chaussidon}, E. and {Cooper}, A. P. and {Dawson}, K. and {de la Macorra}, A. and {Dey}, A. and {Dey}, Biprateep and {Dhungana}, G. and {Eisenstein}, D. J. and {Font-Ribera}, A. and {Forero-Romero}, J. E. and {Gazta\~naga}, E. and {Gontcho A Gontcho}, S. and {Green}, D. and {Honscheid}, K. and {Ishak}, M. and {Kehoe}, R. and {Kirkby}, D. and {Kisner}, T. and {Koposov}, Sergey E. and {Lan}, Ting-Wen and {Landriau}, M. and {Le Guillou}, L. and {Levi}, Michael E. and {Magneville}, C. and {Manser}, Christopher J. and {Martini}, P. and {Meisner}, Aaron M. and {Miquel}, R. and {Moustakas}, J. and {Myers}, Adam D. and {Newman}, Jeffrey A. and {Nie}, Jundan and {Palanque-Delabrouille}, N. and {Percival}, W. J. and {Poppett}, C. and {Prada}, F. and {Raichoor}, A. and {Ravoux}, C. and {Ross}, A. J. and {Schlafly}, E. F. and {Schlegel}, D. and {Schubnell}, M. and {Sharples}, Ray M. and {Tarl\'e}, Gregory and {Weaver}, B. A. and {Y\'eche}, Christophe and {Zhou}, Rongpu and {Zhou}, Zhimin and {Zou}, H.},
  title = {{The Spectroscopic Data Processing Pipeline for the Dark Energy Spectroscopic Instrument}},
  journal = {AJ},
  year = {2023},
  month = {apr},
  volume = {165},
  number = {4},
  eid = {144},
  pages = {144},
  doi = {10.3847/1538-3881/acb212},
  url = {https://doi.org/10.3847/1538-3881/acb212},
  adsurl = {https://ui.adsabs.harvard.edu/search/q=doi%3A%2210.3847%2F1538-3881%2Facb212%22}
}

@article{DESI2024a,
  author = {{DESI Collaboration} and {Adame}, A. G. and {Aguilar}, J. and {Ahlen}, S. and {Alam}, S. and {Aldering}, G. and {Alexander}, D. M. and {Alfarsy}, R. and {Allende Prieto}, C. and {Alvarez}, M. and {Alves}, O. and {Anand}, A. and {Andrade-Oliveira}, F. and {Armengaud}, E. and {Asorey}, J. and {Avila}, S. and {Aviles}, A. and {Bailey}, S. and {Balaguera-Antol{\'\i}nez}, A. and {Ballester}, O. and {Baltay}, C. and {Bault}, A. and {Bautista}, J. and {Behera}, J. and {Beltran}, S. F. and {BenZvi}, S. and {Beraldo e Silva}, L. and {Bermejo-Climent}, J. R. and {Berti}, A. and {Besuner}, R. and {Beutler}, F. and {Bianchi}, D. and {Blake}, C. and {Blum}, R. and {Bolton}, A. S. and {Brieden}, S. and {Brodzeller}, A. and {Brooks}, D. and {Brown}, Z. and {Buckley-Geer}, E. and {Burtin}, E. and {Cabayol-Garcia}, L. and {Cai}, Z. and {Canning}, R. and {Cardiel-Sas}, L. and {Carnero Rosell}, A. and {Castander}, F. J. and {Cervantes-Cota}, J. L. and {Chabanier}, S. and {Chaussidon}, E. and {Chaves-Montero}, J. and {Chen}, S. and {Chen}, X. and {Chuang}, C. and {Claybaugh}, T. and {Cole}, S. and {Cooper}, A. P. and {Cuceu}, A. and {Davis}, T. M. and {Dawson}, K. and {de Belsunce}, R. and {de la Cruz}, R. and {de la Macorra}, A. and {de Mattia}, A. and {Demina}, R. and {Demirbozan}, U. and {DeRose}, J. and {Dey}, A. and {Dey}, B. and {Dhungana}, G. and {Ding}, J. and {Ding}, Z. and {Doel}, P. and {Doshi}, R. and {Douglass}, K. and {Edge}, A. and {Eftekharzadeh}, S. and {Eisenstein}, D. J. and {Elliott}, A. and {Escoffier}, S. and {Fagrelius}, P. and {Fan}, X. and {Fanning}, K. and {Fawcett}, V. A. and {Ferraro}, S. and {Ereza}, J. and {Flaugher}, B. and {Font-Ribera}, A. and {Forero-S\'anchez}, D. and {Forero-Romero}, J. E. and {Frenk}, C. S. and {G\"ansicke}, B. T. and {Garc{\'\i}a}, L. \'A. and {Garc{\'\i}a-Bellido}, J. and {Garcia-Quintero}, C. and {Garrison}, L. H. and {Gil-Mar{\'\i}n}, H. and {Golden-Marx}, J. and {Gontcho A Gontcho}, S. and {Gonzalez-Morales}, A. X. and {Gonzalez-Perez}, V. and {Gordon}, C. and {Graur}, O. and {Green}, D. and {Gruen}, D. and {Guy}, J. and {Hadzhiyska}, B. and {Hahn}, C. and {Han}, J. J. and {Hanif}, M. M. S. and {Herrera-Alcantar}, H. K. and {Honscheid}, K. and {Hou}, J. and {Howlett}, C. and {Huterer}, D. and {Ir\v{s}i\v{c}}, V. and {Ishak}, M. and {Jana}, A. and {Jiang}, L. and {Jimenez}, J. and {Jing}, Y. P. and {Joudaki}, S. and {Jullo}, E. and {Joyce}, R. and {Juneau}, S. and {Kizhuprakkat}, N. and {Kara\c{c}ayl{\i}}, N. G. and {Karim}, T. and {Kehoe}, R. and {Kent}, S. and {Khederlarian}, A. and {Kim}, S. and {Kirkby}, D. and {Kisner}, T. and {Kitaura}, F. and {Kneib}, J. and {Koposov}, S. E. and {Kov\'acs}, A. and {Kremin}, A. and {Krolewski}, A. and {LHuillier}, B. and {Lahav}, O. and {Lambert}, A. and {Lamman}, C. and {Lan}, T.-W. and {Landriau}, M. and {Lang}, D. and {Lange}, J. U. and {Lasker}, J. and {Le Guillou}, L. and {Leauthaud}, A. and {Levi}, M. E. and {Li}, T. S. and {Linder}, E. and {Lyons}, A. and {Magneville}, C. and {Manera}, M. and {Manser}, C. J. and {Margala}, D. and {Martini}, P. and {McDonald}, P. and {Medina}, G. E. and {Medina-Varela}, L. and {Meisner}, A. and {Mena-Fern\'andez}, J. and {Meneses-Rizo}, J. and {Mezcua}, M. and {Miquel}, R. and {Montero-Camacho}, P. and {Moon}, J. and {Moore}, S. and {Moustakas}, J. and {Mueller}, E. and {Mundet}, J. and {Mu\~noz-Guti\'errez}, A. and {Myers}, A. D. and {Nadathur}, S. and {Napolitano}, L. and {Neveux}, R. and {Newman}, J. A. and {Nie}, J. and {Niz}, G. and {Norberg}, P. and {Noriega}, H. E. and {Paillas}, E. and {Palanque-Delabrouille}, N. and {Palmese}, A. and {Zhiwei}, P. and {Parkinson}, D. and {Penmetsa}, S. and {Percival}, W. J. and {P\'erez-Fern\'andez}, A. and {P\'erez-R\`afols}, I. and {Pieri}, M. and {Poppett}, C. and {Porredon}, A. and {Prada}, F. and {Pucha}, R. and {Raichoor}, A. and {Ram{\'\i}rez-P\'erez}, C. and {Ramirez-Solano}, S. and {Rashkovetskyi}, M. and {Ravoux}, C. and {Rocher}, A. and {Rockosi}, C. and {Ross}, A. J. and {Rossi}, G. and {Ruggeri}, R. and {Ruhlmann-Kleider}, V. and {Sabiu}, C. G. and {Said}, K. and {Saintonge}, A. and {Samushia}, L. and {Sanchez}, E. and {Saulder}, C. and {Schaan}, E. and {Schlafly}, E. F. and {Schlegel}, D. and {Scholte}, D. and {Schubnell}, M. and {Seo}, H. and {Shafieloo}, A. and {Sharples}, R. and {Sheu}, W. and {Silber}, J. and {Sinigaglia}, F. and {Siudek}, M. and {Slepian}, Z. and {Smith}, A. and {Sprayberry}, D. and {Stephey}, L. and {Su\'arez-P\'erez}, J. and {Sun}, Z. and {Tan}, T. and {Tarl\'e}, G. and {Tojeiro}, R. and {Ure\~na-L\'opez}, L. A. and {Vaisakh}, R. and {Valcin}, D. and {Valdes}, F. and {Valluri}, M. and {Vargas-Maga\~na}, M. and {Variu}, A. and {Verde}, L. and {Walther}, M. and {Wang}, B. and {Wang}, M. S. and {Weaver}, B. A. and {Weaverdyck}, N. and {Wechsler}, R. H. and {White}, M. and {Xie}, Y. and {Yang}, J. and {Y\`eche}, C. and {Yu}, J. and {Yuan}, S. and {Zhang}, H. and {Zhang}, Z. and {Zhao}, C. and {Zheng}, Z. and {Zhou}, R. and {Zhou}, Z. and {Zou}, H. and {Zou}, S. and {Zu}, Y.},
  title = {{Validation of the Scientific Program for the Dark Energy Spectroscopic Instrument}},
  journal = {AJ},
  year = {2024},
  month = {feb},
  volume = {167},
  number = {2},
  eid = {62},
  pages = {62},
  doi = {10.3847/1538-3881/ad0b08},
  url = {https://doi.org/10.3847/1538-3881/ad0b08},
  adsurl = {https://ui.adsabs.harvard.edu/search/q=doi%3A%2210.3847%2F1538-3881%2Fad0b08%22}
}

@article{DESI2024b,
  author = {{Hahn}, ChangHoon and {Wilson}, Michael J. and {Ruiz-Macias}, Omar and {Cole}, Shaun and {Weinberg}, David H. and {Moustakas}, John and {Kremin}, Anthony and {Tinker}, Jeremy L. and {Smith}, Alex and {Wechsler}, Risa H. and {Ahlen}, Steven and {Alam}, Shadab and {Bailey}, Stephen and {Brooks}, David and {Cooper}, Andrew P. and {Davis}, Tamara M. and {Dawson}, Kyle and {Dey}, Arjun and {Dey}, Biprateep and {Eftekharzadeh}, Sarah and {Eisenstein}, Daniel J. and {Fanning}, Kevin and {Forero-Romero}, Jaime E. and {Frenk}, Carlos S. and {Gazta\~naga}, Enrique and {A Gontcho}, Satya Gontcho and {Guy}, Julien and {Honscheid}, Klaus and {Ishak}, Mustapha and {Juneau}, St\'ephanie and {Kehoe}, Robert and {Kisner}, Theodore and {Lan}, Ting-Wen and {Landriau}, Martin and {Le Guillou}, Laurent and {Levi}, Michael E. and {Magneville}, Christophe and {Martini}, Paul and {Meisner}, Aaron and {Myers}, Adam D. and {Nie}, Jundan and {Norberg}, Peder and {Palanque-Delabrouille}, Nathalie and {Percival}, Will J. and {Poppett}, Claire and {Prada}, Francisco and {Raichoor}, Anand and {Ross}, Ashley J. and {Gaines}, Sasha and {Saulder}, Christoph and {Schlafly}, Eddie and {Schlegel}, David and {Sierra-Porta}, David and {Tarle}, Gregory and {Weaver}, Benjamin A. and {Y\`eche}, Christophe and {Zarrouk}, Pauline and {Zhou}, Rongpu and {Zhou}, Zhimin and {Zou}, Hu},
  title = {{The DESI Bright Galaxy Survey: Final Target Selection, Design, and Validation}},
  journal = {AJ},
  year = {2023},
  month = {jun},
  volume = {165},
  number = {6},
  eid = {253},
  pages = {253},
  doi = {10.3847/1538-3881/accff8},
  url = {https://doi.org/10.3847/1538-3881/accff8},
  adsurl = {https://ui.adsabs.harvard.edu/search/q=doi%3A%2210.3847%2F1538-3881%2Faccff8%22}
}

@article{Hahn2023,
  author = {{Hahn}, ChangHoon and {Wilson}, Michael J. and {Ruiz-Macias}, Omar and {Cole}, Shaun and {Weinberg}, David H. and {Moustakas}, John and {Kremin}, Anthony and {Tinker}, Jeremy L. and {Smith}, Alex and {Wechsler}, Risa H. and {Ahlen}, Steven and {Alam}, Shadab and {Bailey}, Stephen and {Brooks}, David and {Cooper}, Andrew P. and {Davis}, Tamara M. and {Dawson}, Kyle and {Dey}, Arjun and {Dey}, Biprateep and {Eftekharzadeh}, Sarah and {Eisenstein}, Daniel J. and {Fanning}, Kevin and {Forero-Romero}, Jaime E. and {Frenk}, Carlos S. and {Gazta\~naga}, Enrique and {A Gontcho}, Satya Gontcho and {Guy}, Julien and {Honscheid}, Klaus and {Ishak}, Mustapha and {Juneau}, St\'ephanie and {Kehoe}, Robert and {Kisner}, Theodore and {Lan}, Ting-Wen and {Landriau}, Martin and {Le Guillou}, Laurent and {Levi}, Michael E. and {Magneville}, Christophe and {Martini}, Paul and {Meisner}, Aaron and {Myers}, Adam D. and {Nie}, Jundan and {Norberg}, Peder and {Palanque-Delabrouille}, Nathalie and {Percival}, Will J. and {Poppett}, Claire and {Prada}, Francisco and {Raichoor}, Anand and {Ross}, Ashley J. and {Gaines}, Sasha and {Saulder}, Christoph and {Schlafly}, Eddie and {Schlegel}, David and {Sierra-Porta}, David and {Tarle}, Gregory and {Weaver}, Benjamin A. and {Y\`eche}, Christophe and {Zarrouk}, Pauline and {Zhou}, Rongpu and {Zhou}, Zhimin and {Zou}, Hu},
  title = {{The DESI Bright Galaxy Survey: Final Target Selection, Design, and Validation}},
  journal = {AJ},
  year = {2023},
  month = {jun},
  volume = {165},
  number = {6},
  eid = {253},
  pages = {253},
  doi = {10.3847/1538-3881/accff8},
  url = {https://doi.org/10.3847/1538-3881/accff8},
  adsurl = {https://ui.adsabs.harvard.edu/search/q=doi%3A%2210.3847%2F1538-3881%2Faccff8%22}
}

@article{Ruel2014,
  author = {{Ruel}, J. and {Bazin}, G. and {Bayliss}, M. and {Brodwin}, M. and {Foley}, R. J. and {Stalder}, B. and {Aird}, K. A. and {Armstrong}, R. and {Ashby}, M. L. N. and {Bautz}, M. and {Benson}, B. A. and {Bleem}, L. E. and {Bocquet}, S. and {Carlstrom}, J. E. and {Chang}, C. L. and {Chapman}, S. C. and {Cho}, H. M. and {Clocchiatti}, A. and {Crawford}, T. M. and {Crites}, A. T. and {de Haan}, T. and {Desai}, S. and {Dobbs}, M. A. and {Dudley}, J. P. and {Forman}, W. R. and {George}, E. M. and {Gladders}, M. D. and {Gonzalez}, A. H. and {Halverson}, N. W. and {Harrington}, N. L. and {High}, F. W. and {Holder}, G. P. and {Holzapfel}, W. L. and {Hrubes}, J. D. and {Jones}, C. and {Joy}, M. and {Keisler}, R. and {Knox}, L. and {Lee}, A. T. and {Leitch}, E. M. and {Liu}, J. and {Lueker}, M. and {Luong-Van}, D. and {Mantz}, A. and {Marrone}, D. P. and {McDonald}, M. and {McMahon}, J. J. and {Mehl}, J. and {Meyer}, S. S. and {Mocanu}, L. and {Mohr}, J. J. and {Montroy}, T. E. and {Murray}, S. S. and {Natoli}, T. and {Nurgaliev}, D. and {Padin}, S. and {Plagge}, T. and {Pryke}, C. and {Reichardt}, C. L. and {Rest}, A. and {Ruhl}, J. E. and {Saliwanchik}, B. R. and {Saro}, A. and {Sayre}, J. T. and {Schaffer}, K. K. and {Shaw}, L. and {Shirokoff}, E. and {Song}, J. and {\v{S}uhada}, R. and {Spieler}, H. G. and {Stanford}, S. A. and {Staniszewski}, Z. and {Starsk}, A. A. and {Story}, K. and {Stubbs}, C. W. and {van Engelen}, A. and {Vanderlinde}, K. and {Vieira}, J. D. and {Vikhlinin}, A. and {Williamson}, R. and {Zahn}, O. and {Zenteno}, A.},
  title = {{OPTICAL SPECTROSCOPY AND VELOCITY DISPERSIONS OF GALAXY CLUSTERS FROM THE SPT-SZ SURVEY}},
  journal = {ApJ},
  year = {2014},
  month = {aug},
  volume = {792},
  number = {1},
  eid = {45},
  pages = {45},
  doi = {10.1088/0004-637X/792/1/45},
  url = {https://doi.org/10.1088/0004-637X/792/1/45},
  adsurl = {https://ui.adsabs.harvard.edu/search/q=doi%3A%2210.1088%2F0004-637X%2F792%2F1%2F45%22}
}

@article{Capasso2019,
  author = {{Capasso}, R. and {Mohr}, J. J. and {Saro}, A. and {Biviano}, A. and {Clerc}, N. and {Finoguenov}, A. and {Grandis}, S. and {Collins}, C. and {Erfanianfar}, G. and {Damsted}, S. and {Kirkpatrick}, C. and {Kukkola}, A.},
  title = {{Mass calibration of the CODEX cluster sample using SPIDERS spectroscopy {\textendash} I. The richness{\textendash}mass relation}},
  journal = {MNRAS},
  year = {2019},
  month = {jun},
  volume = {486},
  number = {2},
  pages = {1594--1607},
  doi = {10.1093/mnras/stz931},
  url = {https://doi.org/10.1093/mnras/stz931},
  adsurl = {https://ui.adsabs.harvard.edu/search/q=doi%3A%2210.1093%2Fmnras%2Fstz931%22}
}

@article{Brammer2008,
  author = {Brammer, G. B. and van Dokkum, P. G. and Coppi, P.},
  title = {{EAZY: A Fast, Public Photometric Redshift Code}},
  journal = {ApJ},
  year = {2008},
  month = {oct},
  volume = {686},
  number = {2},
  pages = {1503--1513},
  doi = {10.1086/591786},
  url = {https://doi.org/10.1086/591786},
  adsurl = {https://ui.adsabs.harvard.edu/search/q=doi%3A%2210.1086%2F591786%22}
}

@article{Williams2009,
  author = {Williams, R. J. and Quadri, R. F. and Franx, M. and van Dokkum, P. and Labb{\'e}, I.},
  title = {{Detection of Quiescent Galaxies in a Bicolor Sequence from z = 0--2}},
  journal = {ApJ},
  year = {2009},
  month = {feb},
  volume = {691},
  number = {2},
  pages = {1879--1895},
  doi = {10.1088/0004-637X/691/2/1879},
  url = {https://doi.org/10.1088/0004-637X/691/2/1879},
  adsurl = {https://ui.adsabs.harvard.edu/search/q=doi%3A%2210.1088%2F0004-637X%2F691%2F2%2F1879%22}
}

@article{Muzzin2013,
  author = {{Muzzin}, Adam and {Marchesini}, Danilo and {Stefanon}, Mauro and {Franx}, Marijn and {McCracken}, Henry J. and {Milvang-Jensen}, Bo and {Dunlop}, James S. and {Fynbo}, J. P. U. and {Brammer}, Gabriel and {Labb\'e}, Ivo and {van Dokkum}, Pieter G.},
  title = {{The Evolution of the Stellar Mass Functions of Star-forming and Quiescent Galaxies to $z = 4$ from the COSMOS/UltraVISTA Survey}},
  journal = {ApJ},
  year = {2013},
  month = {oct},
  volume = {777},
  number = {1},
  eid = {18},
  pages = {18},
  doi = {10.1088/0004-637X/777/1/18},
  url = {https://doi.org/10.1088/0004-637X/777/1/18},
  adsurl = {https://ui.adsabs.harvard.edu/search/q=doi%3A%2210.1088%2F0004-637X%2F777%2F1%2F18%22}
}

@article{Boquien2019,
  author = {{Boquien}, M. and {Burgarella}, D. and {Roehlly}, Y. and {Buat}, V. and {Ciesla}, L. and {Corre}, D. and {Inoue}, A. K. and {Salas}, H.},
  title = {{CIGALE: a python Code Investigating GALaxy Emission}},
  journal = {Astronomy \& Astrophysics},
  year = {2019},
  month = {feb},
  volume = {622},
  pages = {A103},
  doi = {10.1051/0004-6361/201834156},
  url = {https://doi.org/10.1051/0004-6361/201834156},
  adsurl = {https://ui.adsabs.harvard.edu/search/q=doi%3A%2210.1051%2F0004-6361%2F201834156%22}
}

@article{Yang2020,
  author = {{Yang}, G. and {Boquien}, M. and {Buat}, V. and {Burgarella}, D. and {Ciesla}, L. and {Duras}, F. and {Stalevski}, M. and {Brandt}, W. N. and {Papovich}, C.},
  title = {{x-cigale: fitting AGN/galaxy SEDs from X-ray to infrared}},
  journal = {MNRAS},
  year = {2020},
  month = {jan},
  volume = {491},
  number = {1},
  pages = {740--757},
  doi = {10.1093/mnras/stz3001},
  url = {https://doi.org/10.1093/mnras/stz3001},
  adsurl = {https://ui.adsabs.harvard.edu/search/q=doi%3A%2210.1093%2Fmnras%2Fstz3001%22}
}

@article{Yang2022,
  author = {{Yang}, Guang and {Boquien}, M\'ed\'eric and {Brandt}, W. N. and {Buat}, V\'eronique and {Burgarella}, Denis and {Ciesla}, Laure and {Lehmer}, Bret D. and {Ma{\l}ek}, Katarzyna and {Mountrichas}, George and {Papovich}, Casey and {Pons}, Estelle and {Stalevski}, Marko and {Theul\'e}, Patrice and {Zhu}, Shifu},
  title = {{Fitting AGN/Galaxy X-Ray-to-radio SEDs with CIGALE and Improvement of the Code}},
  journal = {ApJ},
  year = {2022},
  month = {mar},
  volume = {927},
  number = {2},
  eid = {192},
  pages = {192},
  doi = {10.3847/1538-4357/ac4971},
  url = {https://doi.org/10.3847/1538-4357/ac4971},
  adsurl = {https://ui.adsabs.harvard.edu/search/q=doi%3A%2210.3847%2F1538-4357%2Fac4971%22}
}

@article{Schlafly2019,
  author = {{Schlafly}, Edward F. and {Meisner}, Aaron M. and {Green}, Gregory M.},
  title = {{The unWISE Catalog: Two Billion Infrared Sources from Five Years of WISE Imaging}},
  journal = {ApJS},
  year = {2019},
  month = {feb},
  volume = {240},
  number = {2},
  eid = {30},
  pages = {30},
  doi = {10.3847/1538-4365/aafbea},
  url = {https://doi.org/10.3847/1538-4365/aafbea},
  adsurl = {https://ui.adsabs.harvard.edu/search/q=doi%3A%2210.3847%2F1538-4365%2Faafbea%22}
}

@article{Bruzual2003,
  author = {Bruzual, G. and Charlot, S.},
  title = {{Stellar Population Synthesis at the Resolution of 2003}},
  journal = {MNRAS},
  year = {2003},
  month = {oct},
  volume = {344},
  number = {4},
  pages = {1000--1028},
  doi = {10.1046/j.1365-8711.2003.06897.x},
  url = {https://doi.org/10.1046/j.1365-8711.2003.06897.x},
  adsurl = {https://ui.adsabs.harvard.edu/search/q=doi%3A%2210.1046%2Fj.1365-8711.2003.06897.x%22}
}

@article{Calzetti2000,
  author = {{Calzetti}, Daniela and {Armus}, Lee and {Bohlin}, Ralph C. and {Kinney}, Anne L. and {Koornneef}, Jan and {StorchiBergmann}, Thaisa},
  title = {{The Dust Content and Opacity of Actively Star-forming Galaxies}},
  journal = {ApJ},
  year = {2000},
  month = {apr},
  volume = {533},
  number = {2},
  pages = {682--695},
  doi = {10.1086/308692},
  url = {https://doi.org/10.1086/308692},
  adsurl = {https://ui.adsabs.harvard.edu/search/q=doi%3A%2210.1086%2F308692%22}
}

@article{Dale2014,
  author = {{Dale}, Daniel A. and {Helou}, George and {Magdis}, Georgios E. and {Armus}, Lee and {D{\'\i}az-Santos}, Tanio and {Shi}, Yong},
  title = {{A Two-parameter Model for the Infrared/Submillimeter/Radio Spectral Energy Distributions of Galaxies and Active Galactic Nuclei}},
  journal = {ApJ},
  year = {2014},
  month = {mar},
  volume = {784},
  number = {1},
  eid = {83},
  pages = {83},
  doi = {10.1088/0004-637X/784/1/83},
  url = {https://doi.org/10.1088/0004-637X/784/1/83},
  adsurl = {https://ui.adsabs.harvard.edu/search/q=doi%3A%2210.1088%2F0004-637X%2F784%2F1%2F83%22}
}

@article{Fritz2006,
  author = {{Fritz}, J. and {Franceschini}, A. and {Hatziminaoglou}, E.},
  title = {{Revisiting the infrared spectra of active galactic nuclei with a new torus emission model}},
  journal = {MNRAS},
  year = {2006},
  month = {mar},
  volume = {366},
  number = {3},
  pages = {767--786},
  doi = {10.1111/j.1365-2966.2006.09866.x},
  url = {https://doi.org/10.1111/j.1365-2966.2006.09866.x},
  adsurl = {https://ui.adsabs.harvard.edu/search/q=doi%3A%2210.1111%2Fj.1365-2966.2006.09866.x%22}
}

@article{Ding2020,
  author = {{Ding}, Xuheng and {Silverman}, John and {Treu}, Tommaso and {Schulze}, Andreas and {Schramm}, Malte and {Birrer}, Simon and {Park}, Daeseong and {Jahnke}, Knud and {Bennert}, Vardha N. and {Kartaltepe}, Jeyhan S. and {Koekemoer}, Anton M. and {Malkan}, Matthew A. and {Sanders}, David},
  title = {{The Mass Relations between Supermassive Black Holes and Their Host Galaxies at $1 < z < 2$ with HST-WFC3}},
  journal = {ApJ},
  year = {2020},
  month = {jan},
  volume = {888},
  number = {1},
  eid = {37},
  pages = {37},
  doi = {10.3847/1538-4357/ab5b90},
  url = {https://doi.org/10.3847/1538-4357/ab5b90},
  adsurl = {https://ui.adsabs.harvard.edu/search/q=doi%3A%2210.3847%2F1538-4357%2Fab5b90%22}
}

@article{Birrer2015,
  author = {Birrer, S. and Amara, A. and Refregier, A.},
  title = {{Gravitational Lens Modeling with Basis Sets}},
  journal = {ApJ},
  year = {2015},
  month = {nov},
  volume = {813},
  number = {2},
  eid = {102},
  pages = {102},
  doi = {10.1088/0004-637X/813/2/102},
  url = {https://doi.org/10.1088/0004-637X/813/2/102},
  adsurl = {https://ui.adsabs.harvard.edu/search/q=doi%3A%2210.1088%2F0004-637X%2F813%2F2%2F102%22}
}

@article{Birrer2018,
  author = {{Birrer}, Simon and {Amara}, Adam},
  title = {{lenstronomy: Multi-purpose gravitational lens modelling software package}},
  journal = {Physics of the Dark Universe},
  year = {2018},
  month = {dec},
  volume = {22},
  pages = {189--201},
  doi = {10.1016/j.dark.2018.11.002},
  url = {https://doi.org/10.1016/j.dark.2018.11.002},
  adsurl = {https://ui.adsabs.harvard.edu/search/q=doi%3A%2210.1016%2Fj.dark.2018.11.002%22}
}

@INPROCEEDINGS{Bertin2011,
       author = {{Bertin}, E.},
        title = "{Automated Morphometry with SExtractor and PSFEx}",
    booktitle = {Astronomical Data Analysis Software and Systems XX},
         year = 2011,
       editor = {{Evans}, I.~N. and {Accomazzi}, A. and {Mink}, D.~J. and {Rots}, A.~H.},
       series = {Astronomical Society of the Pacific Conference Series},
       volume = {442},
        month = jul,
        pages = {435},
       adsurl = {https://ui.adsabs.harvard.edu/abs/2011ASPC..442..435B}
}

@article{Simard2011,
  author = {{Simard}, Luc and {Trevor Mendel}, J. and {Patton}, David R. and {Ellison}, Sara L. and {McConnachie}, Alan W.},
  title = {{A Catalog of Bulge+Disk Decompositions and Updated Photometry for 1.12 Million Galaxies in the Sloan Digital Sky Survey}},
  journal = {ApJS},
  year = {2011},
  month = {sep},
  volume = {196},
  number = {1},
  eid = {11},
  pages = {11},
  doi = {10.1088/0067-0049/196/1/11},
  url = {https://doi.org/10.1088/0067-0049/196/1/11},
  adsurl = {https://ui.adsabs.harvard.edu/search/q=doi%3A%2210.1088%2F0067-0049%2F196%2F1%2F11%22}
}

@article{Meert2015,
  author = {Meert, A. and Vikram, V. and Bernardi, M.},
  title = {{A catalogue of 2D photometric decompositions in the SDSS-DR7 spectroscopic main galaxy sample: preferred models and systematics}},
  journal = {MNRAS},
  year = {2015},
  month = {feb},
  volume = {446},
  number = {4},
  pages = {3943--3974},
  doi = {10.1093/mnras/stu2333},
  url = {https://doi.org/10.1093/mnras/stu2333},
  adsurl = {https://ui.adsabs.harvard.edu/search/q=doi%3A%2210.1093%2Fmnras%2Fstu2333%22}
}

@article{Rowe2015,
  author = {{Rowe}, B.T.P. and {Jarvis}, M. and {Mandelbaum}, R. and {Bernstein}, G.M. and {Bosch}, J. and {Simet}, M. and {Meyers}, J.E. and {Kacprzak}, T. and {Nakajima}, R. and {Zuntz}, J. and {Miyatake}, H. and {Dietrich}, J.P. and {Armstrong}, R. and {Melchior}, P. and {Gill}, M.S.S.},
  title = {{GalSim: The Modular Galaxy Image Simulation Toolkit}},
  journal = {Astronomy and Computing},
  year = {2015},
  month = {apr},
  volume = {10},
  pages = {121--150},
  doi = {10.1016/j.ascom.2015.02.002},
  url = {https://doi.org/10.1016/j.ascom.2015.02.002},
  adsurl = {https://ui.adsabs.harvard.edu/search/q=doi%3A%2210.1016%2Fj.ascom.2015.02.002%22}
}

@article{Bruzual1983,
  author = {Bruzual, G. A.},
  title = {{Spectral evolution of galaxies. I - Early-type systems}},
  journal = {ApJ},
  year = {1983},
  month = {oct},
  volume = {273},
  pages = {105--127},
  doi = {10.1086/161352},
  url = {https://doi.org/10.1086/161352},
  adsurl = {https://ui.adsabs.harvard.edu/search/q=doi%3A%2210.1086%2F161352%22}
}

@article{Balogh1999,
  author = {{Balogh}, Michael L. and {Morris}, S. L. and {Yee}, H. K. C. and {Carlberg}, R. G. and {Ellingson}, E.},
  title = {{Differential Galaxy Evolution in Cluster and Field Galaxies at $z \approx 0.3$}},
  journal = {ApJ},
  year = {1999},
  month = {dec},
  volume = {527},
  number = {1},
  pages = {54--79},
  doi = {10.1086/308056},
  url = {https://doi.org/10.1086/308056},
  adsurl = {https://ui.adsabs.harvard.edu/abs/1999ApJ...527...54B}
}

@BOOK{Osterbrock2006,
       author = {{Osterbrock}, Donald E. and {Ferland}, Gary J.},
        title = "{Astrophysics of gaseous nebulae and active galactic nuclei}",
         year = 2006,
       adsurl = {https://ui.adsabs.harvard.edu/abs/2006agna.book.....O}
}

@article{Dominguez2013,
  author = {{Dom{\'\i}nguez}, A. and {Siana}, B. and {Henry}, A. L. and {Scarlata}, C. and {Bedregal}, A. G. and {Malkan}, M. and {Atek}, H. and {Ross}, N. R. and {Colbert}, J. W. and {Teplitz}, H. I. and {Rafelski}, M. and {McCarthy}, P. and {Bunker}, A. and {Hathi}, N. P. and {Dressler}, A. and {Martin}, C. L. and {Masters}, D.},
  title = {{Dust Extinction from Balmer Decrements of Star-forming Galaxies at $0.75 \leq z \leq 1.5$ with Hubble Space Telescope/Wide-Field-Camera 3 Spectroscopy from the WFC3 Infrared Spectroscopic Parallel Survey}},
  journal = {ApJ},
  year = {2013},
  month = {jan},
  volume = {763},
  number = {2},
  eid = {145},
  pages = {145},
  doi = {10.1088/0004-637X/763/2/145},
  url = {https://doi.org/10.1088/0004-637X/763/2/145},
  adsurl = {https://ui.adsabs.harvard.edu/search/q=doi%3A%2210.1088%2F0004-637X%2F763%2F2%2F145%22}
}

@article{Kennicutt1998,
  author = {Kennicutt, R. C.},
  title = {{Star Formation in Galaxies Along the Hubble Sequence}},
  journal = {ARA\&A},
  year = {1998},
  month = {sep},
  volume = {36},
  number = {1},
  pages = {189--231},
  doi = {10.1146/annurev.astro.36.1.189},
  url = {https://doi.org/10.1146/annurev.astro.36.1.189},
  adsurl = {https://ui.adsabs.harvard.edu/search/q=doi%3A%2210.1146%2Fannurev.astro.36.1.189%22}
}

@article{Hao2011,
  author = {{Hao}, Cai-Na and {Kennicutt}, Robert C. and {Johnson}, Benjamin D. and {Calzetti}, Daniela and {Dale}, Daniel A. and {Moustakas}, John},
  title = {{Dust-corrected Star Formation Rates of Galaxies. II. Combinations of Ultraviolet and Infrared Tracers}},
  journal = {ApJ},
  year = {2011},
  month = {nov},
  volume = {741},
  number = {2},
  eid = {124},
  pages = {124},
  doi = {10.1088/0004-637X/741/2/124},
  url = {https://doi.org/10.1088/0004-637X/741/2/124},
  adsurl = {https://ui.adsabs.harvard.edu/search/q=doi%3A%2210.1088%2F0004-637X%2F741%2F2%2F124%22}
}

@article{Murphy2011,
  author = {{Murphy}, E. J. and {Condon}, J. J. and {Schinnerer}, E. and {Kennicutt}, R. C. and {Calzetti}, D. and {Armus}, L. and {Helou}, G. and {Turner}, J. L. and {Aniano}, G. and {Beir\~ao}, P. and {Bolatto}, A. D. and {Brandl}, B. R. and {Croxall}, K. V. and {Dale}, D. A. and {Meyer}, J. L. Donovan and {Draine}, B. T. and {Engelbracht}, C. and {Hunt}, L. K. and {Hao}, C.-N. and {Koda}, J. and {Roussel}, H. and {Skibba}, R. and {Smith}, J.-D. T.},
  title = {{Calibrating Extinction-free Star Formation Rate Diagnostics with 33 GHz Free-Free Emission in NGC 6946}},
  journal = {ApJ},
  year = {2011},
  month = {aug},
  volume = {737},
  number = {2},
  eid = {67},
  pages = {67},
  doi = {10.1088/0004-637X/737/2/67},
  url = {https://doi.org/10.1088/0004-637X/737/2/67},
  adsurl = {https://ui.adsabs.harvard.edu/search/q=doi%3A%2210.1088%2F0004-637X%2F737%2F2%2F67%22}
}

@article{Larson1980,
  author = {Larson, R. B. and Tinsley, B. M. and Caldwell, C. N.},
  title = {{The evolution of disk galaxies and the origin of S0 galaxies}},
  journal = {ApJ},
  year = {1980},
  month = {may},
  volume = {237},
  pages = {692--707},
  doi = {10.1086/157917},
  url = {https://doi.org/10.1086/157917},
  adsurl = {https://ui.adsabs.harvard.edu/search/q=doi%3A%2210.1086%2F157917%22}
}

@article{Balogh2000,
  author = {Balogh, M. L. and Navarro, J. F. and Morris, S. L.},
  title = {{The Origin of Star Formation Gradients in Rich Galaxy Clusters}},
  journal = {ApJ},
  year = {2000},
  month = {sep},
  volume = {540},
  number = {1},
  pages = {113--121},
  doi = {10.1086/309323},
  url = {https://doi.org/10.1086/309323},
  adsurl = {https://ui.adsabs.harvard.edu/search/q=doi%3A%2210.1086%2F309323%22}
}

@article{Peng2012,
  author = {{Peng}, Ying-jie and {Lilly}, Simon J. and {Renzini}, Alvio and {Carollo}, C. Marcella},
  title = {{MASS AND ENVIRONMENT AS DRIVERS OF GALAXY EVOLUTION. II. THE QUENCHING OF SATELLITE GALAXIES AS THE ORIGIN OF ENVIRONMENTAL EFFECTS}},
  journal = {ApJ},
  year = {2012},
  month = {sep},
  volume = {757},
  number = {1},
  eid = {4},
  pages = {4},
  doi = {10.1088/0004-637X/757/1/4},
  url = {https://doi.org/10.1088/0004-637X/757/1/4},
  adsurl = {https://ui.adsabs.harvard.edu/abs/2012ApJ...757....4P}
}

@article{Wetzel2013,
  author = {{Wetzel}, Andrew R. and {Tinker}, Jeremy L. and {Conroy}, Charlie and {van den Bosch}, Frank C.},
  title = {{Galaxy evolution in groups and clusters: satellite star formation histories and quenching time-scales in a hierarchical Universe}},
  journal = {MNRAS},
  year = {2013},
  month = {jun},
  volume = {432},
  number = {1},
  pages = {336--358},
  doi = {10.1093/mnras/stt469},
  url = {https://doi.org/10.1093/mnras/stt469},
  adsurl = {https://ui.adsabs.harvard.edu/abs/2013MNRAS.432..336W}
}

@article{Bell2004,
  author = {{Bell}, Eric F. and {Wolf}, Christian and {Meisenheimer}, Klaus and {Rix}, HansWalter and {Borch}, Andrea and {Dye}, Simon and {Kleinheinrich}, Martina and {Wisotzki}, Lutz and {McIntosh}, Daniel H.},
  title = {{Nearly 5000 Distant Early-Type Galaxies in COMBO-17: A Red Sequence and Its Evolution since $z \sim 1$}},
  journal = {ApJ},
  year = {2004},
  month = {jun},
  volume = {608},
  number = {2},
  pages = {752--767},
  doi = {10.1086/420778},
  url = {https://doi.org/10.1086/420778},
  adsurl = {https://ui.adsabs.harvard.edu/search/q=doi%3A%2210.1086%2F420778%22}
}

@article{Martin2007,
  author = {{Martin}, D. Christopher and {Wyder}, Ted K. and {Schiminovich}, David and {Barlow}, Tom A. and {Forster}, Karl and {Friedman}, Peter G. and {Morrissey}, Patrick and {Neff}, Susan G. and {Seibert}, Mark and {Small}, Todd and {Welsh}, Barry Y. and {Bianchi}, Luciana and {Donas}, Jose and {Heckman}, Timothy M. and {Lee}, YoungWook and {Madore}, Barry F. and {Milliard}, Bruno and {Rich}, R. Michael and {Szalay}, Alex S. and {Yi}, Sukyoung K.},
  title = {{The UV-Optical Color Magnitude Diagram. III. Constraints on Evolution from the Blue to Red Sequence}},
  journal = {ApJS},
  year = {2007},
  month = {dec},
  volume = {173},
  number = {2},
  pages = {342--356},
  doi = {10.1086/516639},
  url = {https://doi.org/10.1086/516639},
  adsurl = {https://ui.adsabs.harvard.edu/search/q=doi%3A%2210.1086%2F516639%22}
}

@article{Schawinski2014,
  author = {{Schawinski}, Kevin and {Urry}, C. Megan and {Simmons}, Brooke D. and {Fortson}, Lucy and {Kaviraj}, Sugata and {Keel}, William C. and {Lintott}, Chris J. and {Masters}, Karen L. and {Nichol}, Robert C. and {Sarzi}, Marc and {Skibba}, Ramin and {Treister}, Ezequiel and {Willett}, Kyle W. and {Wong}, O. Ivy and {Yi}, Sukyoung K.},
  title = {{The green valley is a red herring: Galaxy Zoo reveals two evolutionary pathways towards quenching of star formation in early- and late-type galaxies}},
  journal = {MNRAS},
  year = {2014},
  month = {may},
  volume = {440},
  number = {1},
  pages = {889--907},
  doi = {10.1093/mnras/stu327},
  url = {https://doi.org/10.1093/mnras/stu327},
  adsurl = {https://ui.adsabs.harvard.edu/search/q=doi%3A%2210.1093%2Fmnras%2Fstu327%22}
}

@article{Poggianti1999,
  author = {{Poggianti}, Bianca M. and {Smail}, Ian and {Dressler}, Alan and {Couch}, Warrick J. and {Barger}, Amy J. and {Butcher}, Harvey and {Ellis}, Richard S. and {Oemler, Jr.}, Augustus},
  title = {{The Star Formation Histories of Galaxies in Distant Clusters}},
  journal = {ApJ},
  year = {1999},
  month = {jun},
  volume = {518},
  number = {2},
  pages = {576--593},
  doi = {10.1086/307322},
  url = {https://doi.org/10.1086/307322},
  adsurl = {https://ui.adsabs.harvard.edu/search/q=doi%3A%2210.1086%2F307322%22}
}

@article{Koopmann2004,
  author = {Koopmann, Rebecca A. and Kenney, Jeffrey D. P.},
  title = {{H$\alpha$ Morphologies and Environmental Effects in Virgo Cluster Spiral Galaxies}},
  journal = {ApJ},
  year = {2004},
  month = {oct},
  volume = {613},
  number = {2},
  pages = {866--885},
  doi = {10.1086/423191},
  url = {https://doi.org/10.1086/423191},
  adsurl = {https://ui.adsabs.harvard.edu/search/q=doi%3A%2210.1086%2F423191%22}
}

@article{Cibinel2013,
  author = {{Cibinel}, A. and {Carollo}, C. M. and {Lilly}, S. J. and {Miniati}, F. and {Silverman}, J. D. and {van Gorkom}, J. H. and {Cameron}, E. and {Finoguenov}, A. and {Norberg}, P. and {Peng}, Y. and {Pipino}, A. and {Rudick}, C. S.},
  title = {{The Zurich Environmental Study (ZENS) of Galaxies in Groups along the Cosmic Web. II. Galaxy Structural Measurements and the Concentration of Morphologically Classified Satellites in Diverse Environments}},
  journal = {ApJ},
  year = {2013},
  month = {sep},
  volume = {776},
  number = {2},
  eid = {72},
  pages = {72},
  doi = {10.1088/0004-637X/776/2/72},
  url = {https://doi.org/10.1088/0004-637X/776/2/72},
  adsurl = {https://ui.adsabs.harvard.edu/search/q=doi%3A%2210.1088%2F0004-637X%2F776%2F2%2F72%22}
}

@article{Boselli2006,
  author = {Boselli, Alessandro and Gavazzi, Giuseppe},
  title = {{Environmental Effects on Late-Type Galaxies in Nearby Clusters}},
  journal = {PASP},
  year = {2006},
  month = {apr},
  volume = {118},
  number = {842},
  pages = {517--559},
  doi = {10.1086/500691},
  url = {https://doi.org/10.1086/500691},
  adsurl = {https://ui.adsabs.harvard.edu/search/q=doi%3A%2210.1086%2F500691%22}
}

@article{Fillingham2015,
  author = {{Fillingham}, Sean P. and {Cooper}, Michael C. and {Wheeler}, Coral and {Garrison-Kimmel}, Shea and {Boylan-Kolchin}, Michael and {Bullock}, James S.},
  title = {{Taking Care of Business in a Flash: Constraining the Timescale for Low-Mass Satellite Quenching with ELVIS}},
  journal = {MNRAS},
  year = {2015},
  month = {dec},
  volume = {454},
  number = {2},
  pages = {2039--2049},
  doi = {10.1093/mnras/stv2058},
  url = {https://doi.org/10.1093/mnras/stv2058},
  adsurl = {https://ui.adsabs.harvard.edu/search/q=doi%3A%2210.1093%2Fmnras%2Fstv2058%22}
}

@article{Woo2015,
  author = {{Woo}, Joanna and {Dekel}, Avishai and {Faber}, S. M. and {Koo}, David C.},
  title = {{Two conditions for galaxy quenching: compact centres and massive haloes}},
  journal = {MNRAS},
  year = {2015},
  month = {mar},
  volume = {448},
  number = {1},
  pages = {237--251},
  doi = {10.1093/mnras/stu2755},
  url = {https://doi.org/10.1093/mnras/stu2755},
  adsurl = {https://ui.adsabs.harvard.edu/search/q=doi%3A%2210.1093%2Fmnras%2Fstu2755%22}
}

@article{Kauffmann2004,
  author = {{Kauffmann}, Guinevere and {White}, Simon D. M. and {Heckman}, Timothy M. and {M{\'e}nard}, Brice and {Brinchmann}, Jarle and {Charlot}, St{\'e}phane and {Tremonti}, Christy and {Brinkmann}, Jon},
  title = {{The environmental dependence of the relations between stellar mass, structure, star formation and nuclear activity in galaxies}},
  journal = {MNRAS},
  year = {2004},
  month = {aug},
  volume = {353},
  number = {3},
  pages = {713--731},
  doi = {10.1111/j.1365-2966.2004.08117.x},
  url = {https://doi.org/10.1111/j.1365-2966.2004.08117.x},
  adsurl = {https://ui.adsabs.harvard.edu/abs/2004MNRAS.353..713K}
}

@article{Hogg2004,
  author = {{Hogg}, David W. and {Blanton}, Michael R. and {Brinchmann}, Jarle and {Eisenstein}, Daniel J. and {Schlegel}, David J. and {Gunn}, James E. and {McKay}, Timothy A. and {Rix}, Hans-Walter and {Bahcall}, Neta A. and {Brinkmann}, Jon and {Meiksin}, Avery},
  title = {{The Dependence on Environment of the Color-Magnitude Relation of Galaxies}},
  journal = {ApJ},
  year = {2004},
  month = {jan},
  volume = {601},
  number = {1},
  pages = {L29--L32},
  doi = {10.1086/381749},
  url = {https://doi.org/10.1086/381749},
  adsurl = {https://ui.adsabs.harvard.edu/abs/2004ApJ...601L..29H}
}

@article{Cooper2012,
  author = {{Cooper}, Michael C. and {Griffith}, Roger L. and {Newman}, Jeffrey A. and {Coil}, Alison L. and {Davis}, Marc and {Dutton}, Aaron A. and {Faber}, S. M. and {Guhathakurta}, Puragra and {Koo}, David C. and {Lotz}, Jennifer M. and {Weiner}, Benjamin J. and {Willmer}, Christopher N. A. and {Yan}, Renbin},
  title = {{The DEEP3 Galaxy Redshift Survey: the impact of environment on the size evolution of massive early-type galaxies at intermediate redshift}},
  journal = {MNRAS},
  year = {2012},
  month = {feb},
  volume = {419},
  number = {4},
  pages = {3018--3027},
  doi = {10.1111/j.1365-2966.2011.19938.x},
  url = {https://doi.org/10.1111/j.1365-2966.2011.19938.x},
  adsurl = {https://ui.adsabs.harvard.edu/abs/2012MNRAS.419.3018C}
}

@article{PaulinoAfonso2020,
  author = {{Paulino-Afonso}, Ana and {Sobral}, David and {Darvish}, Behnam and {Ribeiro}, Bruno and {Smail}, Ian and {Best}, Philip and {Stroe}, Andra and {Cairns}, Joseph},
  title = {{VIS 3 COS: III. Environmental effects on the star formation histories of galaxies at z \ensuremath{\sim} 0.8 seen in [O II], H \ensuremath{\delta} , and D n 4000}},
  journal = {Astronomy \& Astrophysics},
  year = {2020},
  month = {jan},
  volume = {633},
  eid = {A70},
  pages = {A70},
  doi = {10.1051/0004-6361/201834244},
  url = {https://doi.org/10.1051/0004-6361/201834244},
  adsurl = {https://ui.adsabs.harvard.edu/abs/2020A&A...633A..70P}
}

@article{Maltby2010,
  author = {{Maltby}, David T. and {Arag{\'o}n-Salamanca}, Alfonso and {Gray}, Meghan E. and {Barden}, Marco and {H{\"a}u{\ss}ler}, Boris and {Wolf}, Christian and {Peng}, Chien Y. and {Jahnke}, Knud and {McIntosh}, Daniel H. and {B{\"o}hm}, Asmus and {van Kampen}, Eelco},
  title = {{The environmental dependence of the stellar-mass-size relation in STAGES galaxies}},
  journal = {MNRAS},
  year = {2010},
  month = {feb},
  volume = {402},
  number = {1},
  pages = {282--294},
  doi = {10.1111/j.1365-2966.2009.15953.x},
  url = {https://doi.org/10.1111/j.1365-2966.2009.15953.x},
  adsurl = {https://ui.adsabs.harvard.edu/abs/2010MNRAS.402..282M}
}

@article{Rettura2010,
  author = {{Rettura}, Alessandro and {Rosati}, Piero and {Nonino}, Mario and {Fosbury}, Robert A. E. and {Gobat}, Raphael and {Menci}, Nicola and {Strazzullo}, Veronica and {Mei}, Simona and {Demarco}, Ricardo and {Ford}, Holland C.},
  title = {{Formation Epochs, Star Formation Histories, and Sizes of Massive Early-Type Galaxies in Cluster and Field Environments at z=1.2: Insights from the Rest-frame Ultraviolet}},
  journal = {ApJ},
  year = {2010},
  month = {jan},
  volume = {709},
  number = {1},
  pages = {512--524},
  doi = {10.1088/0004-637X/709/1/512},
  url = {https://doi.org/10.1088/0004-637X/709/1/512},
  adsurl = {https://ui.adsabs.harvard.edu/abs/2010ApJ...709..512R}
}

@article{Morishita2017,
  author = {{Morishita}, Takahiro and {Abramson}, Louis E. and {Treu}, Tommaso and {Vulcani}, Benedetta and {Schmidt}, Kasper B. and {Dressler}, Alan and {Poggianti}, Bianca M. and {Malkan}, Matthew A. and {Wang}, Xin and {Huang}, Kuang-Han and {Trenti}, Michele and {Brada{\v c}}, Maru{\v s}a and {Hoag}, Austin},
  title = {{The Grism Lens-Amplified Survey from Space (GLASS). IX. The Dual Origin of Low-mass Cluster Galaxies as Revealed by New Structural Analyses}},
  journal = {ApJ},
  year = {2017},
  month = {feb},
  volume = {835},
  number = {2},
  eid = {254},
  pages = {254},
  doi = {10.3847/1538-4357/835/2/254},
  url = {https://doi.org/10.3847/1538-4357/835/2/254},
  adsurl = {https://ui.adsabs.harvard.edu/abs/2017ApJ...835..254M}
}

@article{Cebrian2014,
  author = {{Cebri{\'a}n}, Mar{\'i}a and {Trujillo}, Ignacio},
  title = {{The effect of the environment on the stellar mass-size relationship for present-day galaxies}},
  journal = {MNRAS},
  year = {2014},
  month = {oct},
  volume = {444},
  number = {1},
  pages = {682--699},
  doi = {10.1093/mnras/stu1375},
  url = {https://doi.org/10.1093/mnras/stu1375},
  adsurl = {https://ui.adsabs.harvard.edu/abs/2014MNRAS.444..682C}
}

@article{Matharu2019,
  author = {{Matharu}, Jasleen and {Muzzin}, Adam and {Brammer}, Gabriel B. and {van der Burg}, Remco F. J. and {Auger}, Matthew W. and {Hewett}, Paul C. and {van der Wel}, Arjen and {van Dokkum}, Pieter and {Balogh}, Michael and {Chan}, Jeffrey C. C. and {Demarco}, Ricardo and {Marchesini}, Danilo and {Nelson}, Erica J. and {Noble}, Allison G. and {Wilson}, Gillian and {Yee}, Howard K. C.},
  title = {{HST/WFC3 grism observations of $z \sim 1$ clusters: the cluster versus field stellar mass-size relation and evidence for size growth of quiescent galaxies from minor mergers}},
  journal = {MNRAS},
  year = {2019},
  month = {mar},
  volume = {484},
  number = {1},
  pages = {595--617},
  doi = {10.1093/mnras/sty3465},
  url = {https://doi.org/10.1093/mnras/sty3465},
  adsurl = {https://ui.adsabs.harvard.edu/abs/2019MNRAS.484..595M}
}

@article{Matharu2021,
  author = {{Matharu}, Jasleen and {Muzzin}, Adam and {Brammer}, Gabriel B. and {Nelson}, Erica J. and {Auger}, Matthew W. and {Hewett}, Paul C. and {van der Burg}, Remco F. J. and {Balogh}, Michael and {Demarco}, Ricardo and {Marchesini}, Danilo and {Noble}, Allison G. and {Rudnick}, Gregory and {van der Wel}, Arjen and {Wilson}, Gillian and {Yee}, Howard K. C.},
  title = {{HST/WFC3 Grism Observations of z~1 Clusters: Evidence for Rapid Outside-in Environmental Quenching from Spatially Resolved Halpha Maps}},
  journal = {ApJ},
  year = {2021},
  month = {dec},
  volume = {923},
  number = {2},
  eid = {222},
  pages = {222},
  doi = {10.3847/1538-4357/ac26c3},
  url = {https://doi.org/10.3847/1538-4357/ac26c3},
  adsurl = {https://ui.adsabs.harvard.edu/abs/2021ApJ...923..222M}
}

@article{Xu2023,
  author = {{Xu}, Can and {Wang}, Tao and {Gu}, Qiusheng and {Zanella}, Anita and {Xu}, Ke and {Sun}, Hanwen and {Strazzullo}, Veronica and {Valentino}, Francesco and {Gobat}, Raphael and {Daddi}, Emanuele and {Elbaz}, David and {Xiao}, Mengyuan and {Lu}, Shiying and {Zhou}, Luwenjia},
  title = {{Accelerated Structural Evolution of Galaxies in a Starbursting Cluster at z = 2.51}},
  journal = {ApJL},
  year = {2023},
  month = {jul},
  volume = {951},
  number = {1},
  eid = {L21},
  pages = {L21},
  doi = {10.3847/2041-8213/acdb5a},
  url = {https://doi.org/10.3847/2041-8213/acdb5a},
  archivePrefix = {arXiv},
  eprint = {2307.05202},
  primaryClass = {astro-ph.GA},
  adsurl = {https://ui.adsabs.harvard.edu/abs/2023ApJ...951L..21X}
}

@article{Bamford2009,
  author = {{Bamford}, Steven P. and {Nichol}, Robert C. and {Baldry}, Ivan K. and {Land}, Kate and {Lintott}, Chris J. and {Schawinski}, Kevin and {Slosar}, An{\v z}e and {Szalay}, Alexander S. and {Thomas}, Daniel and {Torki}, Mehri and {Andreescu}, Dan and {Edmondson}, Edward M. and {Miller}, Christopher J. and {Murray}, Phil and {Raddick}, M. Jordan and {Vandenberg}, Jan},
  title = {{Galaxy Zoo: the dependence of morphology and colour on environment}},
  journal = {MNRAS},
  year = {2009},
  month = {mar},
  volume = {393},
  number = {4},
  pages = {1324--1352},
  doi = {10.1111/j.1365-2966.2008.14252.x},
  url = {https://doi.org/10.1111/j.1365-2966.2008.14252.x},
  adsurl = {https://ui.adsabs.harvard.edu/abs/2009MNRAS.393.1324B}
}

@article{Skibba2009,
  author = {{Skibba}, Ramin A. and {Bamford}, Steven P. and {Nichol}, Robert C. and {Lintott}, Chris J. and {Andreescu}, Dan and {Edmondson}, Edward M. and {Murray}, Phil and {Raddick}, M. Jordan and {Schawinski}, Kevin and {Slosar}, An{\v z}e and {Szalay}, Alexander S. and {Thomas}, Daniel and {Vandenberg}, Jan},
  title = {{Galaxy Zoo: disentangling the environmental dependence of morphology and colour}},
  journal = {MNRAS},
  year = {2009},
  month = {oct},
  volume = {399},
  number = {2},
  pages = {966--982},
  doi = {10.1111/j.1365-2966.2009.15334.x},
  url = {https://doi.org/10.1111/j.1365-2966.2009.15334.x},
  adsurl = {https://ui.adsabs.harvard.edu/abs/2009MNRAS.399..966S}
}

@article{Bluck2019,
  author = {{Bluck}, Asa F. L. and {Bottrell}, Connor and {Teimoorinia}, Hossen and {Henriques}, Bruno M. B. and {Mendel}, J. Trevor and {Ellison}, Sara L. and {Thanjavur}, Karun and {Simard}, Luc and {Patton}, David R. and {Conselice}, Christopher J. and {Moreno}, Jorge and {Woo}, Joanna},
  title = {{What shapes a galaxy? - unraveling the role of mass, environment, and star formation in forming galactic structure}},
  journal = {MNRAS},
  year = {2019},
  month = {may},
  volume = {485},
  number = {1},
  pages = {666--696},
  doi = {10.1093/mnras/stz363},
  url = {https://doi.org/10.1093/mnras/stz363},
  adsurl = {https://ui.adsabs.harvard.edu/abs/2019MNRAS.485..666B}
}

@article{Astropy2013,
  author = {{Astropy Collaboration} and {Robitaille}, Thomas P. and {Tollerud}, Erik J. and {Greenfield}, Perry and {Droettboom}, Michael and {Bray}, Erik and {Aldcroft}, Tom and {Davis}, Matt and {Ginsburg}, Adam and {Price-Whelan}, Adrian M. and {Kerzendorf}, Wolfgang E. and {Conley}, Alexander and {Crighton}, Neil and {Barbary}, Kyle and {Muna}, Demitri and {Ferguson}, Henry and {Grollier}, Fr\'ed\'eric and {Parikh}, Madhura M. and {Nair}, Prasanth H. and {G\"unther}, Hans M. and {Deil}, Christoph and {Woillez}, Julien and {Conseil}, Simon and {Kramer}, Roban and {Turner}, James E. H. and {Singer}, Leo and {Fox}, Ryan and {Weaver}, Benjamin A. and {Zabalza}, Victor and {Edwards}, Zachary I. and {Azalee Bostroem}, K. and {Burke}, D. J. and {Casey}, Andrew R. and {Crawford}, Steven M. and {Dencheva}, Nadia and {Ely}, Justin and {Jenness}, Tim and {Labrie}, Kathleen and {Lim}, Pey Lian and {Pierfederici}, Francesco and {Pontzen}, Andrew and {Ptak}, Andy and {Refsdal}, Brian and {Servillat}, Mathieu and {Streicher}, Ole},
  title = {{Astropy: A community Python package for astronomy}},
  journal = {Astronomy \& Astrophysics},
  year = {2013},
  month = {oct},
  volume = {558},
  pages = {A33},
  doi = {10.1051/0004-6361/201322068},
  url = {https://doi.org/10.1051/0004-6361/201322068},
  adsurl = {https://ui.adsabs.harvard.edu/search/q=doi%3A%2210.1051%2F0004-6361%2F201322068%22}
}

@article{Astropy2018,
  author = {{Astropy Collaboration} and {Price-Whelan}, A. M. and {Sip\H{o}cz}, B. M. and {G\"unther}, H. M. and {Lim}, P. L. and {Crawford}, S. M. and {Conseil}, S. and {Shupe}, D. L. and {Craig}, M. W. and {Dencheva}, N. and {Ginsburg}, A. and {VanderPlas}, J. T. and {Bradley}, L. D. and {P\'erez-Su\'arez}, D. and {de Val-Borro}, M. and {(Primary Paper Contributors)} and {Aldcroft}, T. L. and {Cruz}, K. L. and {Robitaille}, T. P. and {Tollerud}, E. J. and {(Astropy Coordination Committee)} and {Ardelean}, C. and {Babej}, T. and {Bach}, Y. P. and {Bachetti}, M. and {Bakanov}, A. V. and {Bamford}, S. P. and {Barentsen}, G. and {Barmby}, P. and {Baumbach}, A. and {Berry}, K. L. and {Biscani}, F. and {Boquien}, M. and {Bostroem}, K. A. and {Bouma}, L. G. and {Brammer}, G. B. and {Bray}, E. M. and {Breytenbach}, H. and {Buddelmeijer}, H. and {Burke}, D. J. and {Calderone}, G. and {Rodr{\'\i}guez}, J. L. Cano and {Cara}, M. and {Cardoso}, J. V. M. and {Cheedella}, S. and {Copin}, Y. and {Corrales}, L. and {Crichton}, D. and {DAvella}, D. and {Deil}, C. and {Depagne}, \'E. and {Dietrich}, J. P. and {Donath}, A. and {Droettboom}, M. and {Earl}, N. and {Erben}, T. and {Fabbro}, S. and {Ferreira}, L. A. and {Finethy}, T. and {Fox}, R. T. and {Garrison}, L. H. and {Gibbons}, S. L. J. and {Goldstein}, D. A. and {Gommers}, R. and {Greco}, J. P. and {Greenfield}, P. and {Groener}, A. M. and {Grollier}, F. and {Hagen}, A. and {Hirst}, P. and {Homeier}, D. and {Horton}, A. J. and {Hosseinzadeh}, G. and {Hu}, L. and {Hunkeler}, J. S. and {Ivezi\'c}, \v{Z}. and {Jain}, A. and {Jenness}, T. and {Kanarek}, G. and {Kendrew}, S. and {Kern}, N. S. and {Kerzendorf}, W. E. and {Khvalko}, A. and {King}, J. and {Kirkby}, D. and {Kulkarni}, A. M. and {Kumar}, A. and {Lee}, A. and {Lenz}, D. and {Littlefair}, S. P. and {Ma}, Z. and {Macleod}, D. M. and {Mastropietro}, M. and {McCully}, C. and {Montagnac}, S. and {Morris}, B. M. and {Mueller}, M. and {Mumford}, S. J. and {Muna}, D. and {Murphy}, N. A. and {Nelson}, S. and {Nguyen}, G. H. and {Ninan}, J. P. and {N\"othe}, M. and {Ogaz}, S. and {Oh}, S. and {Parejko}, J. K. and {Parley}, N. and {Pascual}, S. and {Patil}, R. and {Patil}, A. A. and {Plunkett}, A. L. and {Prochaska}, J. X. and {Rastogi}, T. and {Janga}, V. Reddy and {Sabater}, J. and {Sakurikar}, P. and {Seifert}, M. and {Sherbert}, L. E. and {Sherwood-Taylor}, H. and {Shih}, A. Y. and {Sick}, J. and {Silbiger}, M. T. and {Singanamalla}, S. and {Singer}, L. P. and {Sladen}, P. H. and {Sooley}, K. A. and {Sornarajah}, S. and {Streicher}, O. and {Teuben}, P. and {Thomas}, S. W. and {Tremblay}, G. R. and {Turner}, J. E. H. and {Terr\'on}, V. and {Kerkwijk}, M. H. van and {de la Vega}, A. and {Watkins}, L. L. and {Weaver}, B. A. and {Whitmore}, J. B. and {Woillez}, J. and {Zabalza}, V. and {(Astropy Contributors)}},
  title = {{The Astropy Project: Building an Open-science Project and Status of the v2.0 Core Package}},
  journal = {AJ},
  year = {2018},
  month = {sep},
  volume = {156},
  number = {3},
  eid = {123},
  pages = {123},
  doi = {10.3847/1538-3881/aabc4f},
  url = {https://doi.org/10.3847/1538-3881/aabc4f},
  adsurl = {https://ui.adsabs.harvard.edu/search/q=doi%3A%2210.3847%2F1538-3881%2Faabc4f%22}
}

@article{Astropy2022,
  author = {{Astropy Collaboration} and {Price-Whelan}, Adrian M. and {Lim}, Pey Lian and {Earl}, Nicholas and {Starkman}, Nathaniel and {Bradley}, Larry and {Shupe}, David L. and {Patil}, Aarya A. and {Corrales}, Lia and {Brasseur}, C. E. and {N\"othe}, Maximilian and {Donath}, Axel and {Tollerud}, Erik and {Morris}, Brett M. and {Ginsburg}, Adam and {Vaher}, Eero and {Weaver}, Benjamin A. and {Tocknell}, James and {Jamieson}, William and {van Kerkwijk}, Marten H. and {Robitaille}, Thomas P. and {Merry}, Bruce and {Bachetti}, Matteo and {G\"unther}, H. Moritz and {Paper Authors} and {Aldcroft}, Thomas L. and {Alvarado-Montes}, Jaime A. and {Archibald}, Anne M. and {B\'odi}, Attila and {Bapat}, Shreyas and {Barentsen}, Geert and {Baz\'an}, Juanjo and {Biswas}, Manish and {Boquien}, M\'ed\'eric and {Burke}, D. J. and {Cara}, Daria and {Cara}, Mihai and {Conroy}, Kyle E and {Conseil}, Simon and {Craig}, Matthew W. and {Cross}, Robert M. and {Cruz}, Kelle L. and {DEugenio}, Francesco and {Dencheva}, Nadia and {Devillepoix}, Hadrien A. R. and {Dietrich}, J\"org P. and {Eigenbrot}, Arthur Davis and {Erben}, Thomas and {Ferreira}, Leonardo and {Foreman-Mackey}, Daniel and {Fox}, Ryan and {Freij}, Nabil and {Garg}, Suyog and {Geda}, Robel and {Glattly}, Lauren and {Gondhalekar}, Yash and {Gordon}, Karl D. and {Grant}, David and {Greenfield}, Perry and {Groener}, Austen M. and {Guest}, Steve and {Gurovich}, Sebastian and {Handberg}, Rasmus and {Hart}, Akeem and {Hatfield-Dodds}, Zac and {Homeier}, Derek and {Hosseinzadeh}, Griffin and {Jenness}, Tim and {Jones}, Craig K. and {Joseph}, Prajwel and {Kalmbach}, J. Bryce and {Karamehmetoglu}, Emir and {Ka{\l}uszy\'nski}, Miko{\l}aj and {Kelley}, Michael S. P. and {Kern}, Nicholas and {Kerzendorf}, Wolfgang E. and {Koch}, Eric W. and {Kulumani}, Shankar and {Lee}, Antony and {Ly}, Chun and {Ma}, Zhiyuan and {MacBride}, Conor and {Maljaars}, Jakob M. and {Muna}, Demitri and {Murphy}, N. A. and {Norman}, Henrik and {OSteen}, Richard and {Oman}, Kyle A. and {Pacifici}, Camilla and {Pascual}, Sergio and {Pascual-Granado}, J. and {Patil}, Rohit R. and {Perren}, Gabriel I and {Pickering}, Timothy E. and {Rastogi}, Tanuj and {Roulston}, Benjamin R. and {Ryan}, Daniel F and {Rykoff}, Eli S. and {Sabater}, Jose and {Sakurikar}, Parikshit and {Salgado}, Jes\'us and {Sanghi}, Aniket and {Saunders}, Nicholas and {Savchenko}, Volodymyr and {Schwardt}, Ludwig and {Seifert-Eckert}, Michael and {Shih}, Albert Y. and {Jain}, Anany Shrey and {Shukla}, Gyanendra and {Sick}, Jonathan and {Simpson}, Chris and {Singanamalla}, Sudheesh and {Singer}, Leo P. and {Singhal}, Jaladh and {Sinha}, Manodeep and {Sip\H{o}cz}, Brigitta M. and {Spitler}, Lee R. and {Stansby}, David and {Streicher}, Ole and {\v{S}umak}, Jani and {Swinbank}, John D. and {Taranu}, Dan S. and {Tewary}, Nikita and {Tremblay}, Grant R. and {Val-Borro}, Miguel de and {Van Kooten}, Samuel J. and {Vasovi\'c}, Zlatan and {Verma}, Shresth and {de Miranda Cardoso}, Jos\'e Vin{\'\i}cius and {Williams}, Peter K. G. and {Wilson}, Tom J. and {Winkel}, Benjamin and {Wood-Vasey}, W. M. and {Xue}, Rui and {Yoachim}, Peter and {Zhang}, Chen and {Zonca}, Andrea and {Astropy Project Contributors}},
  title = {{The Astropy Project: Sustaining and Growing a Community-oriented Open-source Project and the Latest Major Release (v5.0) of the Core Package}},
  journal = {ApJ},
  year = {2022},
  month = {aug},
  volume = {935},
  number = {2},
  eid = {167},
  pages = {167},
  doi = {10.3847/1538-4357/ac7c74},
  url = {https://doi.org/10.3847/1538-4357/ac7c74},
  adsurl = {https://ui.adsabs.harvard.edu/search/q=doi%3A%2210.3847%2F1538-4357%2Fac7c74%22}
}

@article{Harris2020,
  author = {{Harris}, Charles R. and {Millman}, K. Jarrod and {van der Walt}, St\'efan J. and {Gommers}, Ralf and {Virtanen}, Pauli and {Cournapeau}, David and {Wieser}, Eric and {Taylor}, Julian and {Berg}, Sebastian and {Smith}, Nathaniel J. and {Kern}, Robert and {Picus}, Matti and {Hoyer}, Stephan and {van Kerkwijk}, Marten H. and {Brett}, Matthew and {Haldane}, Allan and {del R{\'\i}o}, Jaime Fern\'andez and {Wiebe}, Mark and {Peterson}, Pearu and {G\'erard-Marchant}, Pierre and {Sheppard}, Kevin and {Reddy}, Tyler and {Weckesser}, Warren and {Abbasi}, Hameer and {Gohlke}, Christoph and {Oliphant}, Travis E.},
  title = {{Array programming with NumPy}},
  journal = {Nature},
  year = {2020},
  month = {sep},
  volume = {585},
  number = {7825},
  pages = {357--362},
  doi = {10.1038/s41586-020-2649-2},
  url = {https://doi.org/10.1038/s41586-020-2649-2},
  adsurl = {https://ui.adsabs.harvard.edu/search/q=doi%3A%2210.1038%2Fs41586-020-2649-2%22}
}

@article{Virtanen2020,
  author = {{Virtanen}, Pauli and {Gommers}, Ralf and {Oliphant}, Travis E. and {Haberland}, Matt and {Reddy}, Tyler and {Cournapeau}, David and {Burovski}, Evgeni and {Peterson}, Pearu and {Weckesser}, Warren and {Bright}, Jonathan and {van der Walt}, St\'efan J. and {Brett}, Matthew and {Wilson}, Joshua and {Millman}, K. Jarrod and {Mayorov}, Nikolay and {Nelson}, Andrew R. J. and {Jones}, Eric and {Kern}, Robert and {Larson}, Eric and {Carey}, C. J. and {Polat}, \.Ilhan and {Feng}, Yu and {Moore}, Eric W. and {VanderPlas}, Jake and {Laxalde}, Denis and {Perktold}, Josef and {Cimrman}, Robert and {Henriksen}, Ian and {Quintero}, E. A. and {Harris}, Charles R. and {Archibald}, Anne M. and {Ribeiro}, Ant\^onio H. and {Pedregosa}, Fabian and {van Mulbregt}, Paul and {SciPy 1.0 Contributors} and {Vijaykumar}, Aditya and {Bardelli}, Alessandro Pietro and {Rothberg}, Alex and {Hilboll}, Andreas and {Kloeckner}, Andreas and {Scopatz}, Anthony and {Lee}, Antony and {Rokem}, Ariel and {Woods}, C. Nathan and {Fulton}, Chad and {Masson}, Charles and {H\"aggstr\"om}, Christian and {Fitzgerald}, Clark and {Nicholson}, David A. and {Hagen}, David R. and {Pasechnik}, Dmitrii V. and {Olivetti}, Emanuele and {Martin}, Eric and {Wieser}, Eric and {Silva}, Fabrice and {Lenders}, Felix and {Wilhelm}, Florian and {Young}, G. and {Price}, Gavin A. and {Ingold}, Gert-Ludwig and {Allen}, Gregory E. and {Lee}, Gregory R. and {Audren}, Herv\'e and {Probst}, Irvin and {Dietrich}, J\"org P. and {Silterra}, Jacob and {Webber}, James T and {Slavi\v{c}}, Janko and {Nothman}, Joel and {Buchner}, Johannes and {Kulick}, Johannes and {Sch\"onberger}, Johannes L. and {de Miranda Cardoso}, Jos\'e Vin{\'\i}cius and {Reimer}, Joscha and {Harrington}, Joseph and {Rodr{\'\i}guez}, Juan Luis Cano and {Nunez-Iglesias}, Juan and {Kuczynski}, Justin and {Tritz}, Kevin and {Thoma}, Martin and {Newville}, Matthew and {K\"ummerer}, Matthias and {Bolingbroke}, Maximilian and {Tartre}, Michael and {Pak}, Mikhail and {Smith}, Nathaniel J. and {Nowaczyk}, Nikolai and {Shebanov}, Nikolay and {Pavlyk}, Oleksandr and {Brodtkorb}, Per A. and {Lee}, Perry and {McGibbon}, Robert T. and {Feldbauer}, Roman and {Lewis}, Sam and {Tygier}, Sam and {Sievert}, Scott and {Vigna}, Sebastiano and {Peterson}, Stefan and {More}, Surhud and {Pudlik}, Tadeusz and {Oshima}, Takuya and {Pingel}, Thomas J. and {Robitaille}, Thomas P. and {Spura}, Thomas and {Jones}, Thouis R. and {Cera}, Tim and {Leslie}, Tim and {Zito}, Tiziano and {Krauss}, Tom and {Upadhyay}, Utkarsh and {Halchenko}, Yaroslav O. and {V\'azquez-Baeza}, Yoshiki},
  title = {{SciPy 1.0: fundamental algorithms for scientific computing in Python}},
  journal = {Nature Methods},
  year = {2020},
  month = {mar},
  volume = {17},
  number = {3},
  pages = {261--272},
  doi = {10.1038/s41592-019-0686-2},
  url = {https://doi.org/10.1038/s41592-019-0686-2},
  adsurl = {https://ui.adsabs.harvard.edu/search/q=doi%3A%2210.1038%2Fs41592-019-0686-2%22}
}

@article{Hunter2007,
  author = {Hunter, J. D.},
  title = {{Matplotlib: A 2D Graphics Environment}},
  journal = {Computing in Science \& Engineering},
  year = {2007},
  volume = {9},
  number = {3},
  pages = {90--95},
  doi = {10.1109/MCSE.2007.55},
  url = {https://doi.org/10.1109/MCSE.2007.55},
  adsurl = {https://ui.adsabs.harvard.edu/search/q=doi%3A%2210.1109%2FMCSE.2007.55%22}
}

@INPROCEEDINGS{Taylor2005,
       author = {{Taylor}, M.~B.},
        title = "{TOPCAT \& STIL: Starlink Table/VOTable Processing Software}",
    booktitle = {Astronomical Data Analysis Software and Systems XIV},
         year = 2005,
       editor = {{Shopbell}, P. and {Britton}, M. and {Ebert}, R.},
       series = {Astronomical Society of the Pacific Conference Series},
       volume = {347},
        month = dec,
        pages = {29},
       adsurl = {https://ui.adsabs.harvard.edu/abs/2005ASPC..347...29T}
}

@article{AragonCalvo2019,
  author = {Arag\'on-Calvo, M. A. and Neyrinck, M. C. and Silk, J.},
  title = {{Galaxy Quenching from Cosmic Web Detachment}},
  journal = {The Open Journal of Astrophysics},
  year = {2019},
  month = {jul},
  volume = {2},
  number = {1},
  pages = {7},
  doi = {10.21105/astro.1607.07881},
  url = {https://doi.org/10.21105/astro.1607.07881},
  adsurl = {https://ui.adsabs.harvard.edu/search/q=doi%3A%2210.21105%2Fastro.1607.07881%22}
}

@article{GalarraEspinosa2023,
  author = {{Gal\'arraga-Espinosa}, Daniela and {Garaldi}, Enrico and {Kauffmann}, Guinevere},
  title = {{Flows around galaxies: I. The dependence of galaxy connectivity on cosmic environments and effects on the star formation rate}},
  journal = {Astronomy \& Astrophysics},
  year = {2023},
  month = {mar},
  volume = {671},
  pages = {A160},
  doi = {10.1051/0004-6361/202244935},
  url = {https://doi.org/10.1051/0004-6361/202244935},
  adsurl = {https://ui.adsabs.harvard.edu/search/q=doi%3A%2210.1051%2F0004-6361%2F202244935%22}
}

@article{Sarron2019,
  author = {{Sarron}, F. and {Adami}, C. and {Durret}, F. and {Laigle}, C.},
  title = {{Pre-processing of galaxies in cosmic filaments around AMASCFI clusters in the CFHTLS}},
  journal = {Astronomy \& Astrophysics},
  year = {2019},
  month = {dec},
  volume = {632},
  pages = {A49},
  doi = {10.1051/0004-6361/201935394},
  url = {https://doi.org/10.1051/0004-6361/201935394},
  adsurl = {https://ui.adsabs.harvard.edu/search/q=doi%3A%2210.1051%2F0004-6361%2F201935394%22}
}

@article{Yesuf2022,
  author = {Yesuf, H. M.},
  title = {{Quenching in the Right Place at the Right Time: Tracing the Shared History of Starbursts, Active Galactic Nuclei, and Poststarburst Galaxies Using Their Structures and Multiscale Environments}},
  journal = {ApJ},
  year = {2022},
  month = {sep},
  volume = {936},
  number = {2},
  eid = {124},
  pages = {124},
  doi = {10.3847/1538-4357/ac83b0},
  url = {https://doi.org/10.3847/1538-4357/ac83b0},
  adsurl = {https://ui.adsabs.harvard.edu/search/q=doi%3A%2210.3847%2F1538-4357%2Fac83b0%22}
}

@article{Woo2013,
  author = {{Woo}, Joanna and {Dekel}, Avishai and {Faber}, S. M. and {Noeske}, Kai and {Koo}, David C. and {Gerke}, Brian F. and {Cooper}, Michael C. and {Salim}, Samir and {Dutton}, Aaron A. and {Newman}, Jeffrey and {Weiner}, Benjamin J. and {Bundy}, Kevin and {Willmer}, Christopher N. A. and {Davis}, Marc and {Yan}, Renbin},
  title = {{Dependence of galaxy quenching on halo mass and distance from its centre}},
  journal = {MNRAS},
  year = {2013},
  month = {feb},
  volume = {428},
  number = {4},
  pages = {3306--3326},
  doi = {10.1093/mnras/sts274},
  url = {https://doi.org/10.1093/mnras/sts274},
  adsurl = {https://ui.adsabs.harvard.edu/search/q=doi%3A%2210.1093%2Fmnras%2Fsts274%22}
}

@article{Woo2017,
  author = {{Woo}, Joanna and {Carollo}, C. M. and {Faber}, S. M. and {Dekel}, Avishai and {Tacchella}, Sandro},
  title = {{Satellite quenching, galaxy inner density and the halo environment}},
  journal = {MNRAS},
  year = {2017},
  month = {jan},
  volume = {464},
  number = {1},
  pages = {1077--1094},
  doi = {10.1093/mnras/stw2403},
  url = {https://doi.org/10.1093/mnras/stw2403},
  adsurl = {https://ui.adsabs.harvard.edu/search/q=doi%3A%2210.1093%2Fmnras%2Fstw2403%22}
}

@article{Mancini2019,
  author = {{Mancini}, Chiara and {Daddi}, Emanuele and {Juneau}, St\'ephanie and {Renzini}, Alvio and {Rodighiero}, Giulia and {Cappellari}, Michele and {Rodr{\'\i}guez-Mu\~noz}, Luc{\'\i}a and {Liu}, Daizhong and {Pannella}, Maurilio and {Baronchelli}, Ivano and {Franceschini}, Alberto and {Bergamini}, Pietro and {DEugenio}, Chiara and {Puglisi}, Annagrazia},
  title = {{Rejuvenated galaxies with very old bulges at the origin of the bending of the main sequence and of the `green valley'}},
  journal = {MNRAS},
  year = {2019},
  month = {oct},
  volume = {489},
  number = {1},
  pages = {1265--1290},
  doi = {10.1093/mnras/stz2130},
  url = {https://doi.org/10.1093/mnras/stz2130},
  adsurl = {https://ui.adsabs.harvard.edu/search/q=doi%3A%2210.1093%2Fmnras%2Fstz2130%22}
}

@article{Seppi2022,
  author = {{Seppi}, R. and {Comparat}, J. and {Bulbul}, E. and {Nandra}, K. and {Merloni}, A. and {Clerc}, N. and {Liu}, T. and {Ghirardini}, V. and {Liu}, A. and {Salvato}, M. and {Sanders}, J. S. and {Wilms}, J. and {Dwelly}, T. and {Dauser}, T. and {K{\"o}nig}, O. and {Ramos-Ceja}, M. E. and {Garrel}, C. and {Reiprich}, T. H.},
  title = {{Detecting clusters of galaxies and active galactic nuclei in an eROSITA all-sky survey digital twin}},
  journal = {Astronomy \& Astrophysics},
  year = {2022},
  month = {sep},
  volume = {665},
  eid = {A78},
  pages = {A78},
  doi = {10.1051/0004-6361/202243824},
  url = {https://doi.org/10.1051/0004-6361/202243824},
  adsurl = {https://ui.adsabs.harvard.edu/abs/2022A&A...665A..78S}
}

@article{Kleinebreil2025,
  author = {{Kleinebreil}, F. and {Grandis}, S. and {Schrabback}, T. and {Ghirardini}, V. and {Chiu}, I. and {Liu}, A. and {Kluge}, M. and {Reiprich}, T. H. and {Artis}, E. and {Bahar}, Y. E. and {Balzer}, F. and {Bulbul}, E. and {Clerc}, N. and {Comparat}, J. and {Garrel}, C. and {Gruen}, D. and {Li}, X. and {Miyatake}, H. and {Miyazaki}, S. and {Ramos-Ceja}, M. E. and {Sanders}, J. and {Seppi}, R. and {Okabe}, N. and {Zhang}, X.},
  title = {{The SRG/eROSITA All-Sky Survey: Weak lensing of eRASS1 galaxy clusters in KiDS-1000 and consistency checks with DES Y3 and HSC-Y3}},
  journal = {Astronomy \& Astrophysics},
  year = {2025},
  month = {mar},
  volume = {695},
  pages = {A216},
  doi = {10.1051/0004-6361/202449599},
  url = {https://doi.org/10.1051/0004-6361/202449599},
  adsurl = {https://ui.adsabs.harvard.edu/search/q=doi%3A%2210.1051%2F0004-6361%2F202449599%22}
}

@article{Chamba2024,
  author = {{Chamba}, Nushkia and {Hayes}, Matthew J. and {{LSST Dark Energy Science Collaboration}}},
  title = {{The impact of environment on size: Galaxies are 50\% smaller in the Fornax Cluster compared to the field}},
  journal = {Astronomy \& Astrophysics},
  year = {2024},
  month = {sep},
  volume = {689},
  eid = {A28},
  pages = {A28},
  doi = {10.1051/0004-6361/202348536},
  url = {https://doi.org/10.1051/0004-6361/202348536},
  adsurl = {https://ui.adsabs.harvard.edu/abs/2024A&A...689A..28C}
}

@article{Balogh2004,
  author = {{Balogh}, Michael L. and {Baldry}, Ivan K. and {Nichol}, Robert and {Miller}, Christopher and {Bower}, Richard and {Glazebrook}, Karl},
  title = {{The Bimodal Galaxy Color Distribution: Dependence on Luminosity and Environment}},
  journal = {ApJ},
  year = {2004},
  month = {nov},
  volume = {615},
  number = {2},
  pages = {L101--L104},
  doi = {10.1086/426079},
  url = {https://doi.org/10.1086/426079},
  adsurl = {https://ui.adsabs.harvard.edu/abs/2004ApJ...615L.101B}
}

@article{Cortese2021,
  author = {{Cortese}, Luca and {Catinella}, Barbara and {Smith}, Rory},
  title = {{The Dawes Review 9: The role of cold gas stripping on the star formation quenching of satellite galaxies}},
  journal = {PASA},
  year = {2021},
  volume = {38},
  eid = {e035},
  pages = {e035},
  doi = {10.1017/pasa.2021.18},
  url = {https://doi.org/10.1017/pasa.2021.18},
  adsurl = {https://ui.adsabs.harvard.edu/abs/2021PASA...38...35C}
}

@article{Jian2018,
  author = {{Jian}, Hung-Yu and {Lin}, Lihwai and {Oguri}, Masamune and {Nishizawa}, Atsushi J. and {Takada}, Masahiro and {More}, Surhud and {Koyama}, Yusei and {Tanaka}, Masayuki and {Komiyama}, Yutaka},
  title = {{First results on the cluster galaxy population from the Subaru Hyper Suprime-Cam survey. I. The role of group or cluster environment in star formation quenching from z = 0.2 to 1.1}},
  journal = {PASJ},
  year = {2018},
  month = {jan},
  volume = {70},
  number = {SP1},
  eid = {S23},
  pages = {S23},
  doi = {10.1093/pasj/psx096},
  url = {https://doi.org/10.1093/pasj/psx096},
  adsurl = {https://ui.adsabs.harvard.edu/abs/2018PASJ...70S..23J}
}

@article{Haines2015,
  author = {{Haines}, C. P. and {Pereira}, M. J. and {Smith}, G. P. and {Egami}, E. and {Babul}, A. and {Finoguenov}, A. and {Ziparo}, F. and {McGee}, S. L. and {Rawle}, T. D. and {Okabe}, N. and {Moran}, S. M.},
  title = {{LoCuSS: The Slow Quenching of Star Formation in Cluster Galaxies and the Need for Pre-processing}},
  journal = {ApJ},
  year = {2015},
  month = {jun},
  volume = {806},
  number = {1},
  eid = {101},
  pages = {101},
  doi = {10.1088/0004-637X/806/1/101},
  url = {https://doi.org/10.1088/0004-637X/806/1/101},
  adsurl = {https://ui.adsabs.harvard.edu/abs/2015ApJ...806..101H}
}

@article{Coenda2018,
  author = {{Coenda}, Valeria and {Mart{\'i}nez}, H{\'e}ctor J. and {Muriel}, Hern{\'a}n},
  title = {{Green valley galaxies as a transition population in different environments}},
  journal = {MNRAS},
  year = {2018},
  month = {feb},
  volume = {473},
  number = {4},
  pages = {5617--5629},
  doi = {10.1093/mnras/stx2707},
  url = {https://doi.org/10.1093/mnras/stx2707},
  adsurl = {https://ui.adsabs.harvard.edu/abs/2018MNRAS.473.5617C}
}

@article{Jian2020,
  author = {{Jian}, Hung-Yu and {Lin}, Lihwai and {Koyama}, Yusei and {Tanaka}, Ichi and {Umetsu}, Keiichi and {Hsieh}, Bau-Ching and {Higuchi}, Yuichi and {Oguri}, Masamune and {More}, Surhud and {Komiyama}, Yutaka and {Kodama}, Tadayuki and {Nishizawa}, Atsushi J. and {Chang}, Yu-Yen},
  title = {{Redshift Evolution of Green Valley Galaxies in Different Environments from the Hyper Suprime-Cam Survey}},
  journal = {ApJ},
  year = {2020},
  month = {may},
  volume = {894},
  number = {2},
  eid = {125},
  pages = {125},
  doi = {10.3847/1538-4357/ab86a8},
  url = {https://doi.org/10.3847/1538-4357/ab86a8},
  adsurl = {https://ui.adsabs.harvard.edu/abs/2020ApJ...894..125J}
}

@article{McNab2021,
  author = {{McNab}, Karen and {Balogh}, Michael L. and {van der Burg}, Remco F. J. and {Forestell}, Anya and {Webb}, Kristi and {Vulcani}, Benedetta and {Rudnick}, Gregory and {Muzzin}, Adam and {Cooper}, Michael C. and {McGee}, Sean and {Biviano}, Andrea and {Cerulo}, Pierluigi and {Chan}, Jeffrey C. C. and {De Lucia}, Gabriella and {Demarco}, Ricardo and {Finoguenov}, Alexis and {Forrest}, Ben and {Golledge}, Caelan and {Jablonka}, Pascale and {Lidman}, Chris and {Nantais}, Julie and {Old}, Lyndsay and {Pintos-Castro}, Irene and {Poggianti}, Bianca and {Reeves}, Andrew M. M. and {Wilson}, Gillian and {Yee}, Howard K. C. and {Zaritsky}, Dennis},
  title = {{The GOGREEN survey: transition galaxies and the evolution of environmental quenching}},
  journal = {MNRAS},
  year = {2021},
  month = {sep},
  volume = {508},
  number = {1},
  pages = {157--174},
  doi = {10.1093/mnras/stab2558},
  url = {https://doi.org/10.1093/mnras/stab2558},
  adsurl = {https://ui.adsabs.harvard.edu/abs/2021MNRAS.508..157M}
}

@article{Sobral2022,
  author = {{Sobral}, David and {van der Wel}, Arjen and {Bezanson}, Rachel and {Bell}, Eric F. and {Muzzin}, Adam and {D'Eugenio}, Francesco and {Darvish}, Behnam and {Gallazzi}, Anna and {Wu}, Po-Feng and {Maseda}, Michael and {Matthee}, Jorryt and {Paulino-Afonso}, Ana and {Straatman}, Caroline and {van Dokkum}, Pieter G.},
  title = {{The LEGA-C of Nature and Nurture in Stellar Populations at z~0.6-1.0: Dn4000 and Hdelta Reveal Different Assembly Histories for Quiescent Galaxies in Different Environments}},
  journal = {ApJ},
  year = {2022},
  month = {feb},
  volume = {926},
  number = {2},
  eid = {117},
  pages = {117},
  doi = {10.3847/1538-4357/ac4419},
  url = {https://doi.org/10.3847/1538-4357/ac4419},
  adsurl = {https://ui.adsabs.harvard.edu/abs/2022ApJ...926..117S}
}

@article{HuertasCompany2013,
  author = {{Huertas-Company}, M. and {Mei}, S. and {Shankar}, F. and {Delaye}, L. and {Raichoor}, A. and {Covone}, G. and {Finoguenov}, A. and {Kneib}, J.-P. and {Le F{\`e}vre}, O. and {Povi{\'c}}, M.},
  title = {{The evolution of the mass-size relation for early-type galaxies from z~1 to the present: dependence on environment, mass range and detailed morphology}},
  journal = {MNRAS},
  year = {2013},
  month = {jan},
  volume = {428},
  number = {2},
  pages = {1715--1742},
  doi = {10.1093/mnras/sts150},
  url = {https://doi.org/10.1093/mnras/sts150},
  adsurl = {https://ui.adsabs.harvard.edu/abs/2013MNRAS.428.1715H}
}

@article{Strazzullo2023,
  author = {{Strazzullo}, Veronica and {Pannella}, Maurilio and {Mohr}, J. J. and {Saro}, A. and {Ashby}, M. L. N. and {Bayliss}, M. B. and {Canning}, R. E. A. and {Floyd}, B. and {Gonzalez}, A. H. and {Khullar}, G. and {Kim}, K. J. and {McDonald}, M. and {Reichardt}, C. L. and {Sharon}, K. and {Somboonpanyakul}, T.},
  title = {{Galaxy populations in the most distant SPT-SZ clusters. II. Galaxy structural properties in massive clusters at 1.4 <~ z <~ 1.7}},
  journal = {Astronomy \& Astrophysics},
  year = {2023},
  month = {jan},
  volume = {669},
  eid = {A131},
  pages = {A131},
  doi = {10.1051/0004-6361/202245268},
  url = {https://doi.org/10.1051/0004-6361/202245268},
  adsurl = {https://ui.adsabs.harvard.edu/abs/2023A&A...669A.131S}
}

@article{Chan2018,
  author = {{Chan}, Jeffrey C. C. and {Beifiori}, Alessandra and {Saglia}, Roberto P. and {Mendel}, J. Trevor and {Stott}, John P. and {Bender}, Ralf and {Galametz}, Audrey and {Wilman}, David J. and {Cappellari}, Michele and {Davies}, Roger L. and {Houghton}, Ryan C. W. and {Prichard}, Laura J. and {Lewis}, Ian J. and {Sharples}, Ray and {Wegner}, Michael},
  title = {{The KMOS Cluster Survey (KCS). II. The Effect of Environment on the Structural Properties of Massive Cluster Galaxies at Redshift 1.39 < z < 1.61}},
  journal = {ApJ},
  year = {2018},
  month = {mar},
  volume = {856},
  number = {1},
  eid = {8},
  pages = {8},
  doi = {10.3847/1538-4357/aaadb4},
  url = {https://doi.org/10.3847/1538-4357/aaadb4},
  adsurl = {https://ui.adsabs.harvard.edu/abs/2018ApJ...856....8C}
}

@article{Andreon2018,
  author = {{Andreon}, Stefano},
  title = {{The cosmic epoch dependence of environmental effects on size evolution of red-sequence early-type galaxies}},
  journal = {Astronomy \& Astrophysics},
  year = {2018},
  month = {sep},
  volume = {617},
  eid = {A53},
  pages = {A53},
  doi = {10.1051/0004-6361/201832627},
  url = {https://doi.org/10.1051/0004-6361/201832627},
  adsurl = {https://ui.adsabs.harvard.edu/abs/2018A&A...617A..53A}
}

@article{Matharu2020,
  author = {{Matharu}, J. and {Muzzin}, A. and {Brammer}, G. B. and {van der Burg}, R. F. J. and {Auger}, M. W. and {Hewett}, P. C. and {Chan}, J. C. C. and {Demarco}, R. and {van Dokkum}, P. and {Marchesini}, D. and {Nelson}, E. J. and {Noble}, A. G. and {Wilson}, G.},
  title = {{HST/WFC3 grism observations of z~1 clusters: evidence for evolution in the mass-size relation of quiescent galaxies from post-starburst galaxies}},
  journal = {MNRAS},
  year = {2020},
  month = {apr},
  volume = {493},
  number = {4},
  pages = {6011--6032},
  doi = {10.1093/mnras/staa610},
  url = {https://doi.org/10.1093/mnras/staa610},
  adsurl = {https://ui.adsabs.harvard.edu/abs/2020MNRAS.493.6011M}
}

@article{Bluck2016,
  author = {{Bluck}, Asa F. L. and {Mendel}, J. Trevor and {Ellison}, Sara L. and {Patton}, David R. and {Simard}, Luc and {Henriques}, Bruno M. B. and {Torrey}, Paul and {Teimoorinia}, Hossen and {Moreno}, Jorge and {Starkenburg}, Else},
  title = {{The impact of galactic properties and environment on the quenching of central and satellite galaxies: a comparison between SDSS, Illustris and L-Galaxies}},
  journal = {MNRAS},
  year = {2016},
  month = {nov},
  volume = {462},
  number = {3},
  pages = {2559--2586},
  doi = {10.1093/mnras/stw1665},
  url = {https://doi.org/10.1093/mnras/stw1665},
  adsurl = {https://ui.adsabs.harvard.edu/abs/2016MNRAS.462.2559B}
}

@article{LacknerGunn2013,
  author = {{Lackner}, C. N. and {Gunn}, J. E.},
  title = {{The effect of environment on discs and bulges}},
  journal = {MNRAS},
  year = {2013},
  month = {jan},
  volume = {428},
  number = {3},
  pages = {2141--2162},
  doi = {10.1093/mnras/sts179},
  url = {https://doi.org/10.1093/mnras/sts179},
  adsurl = {https://ui.adsabs.harvard.edu/abs/2013MNRAS.428.2141L}
}

@BOOK{Sersic1968,
       author = {{Sersic}, Jose Luis},
        title = "{Atlas de Galaxias Australes}",
         year = 1968,
       adsurl = {https://ui.adsabs.harvard.edu/abs/1968adga.book.....S}
}
\bibliographystyle{aasjournalv7}



\end{document}